\documentclass[aps,prd,10pt,notitlepage,nofootinbib,superscriptaddress,showkeys,showpacs]{revtex4-1}

\usepackage{amstext,amsmath,amssymb,amsfonts,bbm}
\usepackage[latin1]{inputenc}
\usepackage{epsfig}
\usepackage{hyperref}
\usepackage{amsthm}
\usepackage{subfigure}
\usepackage{color}
\usepackage{multirow}

\usepackage{latexsym}

\usepackage{cancel}

\newcommand{\bea}{\begin{eqnarray}}	
\newcommand{\eea}{\end{eqnarray}}
\newcommand{\beq}{\begin{equation}}	
\newcommand{\eeq}{\end{equation}}

\newcommand{\cG}{{\mathcal G}}
\newcommand{\cS}{{\mathcal S}}
\newcommand{\cF}{{\mathcal F}}
\newcommand{\bG}{{\partial\mathcal G}}
\newcommand{\tJ}{{\widetilde{J}}}

\newcommand{\cexG}{\mathcal G_{\text{color}}}
\newcommand{\bJ}{ J_{\partial} }

\newcommand{\ren}{\text{ren\,}}

\newcommand{\kin}{\text{kin\,}}

\newcommand{\ext}{\text{ext\,}}

\newcommand{\lra}{ \leftrightarrow } 

\newtheorem{lemma}{Lemma}
\newtheorem{corollary}{Corollary}

\newtheorem{definition}{Definition}
\newtheorem{theorem}{Theorem}

\begin{document}

\title{Two and four-loop $\beta$-functions of rank 4 renormalizable
tensor field theories}

\author{Joseph Ben Geloun}\email{jbengeloun@perimeterinstitute.ca}

\affiliation{Perimeter Institute for Theoretical Physics, 31 Caroline
St, Waterloo, ON, Canada \\
International Chair in Mathematical Physics and Applications, 
ICMPA-UNESCO Chair, 072BP50, Cotonou, Rep. of Benin}

\date{\small\today}

\begin{abstract}
\noindent 
A recent rank 4 tensor field model generating 4D simplicial 
manifolds has been proved to be renormalizable at all orders of 
perturbation theory [arXiv:1111.4997 [hep-th]]. 
The  model is built out of $\phi^6$ ($\phi^6_{(1/2)}$), $\phi^4$ 
($\phi^4_{(1)}$) interactions and an anomalous term ($\phi^4_{(2)}$). 
The $\beta$-functions of this model are evaluated  
 at two and four loops.  We find that the model is asymptotically free in the UV for
both the main $\phi^6_{(1/2)}$ interactions whereas it is safe in the 
$\phi^4_{(1)}$ sector. The remaining anomalous term turns out
to possess a Landau ghost.

\medskip
\noindent Pacs numbers: 11.10.Gh, 04.60.-m, 02.10.Ox
pi-qg-278 and ICMPA/MPA/2012/009 \\

\end{abstract}

\maketitle

\section{Introduction}

The mid 80's has witnessed significant developments 
on quantum gravity (QG) in 2D through matrix models. These models appear to be appropriate candidates achieving a discrete version of the sum of geometries and topologies of surfaces through a sum over random triangulations  
 \cite{Di Francesco:1993nw}. 
One of the main tools
in order to perform analytically the statistical analysis of these
models and their different continuum limits is  
the $1/N$ expansion of t'Hooft. 
In the large $N$ (matrix size) limit,  only dominate in the partition
function planar graphs triangulating  surfaces
of genus zero. Higher dimensional extensions of these 2D models which were naturally called  {\it tensor models} 
with relevance for 3D and 4D gravity, turn out to be a far greater challenge \cite{ambj3dqg,mmgravity,sasa1,Boul,oriti}. 
The crucial $1/N$ expansion providing a control on the topology 
of simplices was missing for models generating simplicial manifolds
 in higher dimensions. In last resort, main results on tensor models then relied on numerics. 

Recently important progresses on this latter point have been made. 
The tensor analogue of the $1/N$ expansion has been found
  \cite{Gur3,GurRiv,Gur4} for a special class of models called colored
discovered by Gurau \cite{color,Gurau:2009tz,Gurau:2010nd}. 
The prominent feature in this expansion is that 
the dominant contributions in the partition function are dual to spheres
thus generalizing surfaces of genus zero in this higher dimensional context (see \cite{Gurau:2011xp} for a review on colored models). 
From this breakthrough, one acknowledges interesting achievements on the statistical
analysis  around tensor models \cite{Bonzom:2011zz,Bonzom:2011ev,Benedetti:2011nn,
Gurau:2011sk,Bonzom:2012sz,Bonzom:2012hw,Bonzom:2012qx} as well as on longstanding mathematical physics questions \cite{Gurau:2011tj,Gurau:2011kk,Gurau:2012ix}. 
These results have given birth to a new framework, the so-called  Tensor Field Theory approach for QG \cite{Rivasseau:2011xg,Rivasseau:2011hm} which combines tensor interactions and quantum field theory propagators
to formulate a Renormalization Group (RG) based scenario for QG in higher dimensions.

One point should be stressed in a straightforward manner:
tensor models of this kind are  combinatorial models
generating topological spaces and, although they should belong 
to the scenario of an emergent theory for gravity, 
their connection with a full-fledged quantization of General Relativity (GR)
is  not well understood at this stage.  Imposing particular conditions on the tensors may convey these models  presently discussed closer to  what can be expected 
from a quantization of topological BF theory \cite{Boul,oriti} 
which after further constraints leads to the quantization of GR. 
Hence, the deeper understanding of these models could be useful for the randomization of geometry.  
Besides, they possess a number of interesting properties worthy to be studied in details. Indeed, in addition of all important features 
aforementioned, this class of tensor models generates, in the correct truncation and for the first time, a renormalizable theory for quantum topology in 3 and 4D \cite{arXiv:1111.4997,BenGeloun:2012pu}. 

The model considered in \cite{arXiv:1111.4997} is a dynamical rank 4 tensor model over $T_4 \equiv U(1)^4$ built with $\phi^6$ and $\phi^4$
interactions (including one anomalous term). It addresses the generation of 4D simplicial (pseudo-)manifolds in an Euclidean path integral formalism. 
The three ingredients of perturbative renormalization 
at all orders \cite{Rivasseau:1991ub} have been identified: (1) A multi-scale analysis showed that slices can be understood as in the ordinary situation: high scales mean high momenta meaning small distances
on the torus; (2) A power counting theorem generalizing
known power countings for the local $\phi^4_4$
and the $\phi^4$ Grosse-Wulkenhaar  matrix model \cite{Grosse:2004yu,Rivasseau:2007ab} 
and (3) a generalized locality principle yielding 
a characterization of the most divergent contributions
which are of the form of terms included in the initial Lagrangian. 
A rank 3 analogue model was investigated in \cite{BenGeloun:2012pu}. 
This last model  also proves to be renormalizable at all orders and, 
by computing its one-loop $\beta$-function, turns out to asymptotically free in the UV. In other words, the latter statement claims that, in the UV limit, the theory describes the dynamics of non interacting three dimensional objects with the sphere topology. 

In this work, we investigate the $\beta$-functions 
related to all coupling constants of the 4D model defined in \cite{arXiv:1111.4997}. Two-loop computations 
are sufficient for some couplings whereas, 
for some other couplings, four-loop calculations
are required in order to understand the UV behaviour
of the model. One needs to go beyond one-loop calculations
in order to understand the RG flows due to the presence of the $\phi^6$ nonlocal interactions.
 We prove that the model is asymptotically free in the UV
that is, there exists a UV fixed manifold associated with this theory
defined by 
\beq
\lambda_{6}  = 0 \qquad\qquad \forall \lambda_{4;1}  \quad\qquad 
\lambda_{4;2} =0 
\eeq
where $\lambda_6$ represents any coupling constant of the $\phi^6$
interactions, $\lambda_{4;1}$  represents any coupling constant
of the $\phi^4$ interactions and
 $\lambda_{4;2}$ the coupling constant the anomalous term 
of the form $(\phi^2)^2$. 
Perturbing the system around this fixed manifold
\beq
\lambda'_{6}  = \lambda_6 + \epsilon \qquad 
\forall\lambda_{4;1} \qquad 
\lambda'_{4;2} =\lambda_{4;2} + \epsilon'
\eeq
for small quantities $\epsilon$ and $\epsilon'$,
then $\lambda'_{6}$, $\lambda_{4;1}$
and $\lambda'_{4;2}$ increase in the infrared (IR).
These are the main results of this paper.

The plan of this paper is as follows: The next section presents
the model and reviews its power counting theorem.
Section III investigates in details the two and four-loop $\beta$-functions 
of the enlarged model incorporating fourteen plus one different couplings associated with all interactions.
A conclusion follows in Section IV and an appendix
gathers the proofs of different lemmas and important steps in the calculations.

\section{The model and its renormalizability: An overview}

This section yields, in a streamlined analysis, a review of the model as defined in \cite{arXiv:1111.4997} and its power counting theorem
which will be used at each step of the rest of the paper. 

Let us consider a fourth rank complex tensor field  over the group
$U(1)$, $\varphi: U(1)^4 \to \mathbb{C}$. This field can be decomposed
in Fourier modes as
\bea
\varphi(h_1,h_2,h_3,h_4)
= \sum_{p_j \in \mathbb{Z}} \varphi_{[p_j]} e^{ip_1 \theta_1} e^{ip_2 \theta_2}
 e^{ip_3 \theta_3} e^{ip_4 \theta_4}
\eea
where  the group elements $h_i \in U(1)$, $\theta_i \in [0,2\pi)$ 
and $[p_j]= [p_1,p_2,p_3,p_4]$ are momentum indices. 
 We will adopt the notation $\varphi_{[p_1,p_2,p_3,p_4]}=$
$ \varphi_{1,2,3,4}$. Note that no symmetry under permutation of arguments 
is assumed for the tensor $\varphi_{[p_j]}$.

The action is defined by the kinetic term given in momentum space as
\bea
S^{\kin,0} =
 \sum_{p_{j}}
\bar\varphi_{1,2,3,4}
\Big(\sum_{s=1}^4 (p_{s})^2 + m^2\Big)\varphi_{1,2,3,4} 
\eea
where the sum is performed over all momentum values $p_j$.
Clearly, such a kinetic term is inferred from a Laplacian dynamics 
acting on the strand index $s$. It could be interesting
to find in which sense the above Laplacian dynamics 
might be related to an Osterwalder-Schrader positivity axiom
\cite{vinc}. 
Other motivations on 
the introduction of such a kinetic term can be found 
in \cite{Geloun:2011cy}. 
The corresponding Gaussian measure 
of covariance $C = (\sum_s p^2_s + m^2)^{-1}$ is noted as $d\mu_C$.

\begin{figure}
 \centering
     \begin{minipage}[t]{.8\textwidth}
      \centering
\includegraphics[angle=0, width=10cm, height=3cm]{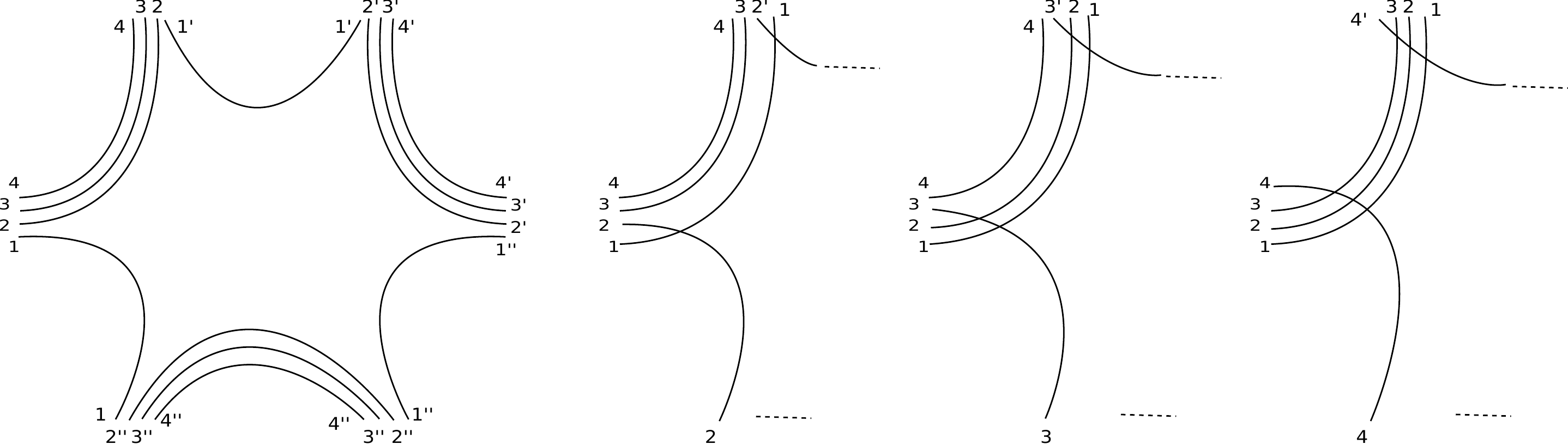} \\
\vspace{0.8cm}
\includegraphics[angle=0, width=14cm, height=3cm]{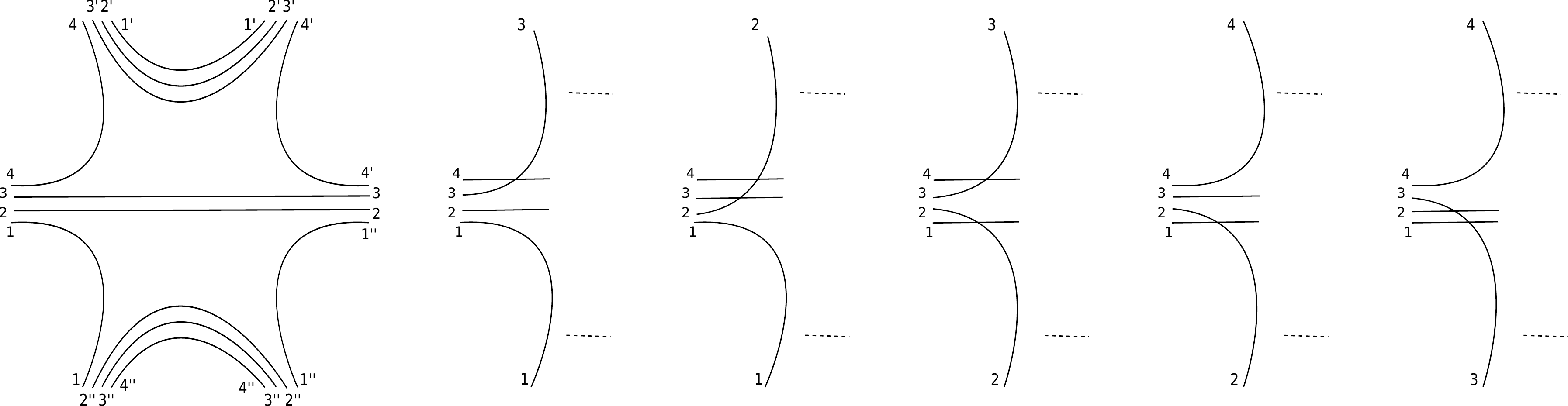}
\vspace{0.5cm}
\caption{ {\small Vertices of the  type $\phi^6_{(1)}$  (top, parametrized by $\rho=1,\dots,4$)  and of the type $\phi^6_{(2)}$ (bottom, parametrized by $\rho\rho'=12,\dots,34$). \label{vert6}}}
\end{minipage}
\put(-310,-12){$\rho=1$}
\put(-222,-12){$\rho=2$}
\put(-162,-12){$\rho=3$}
\put(-103,-12){$\rho=4$}
\put(-373,-118){$\rho\rho'=14$}
\put(-280,-118){$\rho\rho'=13$}
\put(-220,-118){$\rho\rho'=12$}
\put(-160,-118){$\rho\rho'=23$}
\put(-100,-118){$\rho\rho'=24$}
\put(-40,-118){$\rho\rho'=34$}
\end{figure}

\begin{figure}
 \centering
     \begin{minipage}[t]{.8\textwidth}
      \centering
\includegraphics[angle=0, width=14cm, height=3.2cm]{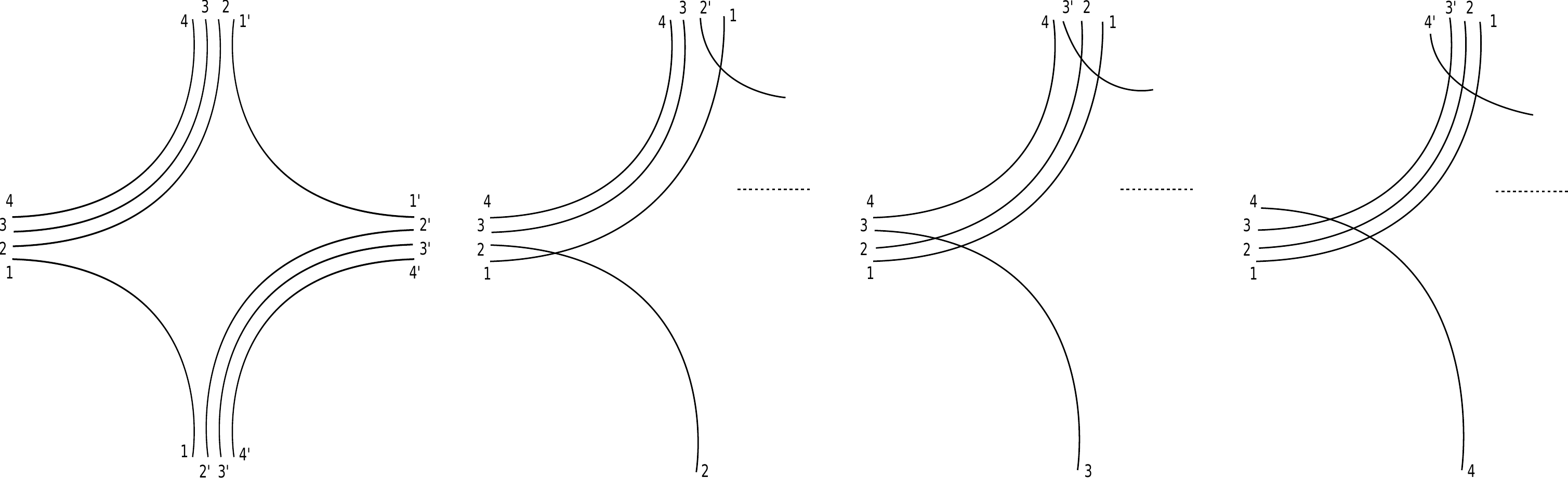}\\
\vspace{1cm}
\includegraphics[angle=0, width=3.5cm, height=3.2cm]{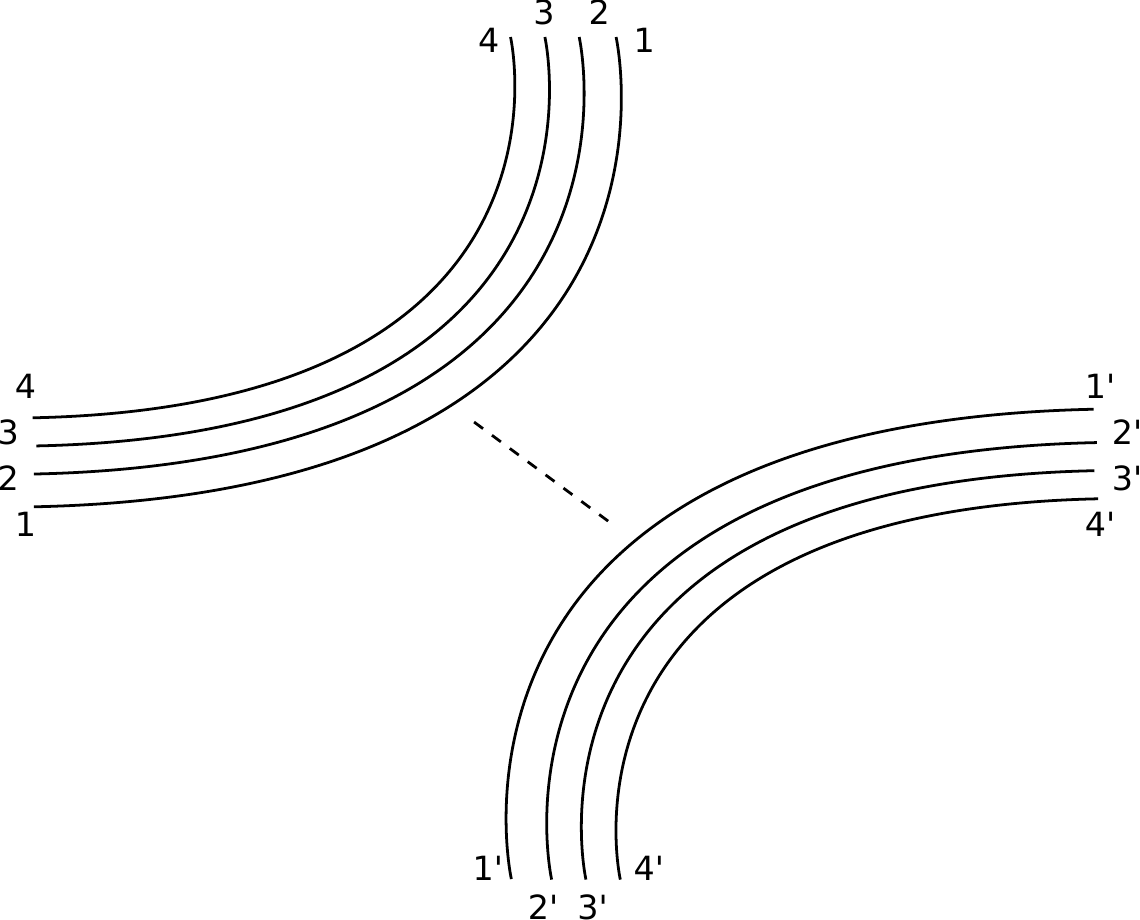}
\caption{ {\small Vertices of the type $\phi^4_{(1)}$ (parameterized by permutations $\rho=1,2,3,4$) and the anomalous term $\phi^4_{(2)} = (\phi^2)^2$ (bottom).
\label{vert4}}}
\end{minipage}
\put(-360,-12){$\rho=1$}
\put(-233,-12){$\rho=2$}
\put(-135,-12){$\rho=3$}
\put(-39,-12){$\rho=4$}
\end{figure}

The interactions of the model are effective interaction terms 
obtained after color integration \cite{Gurau:2011tj}. 
They can be equivalently defined from unsymmetrized tensors as trace invariant
objects \cite{Gurau:2012ix}. The renormalization
requires to keep relevant to marginal terms so that
only the following monomials of order six at most will be significant
\bea \label{S62}
S_{6;1} &=& \sum_{p_{j}}
\varphi_{1,2,3,4} \bar\varphi_{1',2,3,4}\varphi_{1',2',3',4'} 
 \bar\varphi_{1'',2',3',4'} \varphi_{1'',2'',3'',4''} \bar\varphi_{1,2'',3'',4''} 
+ \text{permutations } 
\cr\cr
S_{6;2} &=& \sum_{p_{j}}
\varphi_{1,2,3,4} \bar\varphi_{1',2',3',4}\varphi_{1',2',3',4'} 
\bar\varphi_{1'',2,3,4'}\varphi_{1'',2'',3'',4''} \bar\varphi_{1,2'',3'',4''}
+ \text{permutations } \label{interac6} 
 \\
  S_{4;1} & =&  \sum_{p_{j}} \varphi_{1,2,3,4} \,\bar\varphi_{1',2,3,4}\,\varphi_{1',2',3',4'} \,\bar\varphi_{1,2',3',4'}
+ \text{permutations }  
\label{S41}
\eea
where the sum is over all 24 permutations of the four color indices
giving rise to the present model.  Note that
 several configurations  have to be moded out from these 24 permutations due to both the momentum summations and the vertex color symmetry.
At the end, one ends up with  the following: 

\begin{enumerate}
\item[(i)] 4 inequivalent vertex configurations 
appearing in $S_{6;1}$ and $S_{4;1}$; these will be parameterized by an index $\rho=1,2,3,4$ (see Figure \ref{vert6}, top, for the set of
vertices in $S_{6;1}$ and Figure \ref{vert4}, top, for those which should appear in $S_{4;1}$);

\item[(ii)] 6 inequivalent  vertex 
configurations in $S_{6;2}$; each of these will be parameterized by a double index $\rho\rho'=12,13,14,23,24,34$ (see Figure \ref{vert6}, bottom).
\end{enumerate}

\begin{figure}[ht]
\begin{center}
 \includegraphics[width=3cm, height=0.8cm]{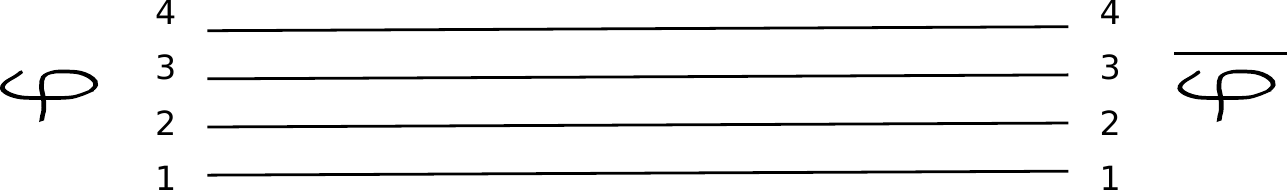}  
\caption{{\small The propagator. }}
\label{fig:prop}
\end{center}
\end{figure}

Feynman graphs have a tensor structure that we describe
now. Fields are represented by 
half lines with four strands and propagators are lines 
with the same structure, see Figure \ref{fig:prop}.  
Vertices become nonlocal objects (see Figure \ref{vert6} and Figure  \ref{vert4}). 
Simplified diagrams will be often used for simplicity. 

The renormalization analysis prescribes
to add to the action another $\phi^4$ type interaction that 
we will refer to as  {\it anomalous term}
of the form
\beq
S_{4;2}=
\left[\sum_{p_j} \bar\varphi_{1,2,3,4} \,
\varphi_{1,2,3,4} \right]
\left[\sum_{p'_j} \bar\varphi_{1',2',3',4'}\, 
\varphi_{1',2',3',4'} \right]   
\label{fifi}
\eeq
Such a term can be generated, for instance, by a contraction from a vertex of the $\phi^6_{(2)}$ type and can be seen as two 
factorized $\phi^2$ vertices (see Figure \ref{vert4}).

An ultraviolet cutoff $\Lambda$ on the propagator is introduced such that $C$ becomes $C^{\Lambda}$. As in ordinary quantum field theory, 
bare and renormalized couplings, the difference of which are coupling constant counterterms denoted by $CT$ are introduced. 
Counterterms $S_{2;1}$ and $S_{2;2}$ should be also
introduced in the bare action to perform the mass and wave function renormalization, respectively. 
The propagator $C$ includes the renormalized mass $m^2$ and
the renormalized wave function $1$.

The action of the model is then defined as
\beq
S^{\Lambda} =  \lambda^{\Lambda}_{6;1}   S_{6;1} +   \lambda^{\Lambda}_{6;2}  S_{6;2} +
\lambda^{\Lambda}_{4;1}  S_{4;1} 
+  \lambda^{\Lambda}_{4;2} S_{4;2} +  CT^{\Lambda}_{2;1} S_{2;1} +  CT^{\Lambda}_{2;2} S_{2;2}  
\label{action}
\eeq
and the partition function is
\beq
Z = \int d\mu_{C^{\Lambda}}  [\varphi]\; e^{- S^{\Lambda}  } 
\label{baremodel}
\eeq
We can define four renormalized coupling constants
$ \lambda^{\text{ren}}_{6;1}, \lambda^{\text{ren}}_{6;2}, \lambda^{\text{ren}}_{4;1}$ and $\lambda^{\text{ren}}_{4;2}$
such that, choosing appropriately  6 counterterms,
 the power series expansion of any Schwinger function of the model expressed
in powers of the renormalized couplings has a finite limit  when removing the cut-off at all orders. 
This statement has been proved in \cite{arXiv:1111.4997} by a multiscale analysis \cite{Rivasseau:1991ub}
and the fine study of the graph topology. 

A central point in the proof of the renormalizability is the reintroduction
of colors in order to get a useful bound on the graph amplitude. A graph $\cG$ admits
a color extension $\cexG$ (obtained uniquely by restoration of colors)
which is itself a rank four tensor graph.  The next stage is to define
ribbon subgraphs lying inside the tensor graph structure and also the notion of
boundary graph encoding mainly the external data. 

\begin{figure}
 \centering
     \begin{minipage}[t]{.8\textwidth}
      \centering
\includegraphics[angle=0, width=12cm, height=5cm]{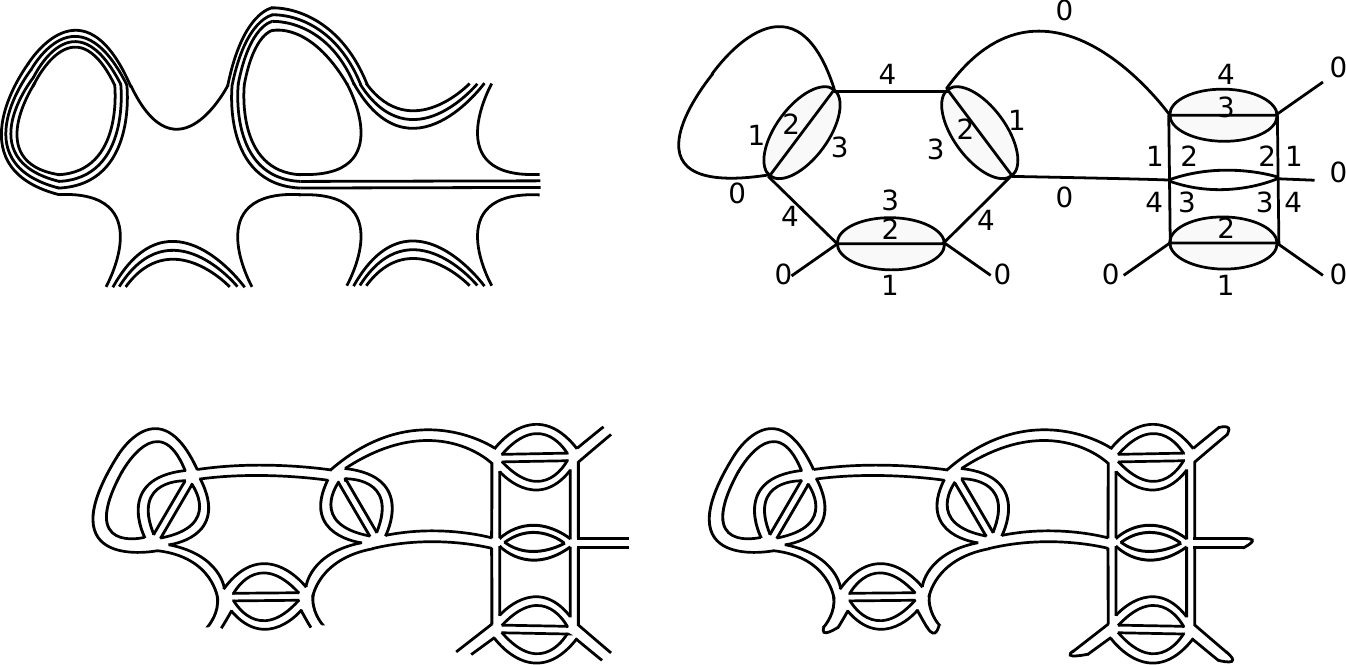}
\vspace{0.1cm}
\caption{ {\small A graph $\cG$, its colored extension $\cexG$ (five valence vertices with colored (half-)lines), 
the jacket subgraph $J$ $(01234)$ of $\cexG$ and its associated 
pinched jacket $\tJ$. \label{cc}}}
\end{minipage}
\put(-288,70){$\cG$}
\put(-110,-12){$\tJ$}
\put(-265,-12){$J$}
\put(-105,70){$\cexG$}
\end{figure}

\begin{figure}
 \centering
     \begin{minipage}[t]{.8\textwidth}
      \centering
\includegraphics[angle=0, width=6cm, height=2cm]{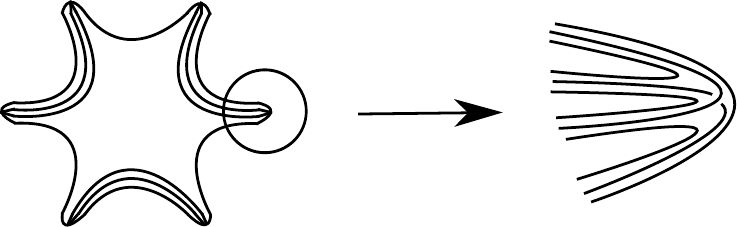}
\vspace{0.1cm}
\caption{ {\small The boundary $\bG$ of $\cG$ (see Fig.\ref{cc}) and its rank 3 
tensor structure.\label{cc2}}}
\end{minipage}
\end{figure}

\begin{definition}
Let $\cG$ be a graph in the rank $4$ theory. 
\begin{enumerate}
\item[(i)] We call colored extension of $\cG$ the unique  graph $\cexG$
obtained after restoring in $\cG$ the former colored theory graph
(see Fig.\ref{cc}).

\item[(ii)]  A jacket $J$ of $\cexG$ is a ribbon subgraph of 
$\cexG$ defined by a color cycle $(0abcd)$ up to a cyclic permutation  (see Fig.\ref{cc}). There are 12
such jackets in $D= 4$ \cite{GurRiv}. 

\item[(iii)] The jacket $\tJ$ is the jacket obtained from 
$J$ after ``pinching'' viz. the procedure consisting in 
closing all external legs present in $J$ (see Fig.\ref{cc}). 
Hence it is always a vacuum graph. 

\item[(iv)]  The boundary $\bG$ of the graph $\cG$
is the closed graph defined by vertices corresponding to 
external legs and by lines corresponding to external strands of $\cG$ 
\cite{Gurau:2009tz} (see Fig.\ref{cc2}). It is, in the present case, a vacuum graph
of the $3$ dimensional colored theory.

\item[(v)]  A boundary jacket $\bJ$ is a jacket of 
$\bG$. There are 3 such boundary jackets in $D=4$.

\end{enumerate}
\end{definition}

Consider a connected graph $\cG$.
Let $V_6$ be its number of $\phi^6$ vertices (of any type)
and $V_4$ its number of $\phi^4_{(1)}$ vertices, 
$V_4'$ its number of vertices of 
type $\phi^4_{(2)}=(\phi^2)^2$, $V_2$ the number of vertices of
the type $\phi^2$ (mass counterterms) and $V'_2$
the number of vertices of the type $(\nabla\phi)^2$ 
(wave function counterterms). Let $L$ be its number of lines and $N_{\ext}$ its number
of external legs. Consider also its colored extension $\cexG$
and its boundary $\bG$. 

Vertices contributing to $V_4'$ are
disconnected from the point of view of their strands. 
We reduce them in order to 
find the power counting with respect to only connected
component graphs.  These types of 
vertices will be therefore considered as a pair of two 2-point vertices $V_2''$, hence $V''_2 = 2V_4' $.

The renormalizability proof involves a power counting theorem
based on a multi-scale analysis. For simplicity here and without
loss of generality, we use the  following monoscale power 
counting: the amplitude of
any connected  (with respect to $V_2''$ and not to $V_4'$) graph $\cG$ is bounded by $K M^{i \omega_d(\cG)}$, where $K$ is a constant
and $\omega_d(\cG)$ is called the divergence degree of $\cG$
which is an integer and can be written 
\bea 
\label{contopformula}
\omega_d(\cG) = -\frac13 \left[ \sum_{J} g_{\tJ} 
-  \sum_{\bJ} g_{\bJ} \right] - (C_{\bG}-1) 
- V_4- 2(V_2 +V_2'') -  \frac12 \left[ N_{\ext}- 6\right]
\eea
where $g_{\tJ} $ and $g_{\bJ}$ are the genus of $\tJ$ 
and $\bJ$, respectively, $C_{\bG}$ is the number
of connected components of the boundary graph $\bG$; the first sum is performed on all closed jackets 
$\tJ$ of $\cexG$  and the second sum is performed on
all boundary jackets $\bJ$ of $\bG$.

The detailed study of the $\omega_d(\cG)$ yields a classification 
of all diverging contributions participating to the RG  flow of coupling
constants. It occurs that $\omega_d(\cG)$ does not
depend on $V_6$. One obtains the following table listing all primitively divergent
graphs:

\begin{center}
\begin{tabular}{lccccc||cc|}
$N_{\ext}$ & $V_2 + V_2''$ & $V_4$ & $\sum_{\bJ} g_{\bJ}$ & $C_{\bG}-1$ & $\sum_{\tJ}  g_{\tJ}$ & $\omega_d(\cG)$  \\
\hline\hline
6 &0 & 0 & 0 & 0& 0& 0\\
\hline
4 & 0 & 0 & 0 & 0 & 0 & 1 \\
4 & 0  & 1 & 0 & 0 & 0 & 0\\
4 & 0 & 0 & 0 & 1 & 0 & 0 \\
\hline
2 & 0 & 0 & 0 & 0 & 0 & 2\\
2 & 0 & 1 & 0 & 0 & 0 & 1\\
2 & 0 & 2 & 0 & 0 & 0 & 0\\
2 & 0 & 0 & 0 & 0 & 6 & 0\\
2 & 1 & 0 & 0 & 0 & 0 & 0\\
\hline\hline
\end{tabular}

\vspace{0.2cm}
Table 1: List of primitively divergent graphs
\end{center}

Since $V_2''=2V_4'$ is always even, the last row of the table
can be forgotten because it mainly involve a graph as a
pure mass renormalization. 

Call  graphs satisfying
$\sum_{\tJ} g_{\tJ}=0$ ``melonic'' graphs or simply ``melons'' \cite{Bonzom:2011zz}. 
Thus, in Table 1, some graphs are melons with melonic boundary, namely
those for which also holds $\sum_{\bJ} g_{\bJ}=0$. 
We are now in position to address the computation of the 
$\beta$-functions of the model.

\section{$\beta$-functions at two and four loops}

The computation of the $\beta$-functions in this model
turns out to be very involved. The method used 
in this work, though somehow lengthy, is efficient enough to deal
with a large number of Feynman graphs and
give a precise result. 

We shall enlarge the space of couplings by assigning to each interaction
 in \eqref{interac6}, \eqref{S41} and \eqref{fifi} a different coupling. 
Only at the end, we will reduce this space of coupling in order to
have the UV behavior of some reduced models. We emphasize that,
at this level, this can be viewed as an artefact in order to distinguish
the different configuration contributing to each of the renormalized coupling constant equation. In short, the combinatorics of the graph configurations can be better addressed in the different coupling setting. From the point of view of renormalization,
the extended model with different coupling constants for interactions
 can be shown to be renormalizable, if at the same time, we enlarge the space of wave function couplings (see the discussion in Subsection 5.3 in \cite{BenGeloun:2012pu} which addresses this issue for a similar tensor
model). 
 
First, we associate to each 
interaction a different coupling constant such that the total
interaction part (without counterterms and omitting to write the cut-off) becomes  after having introduced a symmetry factor for interactions
in  $S_{6;1}$ and $S_{4;1/2}$:
\bea
S = \frac13\sum_{\rho} \lambda_{6;1; \rho} \,S_{6;1;\rho} +  \sum_{\rho\rho'} \lambda_{6;2;\rho\rho'} \, S_{6;2;\rho\rho'} +
 \frac12 \sum_{\rho} \lambda_{4;1;\rho} \, S_{4;1;\rho} 
+\frac12\lambda_{4;2}\, S_{4;2} 
 \eea
where $\rho$ and $\rho\rho'$ are  permutations of indices
as given in Figure \ref{vert6} and Figure \ref{vert4}. Mainly, there are 4 terms in the sum involving $S_{6;1;\rho} $, in the second sum
involving $S_{6;2;\rho\rho'}$, there are 6 terms and, in the last regarding
$S_{4;1;\rho}$, the sum is also performed over 4 terms. 
Note that in the following, we always consider $\lambda_{6;2;\rho\rho'}=\lambda_{6;2;\rho'\rho}$
and $\rho\neq \rho'$.

We are mainly interested in the behaviour of the renormalized
coupling coupling constants $\lambda^\ren_{6;\xi;\rho/\rho\rho'}$,  $\lambda^\ren_{4;1;\rho}$ and $\lambda^\ren_{4;2}$ in the UV. In fact, the determination
of the $\beta$-functions of the $\phi^6_{(\xi)}$ vertices, $\xi=1,2$,
turns out to be crucial for the entire 
analysis. 

Any $\beta$-function, at a certain number of loops, 
is generally computed after the determination
of two ingredients: the wave function renormalization 
and the truncated and amputated one particle irreducible (1PI) 
$N$-point function the external data of which are designed 
in the form of the initial (bare) interaction. 
In the present situation, the wave function renormalization $Z$ can be written 
as 
\bea
Z = 1 - \partial_{b_1^2} \;\Sigma\;\Big|_{b_{1,2,3,4} = 0} \qquad \quad 
\Sigma(b_1,b_2,b_3,b_4) = \big\langle \varphi_{1,2,3,4}
\bar\varphi_{1,2,3,4}\big\rangle^t_{1PI} 
\qquad \quad  
\varphi_{1,2,3,4}= \varphi_{b_1,b_2,b_3,b_4}
\label{wfr}
\eea
where $b_i$ are external momenta and 
$\Sigma$  is the so-called self-energy 
or sum of all amputated 1PI two-point functions. 
The latter will be computed at two loops at first. Note that $\Sigma$ 
should be symmetric in its arguments so that the above derivative
with respect to $b_1^2$ can be replaced by any derivative
 with respect to another argument without loss of generality.

The $\beta$-functions  related to the running of coupling constants are encoded in the following ratios:
\bea
\label{gamma4}
\lambda_{6;\xi;\rho/\rho\rho'}^{\ren} &=& -\frac{\Gamma_{6;\xi;\rho/\rho\rho'}(0,0,0,0,0,0,0,0,0,0,0,0)}{ Z^{3}} 
\qquad \quad 
\lambda_{4;1;\rho}^{\ren} = -\frac{\Gamma_{4;1;\rho}(0,0,0,0,0,0,0,0)}{ Z^{2}} \cr\cr
\lambda_{4;2}^{\ren} &=& -\frac{\Gamma_{4;2}(0,0,0,0,0,0,0,0)}{ Z^{2}}
\eea
where $\Gamma_{6;\xi;\rho/\rho\rho'}(b_1,b_2, b_3,b_4,b_1',b_2', b_3',b_4',b_1'',b_2'', b_3'',b_4'')$
is the sum of amputated 1PI six-point functions or corrections to 
one of the $\phi^6_{(\xi)}$  vertices with 
coupling constant $\lambda_{6;\xi;\rho/\rho\rho'}$ 
and  $\Gamma_{4;1;\rho}(b_1,b_2, b_3,b_4,b_1',b_2', b_3',b_4')$
 the sum of amputated 1PI four-point functions, corrections to
one of the vertex  $\phi^4_{(1)}$ with coupling 
constant $\lambda_{4;1;\rho}$. The same holds
for the last $\phi^4_{(2)}$ interaction.
For instance, for particular vertices $\phi^6_{(1);\rho=1}$, 
 $\phi^6_{(2);\rho\rho'=14}$ and $\phi^4_{(1);\rho=1}$, 
we have 
\begin{eqnarray}
&&\Gamma_{6;1;\rho=1}(b_1,b_2, b_3,b_4,b_1',b_2', b_3',b_4',b_1'',b_2'', b_3'',b_4'') = 
\langle \,\varphi_{1,2,3,4}\, \bar\varphi_{1',2,3,4}\,\varphi_{1',2',3',4'} \,
 \bar\varphi_{1'',2',3',4'}\, \varphi_{1'',2'',3'',4''}\, \bar\varphi_{1,2'',3'',4''} \, \rangle^{t}_{1PI}\cr\cr
&&\Gamma_{6;2;\rho\rho'=14}(b_1,b_2, b_3,b_4,b_1',b_2', b_3',b_4',b_1'',b_2'', b_3'',b_4'') = 
\langle \,\varphi_{1,2,3,4}\, \bar\varphi_{1',2',3',4}\,\varphi_{1',2',3',4'} \,
\bar\varphi_{1'',2,3,4'}\,\varphi_{1'',2'',3'',4''} \,\bar\varphi_{1,2'',3'',4''}\, \rangle^{t}_{1PI}\cr\cr
&&\Gamma_{4;1;\rho=1}(b_1,b_2, b_3,b_4,b_1',b_2', b_3',b_4') = 
\langle \, \varphi_{1,2,3,4} \,\bar\varphi_{1',2,3,4}\,\varphi_{1',2',3',4'} \,\bar\varphi_{1,2',3',4'}  \, \rangle^{t}_{1PI} \cr\cr
&&\Gamma_{4;2}(b_1,b_2, b_3,b_4,b_1',b_2', b_3',b_4') = 
\langle \, \varphi_{1,2,3,4} \,\bar\varphi_{1,2,3,4}\,\varphi_{1',2',3',4'} \,\bar\varphi_{1',2',3',4'}  \, \rangle^{t}_{1PI} 
\label{fopt}
\end{eqnarray}
The remaining cases indexed by $\rho$ and $\rho\rho'$ 
can be easily inferred by permutations. Note that the choice of particular 
external momentum data is justified by the renormalization prescription.

The main results of this paper are captured by the following statements:

\begin{theorem}
\label{theobeta6}
At two loops, the renormalized  coupling constants satisfy
the equations
\bea
\lambda^\ren_{6;1;\rho} &=&   
 \lambda_{6;1;\rho} - 6 \lambda_{6;1;\rho}\Big[ \lambda_{6;1;\rho} - \lambda_{6;1;1} \Big]\, S^1 - 3\lambda_{6;1;\rho}
\Big[\sum_{\rho' \in \{2,3,4\} \setminus \{\rho\}} \Big(\lambda_{6;2;\rho\rho'}-   \lambda_{6;2;1\rho'}\Big)\Big]   [S^1 +  S^{12}] 
+O(\lambda^3)
\label{lamren61}\\
\lambda_{6;2;\rho\rho'}^\ren &=&  
\lambda_{6;2;\rho\rho'}  
- 2\lambda_{6;2;\rho\rho'} [ \lambda_{6;1;\rho} +  \lambda_{6;1;\rho'} - 3 \lambda_{6;1;1}]S^1
- \lambda_{6;2;\rho\rho'}
\Bigg[- \sum_{\bar\rho=2,3,4}\lambda_{6;2;1\bar\rho} \crcr
&+&
 \sum_{\bar\rho \in \{2,3,4\} \setminus \{\rho\} }
(\lambda_{6;2;\rho \bar\rho}-\lambda_{6;2;1\bar\rho})
 +  \sum_{\bar\rho \in \{2,3,4\} \setminus \{\rho'\} }
(\lambda_{6;2;\rho' \bar\rho}  -\lambda_{6;2;1\bar\rho})
\Bigg]  [S^1 +  S^{12}] +O(\lambda^3) 
\crcr
&&
\label{lamren62} \\
S^1 &:=&  \sum_{p_1,\dots, p_6}
\frac{1}{(p_1^2+ p_2^2+ p_3^2+m^2)^2}
\frac{1}{(p_4^2+ p_5^2+ p_6^2+m^2)} 
  \crcr
S^{12} &:=& \sum_{p_1,\dots, p_6}
\frac{1}{(p_1^2+ p_2^2+ p_3^2+m^2)^2}
\frac{1}{(p_1^2 + p_4^2+ p_5^2+ p_6^2+m^2)}  
\label{s0}
 \eea
where $S^1$ and $S^{12}$ are formal log-divergent sums, 
$\rho,\rho' \in \{1,2,3,4\}$, $\rho'\neq \rho$, 
and $O(\lambda^3)$ denotes a sum of O-functions with 
arguments any cubic power of the coupling constants
$O(\lambda_{6;1;\bullet}^3) +
O(\lambda_{6;1;\bullet}^2\lambda_{6;2;\bullet})  + 
O(\lambda_{6;1;\bullet}\lambda_{6;2;\bullet}^2)  + O(\lambda_{6;2;\bullet}^3)$. 

\end{theorem}
\begin{theorem}\label{theo:fourloop}

At a vanishing bare value of all $\lambda_{6;2;\rho\rho'}$,
the renormalized coupling constants at four loops for the 
$\phi^6_{(1)}$ sector satisfy the equations
\bea
\lambda_{6;1;\rho}^\ren &=&
\lambda_{6;1;\rho} + 6\lambda_{6;1;\rho}[\lambda_{6;1;1}-\lambda_{6;1;\rho}]S^1
+  \lambda_{6;1;\rho} 
\Bigg\{ 
4\Big[5\lambda_{6;1;\rho}^2 -3 \lambda_{6;1;1}^2\Big]    \cS^1_{(1)}  
+
6\Big[5\lambda_{6;1;\rho}^2 + \lambda_{6;1;1}^2
-6\lambda_{6;1;1}\lambda_{6;1;\rho} \Big]    \cS^1_{(2)}  
\Bigg\}   \crcr
&&
+ 6  \lambda_{6;1;\rho}\Bigg[
\lambda_{6;1;\rho}\sum_{\rho'\neq \rho}\lambda_{6;1;\rho'}
 -  \lambda_{6;1;1}\sum_{\rho'=2,3,4}\lambda_{6;1;\rho'}
\Bigg] \Big[2\cS^{12}_{(1)} + \cS^{12}_{(2)}\Big]
+ O(\lambda^4) 
\eea
where $O(\lambda^4) $  denotes a $O$-function involving
any quartic product of coupling constants and 
where 
\bea
\cS^{1}_{(1)}  &:=&  \sum_{p_1,\dots, p_{12}} \Big[ 
\frac{1}{(p_1^2+ p_2^2+ p_3^2+m^2)^3}
\frac{1}{(p_4^2+ p_5^2+ p_6^2+m^2)}
\frac{1}{(p_7^2+ p_8^2+ p_9^2+m^2)} 
\frac{1}{(p_{10}^2+ p_{11}^2+ p_{12}^2+m^2)} \Big]  
\crcr
\cS^{1}_{(2)}  &:=& \sum_{p_1,\dots, p_{12}} \Big[ 
\frac{1}{(p_1^2+ p_2^2+ p_3^2+m^2)^2}
\frac{1}{(p_4^2+ p_5^2+ p_6^2+m^2)^2}
\frac{1}{(p_7^2+ p_8^2+ p_9^2+m^2)} 
\frac{1}{(p_{10}^2+ p_{11}^2+ p_{12}^2+m^2)} \Big]  \crcr
\cS^{12}_{(1)}  &:=&  \sum_{p_1,\dots, p_{12}} \Big[ 
\frac{1}{(p_1^2+ p_2^2+ p_3^2+m^2)^3}
\frac{1}{(p_1^2+ p_4^2+ p_5^2+ p_6^2+m^2)}
\frac{1}{(p_1^2+ p_7^2+ p_8^2+ p_9^2+m^2)}\times  \crcr
&& 
\frac{1}{(p_{10}^2+ p_{11}^2+ p_{12}^2+m^2)} \Big]  \crcr
\cS^{12}_{(2)}  &:=&  \sum_{p_1,\dots, p_{12}} \Big[ 
\frac{1}{(p_1^2+ p_2^2+ p_3^2+m^2)^2}
\frac{1}{(p_1^2+ p_4^2+ p_5^2+ p_6^2+m^2)}
\frac{1}{(p_1^2+ p_7^2+ p_8^2+ p_9^2+m^2)} \times\crcr
&&
\frac{1}{(p_{10}^2+ p_{11}^2+ p_{12}^2+m^2)^2} \Big] 
\label{calS}
\eea
\end{theorem}

\begin{corollary}
\label{corobeta4}
At a vanishing bare value of all $\lambda_{6;2;\rho\rho'}$
and at two loops,
the renormalized coupling constants associated with 
the $\phi^4$ interactions satisfy the equations
\bea
&&
\lambda^\ren_{4;1;\rho} = \lambda_{4;1;\rho} 
\label{phi4safe}\\
&&
\lambda^\ren_{4;2} = \lambda_{4;2}  -\lambda_{4;2}^2\, S'^0 + O(\lambda_{4;2}^3)
\qquad
\qquad 
S'^0 = \sum_{p_1,\dots,p_4} \frac{1}{(p_1^2+p^2_2+p^2_3+p^2_4 + m^2)^2}
\label{anom}
\eea
and the first equation \eqref{phi4safe} holds at all orders. 
\end{corollary} 
The rest of the manuscript is devoted to a  proof of these 
claims.

\subsection{Self-energy $\Sigma$ and wave function renormalization $Z$}
\label{subsect:self}

In this section, we will focus on the proof of the next statement:

\begin{lemma}
\label{lem:self}
At two loops, the self-energy $\Sigma$ and wave function renormalization $Z$ are given 
by 
\bea
\Sigma(b_1,b_2,b_3,b_4)&=&  \Sigma^0(b_1,b_2,b_3,b_4)+ \Sigma'(b_2,b_3,b_4) 
\cr\cr
\Sigma^0(b_1,b_2,b_3,b_4)&=&-\lambda_{6;1;1} S^{1}(b_1,b_1)
-\sum_{\rho=2,3,4}\Big[\lambda_{6;2;1\rho} S^{1}(b_1,b_\rho)\Big]
 - \Big[\sum_{\rho=2,3,4}\lambda_{6;2;1\rho} \Big] S^{12}(b_1) + O(\lambda^2)
\label{sig0}\\
Z &=& 1 - \Big[2\lambda_{6;1;1} + \sum_{\rho=2,3,4}\lambda_{6;2;1\rho}\Big] S^1 
 - \Big[\sum_{\rho=2,3,4}\lambda_{6;2;1\rho} \Big]S^{12}  + O(\lambda^2)
\label{wfren}\\
S^{1}(b,b') &:=& 
 \sum_{p_1,\dots, p_6}
\frac{1}{(b^2 + p_1^2+ p_2^2+ p_3^2+m^2)}
\frac{1}{(b'^2 + p_4^2+ p_5^2+ p_6^2+m^2)} 
\label{s1}  \\
 S^{12}(b) &:=&  \sum_{p_1,\dots, p_6}
\frac{1}{(b^2 + p_1^2+ p_2^2+ p_3^2+m^2)}
\frac{1}{(p_1^2 + p_4^2+ p_5^2+ p_6^2+m^2)}  
\nonumber
\eea
where $\Sigma^0$ refers to the sum of contributions useful for the
determination of $Z$ whereas $\Sigma' =\Sigma- \Sigma^0$ consists in the self-energy remaining part which is independent of the variable $b_1$ 
and $O(\lambda^2)$ denotes a sum of O-functions with 
arguments any quadratic power of the coupling constants
$O(\lambda_{6;1;\bullet}^2) +  O(\lambda_{6;2;\bullet}^2) 
+ O(\lambda_{6;1;\bullet}\lambda_{6;2;\bullet})$.

\end{lemma}

\noindent{\bf Proof.} We start by considering the self-energy $\Sigma$ at given external 
momentum data $(b_1,b_2,b_3,b_4)$ which is
\beq
\Sigma(b_1,b_2,b_3,b_4) = \big\langle \varphi_{1,2,3,4}
\bar\varphi_{1,2,3,4}\big\rangle^t_{1PI}  = \sum_{\cG_c} K_{\cG_c} \cS_{\cG_c}(b_1,b_2,b_3,b_4)
\label{sigma}
\eeq
where the sum is performed on all amputated 1PI  two-point
graphs $\cG_c$ truncated at two loops, $K_{\cG_c}$ corresponds
to the combinatorial weight factor given rise to such a graph and 
$\cS_{\cG_c}$ consists in the amplitude of $\cG_c$.

To the self-energy \eqref{sigma} contribute generalized tadpoles  
made with contractions of one vertex and 
which have to be computed from one up to two loops. 
Keeping in mind all divergent 
two-point graphs listed in Table 1 (but not the last line with 
$V_2 + V_2''=1$ which is characterized by the insertion of a special mass two-point vertex that we omit), the possible contributions to $\Sigma$ are of the form of Figure \ref{fig:prototad}
(forgetting a moment the tensor structure): 
\begin{figure}[ht]
\centering
 \begin{minipage}{.8\textwidth}
 \centering
 \includegraphics[width=2.5cm, height=0.8cm]{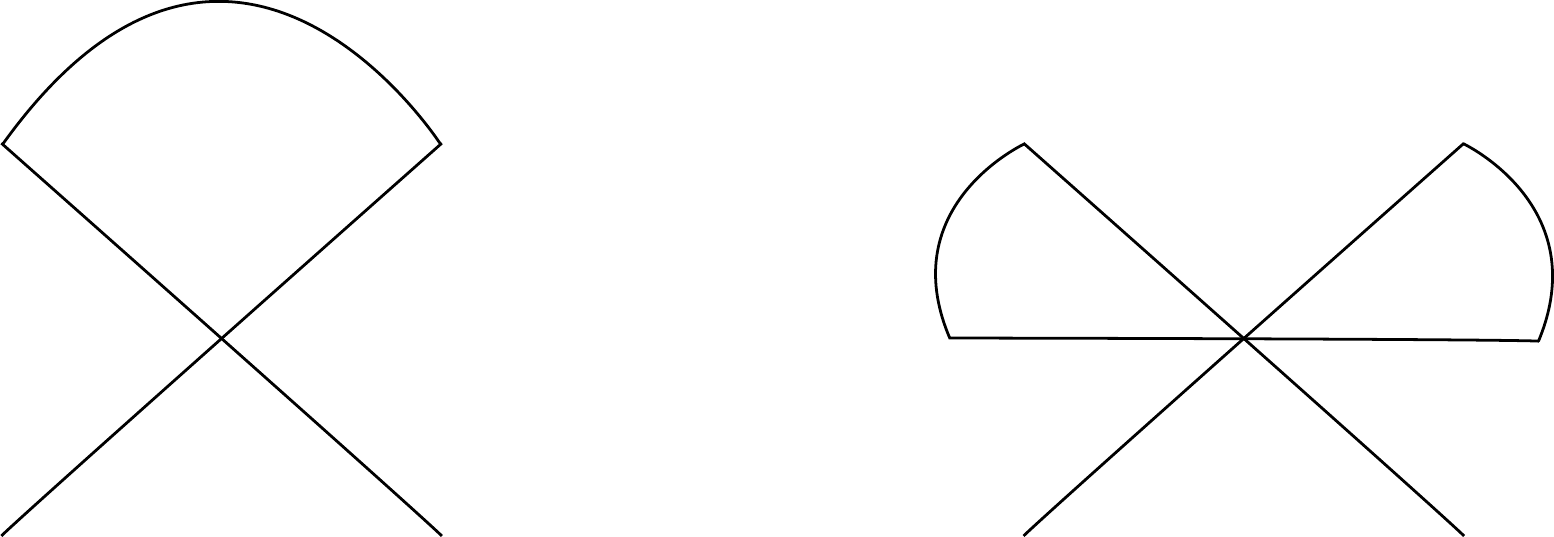}  
\caption{{\small Two  tadpole forms: TA is generated by 
$\phi^4$ vertices  and TB by $\phi^6$ vertices. }}
\label{fig:prototad}
\end{minipage}
\put(-258,3){TA}
\put(-164,3){TB}
\end{figure}

Using now the power counting, all graphs with two external legs including one or more vertices of the type $\phi^4$ should be melonic with melonic boundary. 
Furthermore, a simple inspection shows that graphs such that $V_4=1,2$ (Graph TA) are at most linearly divergent. 
Differentiating their amplitude with respect to an external argument will lead to a convergent contribution which can be neglected for the computation of $Z$. 
Only graphs of the form $V_4=0$, hence of the form TB 
made with a $\phi^6$ vertex should contribute to $Z$ and we will 
focus on them. 
Inside this category of graphs ($V_4=0$), there are graphs for 
which $\sum_{\tJ} g_{\tJ}=6$ and, hence, are log-divergent. These graphs
 should be also forgotten for the same reason given above, namely a differentiation
will make them convergent. Finally, only are significant  melonic graphs with melonic boundary with 
$V_4=0$, characterized by the first  line of Table 1 for $N_{\ext}=2$. These graphs are quadratically divergent.  

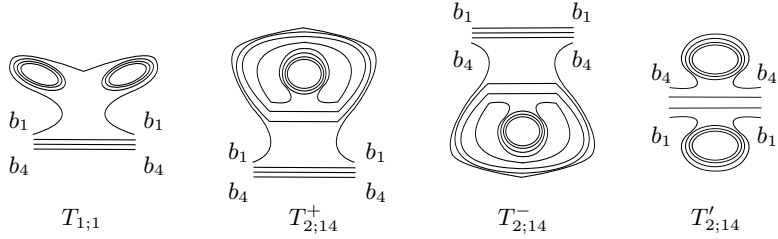
\begin{figure}[ht]
\centering
 \begin{minipage}{.8\textwidth}
 \centering
 \includegraphics[width=10cm, height=2cm]{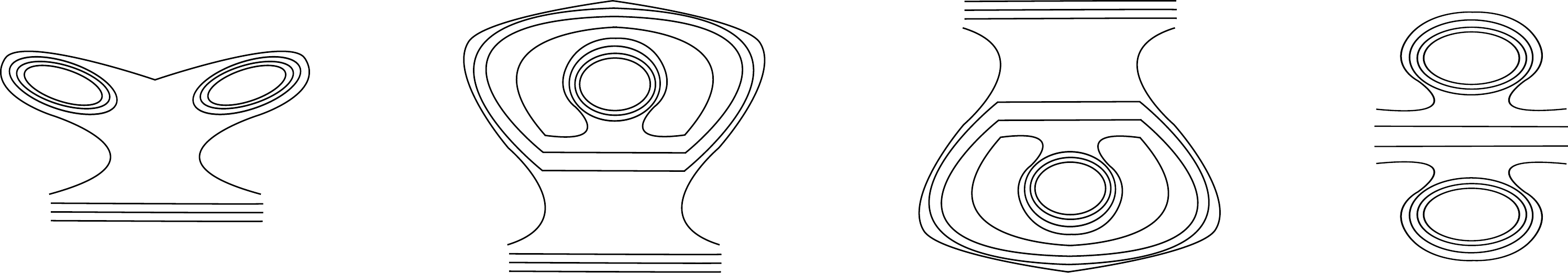}  
\vspace{0.5cm}
\caption{{\small Different tadpoles. }}
\label{fig:T1T2}
\end{minipage}
\put(-328,-25){$T_{1;1}$}
\put(-348,12){$b_{1}$}
\put(-348,-5){$b_{4}$}
\put(-297,12){$b_{1}$}
\put(-297,-5){$b_{4}$}
\put(-242,-25){$T^+_{2;14}$}
\put(-265,0){$b_{1}$}
\put(-265,-15){$b_{4}$}
\put(-213,0){$b_{1}$}
\put(-213,-15){$b_{4}$}
\put(-163,-25){$T^-_{2;14}$}
\put(-180,35){$b_{4}$}
\put(-180,53){$b_{1}$}
\put(-135,35){$b_{4}$}
\put(-135,53){$b_{1}$}
\put(-90,-25){$T'_{2;14}$}
\put(-105,6){$b_{1}$}
\put(-105,30){$b_{4}$}
\put(-65,6){$b_{1}$}
\put(-65,30){$b_{4}$}
\end{figure}

Tadpoles made with $\phi^6$ vertices are of the form given by Figure \ref{fig:T1T2}.
Note that each tadpole should be symmetrized with respect to 
all possible interactions such that one obtains the list of 
graphs $\{ T_{1;\rho}, T^\pm_{2;\rho\rho'},T'_{2;\rho\rho'}\}$ which could contribute to $Z$. $T_{1;\rho}$ graphs
are built out of a vertex of the type $\phi^6_{(1)}$ whereas 
$T^{\pm}_{2;\rho\rho'}$ and $T'_{2;\rho\rho'}$ are built from $\phi^6_{(2)}$. 
We aim at writing the sum of amputated amplitudes of all tadpoles. 
For $T_{1;\rho=1,2,3,4}$ (see $T_{1;1}$ in Figure \ref{fig:T1T2}), we have the following expression:
\beq
A_{T;1}(b_1,b_2,b_3,b_4) =
\sum_{\rho=1,2,3,4} A_{T_{1;\rho}}(b_\rho)   
=\sum_{\rho=1,2,3,4} \Big[-\frac{\lambda_{6;1;\rho}}{3}\Big]
\Big[K_{T;1;\rho}\Big] S^{1}(b_\rho,b_\rho)
\label{t1rho} 
\eeq
where the combinatorial factors are given by 
$K_{T;1;\rho} = 3$ and the formal sum $S^1(b,b')$
definition can be found in \eqref{s1}.
For $T^\pm_{2;\rho\rho'}$ (see $T^\pm_{2;14}$ in Figure \ref{fig:T1T2}), one gets
\bea
&&
A_{T;2}(b_1,b_2,b_3,b_4) = \sum_{\rho=2,3,4} A_{T^+_{2;1\rho}}(b_1) + 
\sum_{\rho=3,4} A_{T^+_{2;2\rho}}(b_2)  +  A_{T^+_{2;34}}(b_3) 
+  \sum_{\rho=1,2,3} A_{T^-_{2;4\rho}}(b_4) + 
\sum_{\rho=1,2} A_{T^-_{2;3\rho}}(b_3)  +  A_{T^-_{2;12}}(b_2) \crcr
&&=\sum_{\rho=2,3,4}
\Big[-\lambda_{6;2;1\rho}\Big]
\Big[K^+_{T;2;1\rho}\Big] \;S^{12}(b_1)
+ 
\sum_{\rho=3,4}
\Big[-\lambda_{6;2;2\rho}\Big]
\Big[K^+_{T;2;2\rho}\Big] \; S^{12}(b_2)
 + 
\Big[-\lambda_{6;2;34}\Big]
\Big[K^+_{T;2;34}\Big] S^{12}(b_3)
 \crcr
&& + \sum_{\rho=1,2,3}
\Big[-\lambda_{6;2;4\rho}\Big]
\Big[K^-_{T;2;4\rho}\Big] \; S^{12}(b_4)
 + 
\sum_{\rho=1,2}
\Big[-\lambda_{6;2;3\rho}\Big]
\Big[K^-_{T;2;3\rho}\Big] \; S^{12}(b_3)
  + 
\Big[-\lambda_{6;2;12}\Big]
\Big[K^-_{T;2;12}\Big] S^{12}(b_2)
\label{22+1} 
\eea
where the combinatorial factors are given by $K^\pm_{T;2;\rho\rho'} = 1$
and  $S^{12}(b)$ is defined in \eqref{s1}.

 One notices that $T^\pm_{2;\rho\rho'}$ correspond, in a sense, 
to tensor graphs generalizing the so-called tadpole up and tadpole down appearing in the context of ribbon graphs for noncommutative field theory \cite{Rivasseau:2007ab}. 
 Note also that, due to the nonlocality, the associated combinatorial  weight has been drastically affected. It reduces to a unique
possibility to built such a graph. 

The sum of the remaining tadpole amplitudes 
$T'_{2;\rho\rho'}$  ($T'_{2;14}$ is given in Figure \ref{fig:T1T2}) is given by 
\bea
A'_{T;2}(b_1,b_2,b_3,b_4) &=& \sum_{\rho\rho'} A_{T'_{2;\rho\rho'}}(b_\rho,b_{\rho'})   \crcr
&=&
 \sum_{\rho=2,3,4}\Big[-\lambda_{6;2;1\rho}\Big]
\Big[K'_{T;2;1\rho}\Big] S^{1}(b_1,b_{\rho})
 + \sum_{\rho\rho' \in \{23,24,34\}  }  \Big[-\lambda_{6;2;\rho\rho'}\Big]
\Big[K'_{T;2;\rho\rho'}\Big] S^{1}(b_\rho,b_{\rho'}) 
 \label{22p} 
\eea
 where $K'_{T;2;\rho\rho'}=1$.

We collect all contributions involving only the variable $b_1$.
In this specific instance, only the amputated amplitudes of $T_{1;1}$,
$T^+_{2;1\rho=2,3,3}$ and of $T'_{2;1\rho=2,3,4}$ involve the external momentum $b_1$. Neglecting the remaining amplitudes, the significant contributions to the wave function renormalization are summed and yield
\beq
\Sigma^0 (b_1,b_2,b_3,b_4)=-\lambda_{6;1;1} S^{1}(b_1,b_{1})
 - \Big[\sum_{\rho=2,3,4}\lambda_{6;2;1\rho} \Big] S^{12}(b_1)
-\sum_{\rho=2,3,4}\Big[\lambda_{6;2;1\rho} S^{1}(b_1,b_{\rho})\Big]
+ O(\lambda^2)
 \label{insig0}
\eeq
The latter \eqref{insig0} can be differentiated as
\beq
-\partial_{b_1^2} \Sigma^0 |_{b_{1,2,3,4} = 0}   = 
 - \Big[2\lambda_{6;1;1} + \sum_{\rho=2,3,4}\lambda_{6;2;1\rho}\Big] S^1 
 - \Big[\sum_{\rho=2,3,4}\lambda_{6;2;1\rho} \Big]S^{12} + O(\lambda^2)
\eeq
where the formal (log-divergent) sums $S^{1,12}$ have been introduced
in \eqref{s0}. 
Finally, one gets the wave function renormalization as
\bea
Z = 1 - \Big[2\lambda_{6;1;1} + \sum_{\rho=2,3,4}\lambda_{6;2;1\rho}\Big] S^1 
 - \Big[\sum_{\rho=2,3,4}\lambda_{6;2;1\rho} \Big]S^{12} + O(\lambda^2)
\eea
and Lemma \ref{lem:self} is proved. 
\qed

\subsection{$\beta_{6;\xi;\rho/\rho\rho'}$-functions at two loops}
\label{sect:6pt}

\begin{figure}[ht]
\centering
 \begin{minipage}{.8\textwidth}
 \centering
 \includegraphics[width=2.5cm, height=0.8cm]{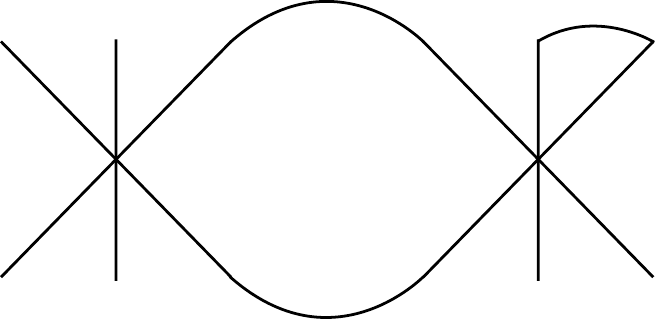}  
\caption{{\small Unique simplified melonic configuration for 1PI six-point functions. }}
\label{fig:protophi6}
\end{minipage}
\end{figure}

Roughly, melonic six-point functions are of the sole diagrammatic form given by 
Figure \ref{fig:protophi6}.
  In an expanded form, six-point function configurations can be divided 
into  three classes whenever contractions are
performed between $\phi^6_{(1)}-\phi^6_{(1)}$, 
$\phi^6_{(1)}-\phi^6_{(2)}$ and $\phi^6_{(2)}-\phi^6_{(2)}$. 
These graphs and their  amplitude 
contribute to different  $\Gamma_{6;\xi;\rho/\rho\rho'}$. 
In the same previous notations, we will use the following statement
\begin{lemma}
\label{lemgam6}
At two loops, the amputated truncated six-point functions at zero external momenta are given by the following
expressions: For $\rho=1,2,3,4,$
\beq
\Gamma_{6;1;\rho} (0,\dots,0)=
- \lambda_{6;1;\rho} +  \lambda_{6;1;\rho}
\Bigg[ 6\,\lambda_{6;1;\rho}S^1 \,  + 3\Big[\sum_{\rho' \in \{1,2,3,4\} \setminus \{\rho\}}\lambda_{6;2;\rho\rho'} \Big]  [S^1 +  S^{12}] \Bigg] + O(\lambda^3) 
\eeq
where $O(\lambda^3)$ stands for a sum of $O$-functions of any cubic power in
the coupling constants,  and for $\rho' \in \{1,2,3,4\} \setminus \{\rho\}$, 
\bea
&&
\Gamma_{6;2;\rho\rho'}(0,\dots,0) = \crcr
&&
-\lambda_{6;2;\rho\rho'} +  \lambda_{6;2;\rho\rho'} \Bigg[
2[ \lambda_{6;1;\rho} +  \lambda_{6;1;\rho'} ]S^1
+
\Big[ \sum_{\bar\rho \in \{1,2,3,4\} \setminus \{\rho\} }\lambda_{6;2;\rho \bar\rho}
 +  \sum_{\bar\rho \in \{1,2,3,4\} \setminus \{\rho'\} }\lambda_{6;2;\rho' \bar\rho}\Big]   [S^1 +  S^{12}] \Bigg] +O(\lambda^3)
\eea

\end{lemma}
\noindent{\bf Proof.} See Appendix \ref{appsect:6pt}. 

We can now proceed to the  

\noindent{\bf Proof of Theorem \ref{theobeta6}.}
Using Lemma \ref{lem:self} and \ref{lemgam6}, the renormalized 
coupling constants $\lambda^\ren_{6;1;\rho}$
 are defined by the ratios
\bea
 \lambda_{6;1;\rho}^\ren &=& -\frac{\Gamma_{6;1;\rho} (0,\dots,0)}{Z^3} \crcr
&= & 
\lambda_{6;1;\rho} -  \lambda_{6;1;\rho}
\Big[ 6\,\lambda_{6;1;\rho}S^1 \,  + 3\sum_{\rho'\neq \rho}\lambda_{6;2;\rho\rho'}  [S^1 +  S^{12}] \Big] 
+ 3\lambda_{6;1;\rho} \Big[2\lambda_{6;1;1} S^1 + 
\sum_{\rho'=2,3,4}\lambda_{6;2;1\rho'}   [S^1 +  S^{12}] \Big] + O(\lambda^3)\cr\cr
&=&  
\lambda_{6;1;\rho}  - 6 \lambda_{6;1;\rho} [\lambda_{6;1;\rho} -\lambda_{6;1;1} ] S^1
- 3 \lambda_{6;1;\rho} \Big[\sum_{\rho'\in \{1,2,3,4\} \setminus \{\rho\}}\lambda_{6;2;\rho\rho'} 
-\sum_{\rho'=2,3,4}\lambda_{6;2;1\rho'} \Big] [S^1 +  S^{12}]   + O(\lambda^3)
\eea
An obvious  simplification leads to \eqref{lamren61}. Focusing on the second sector,  $\lambda^\ren_{6;2;\rho\rho'}$ are determined by the following 
\bea
\lambda_{6;2;\rho\rho'}^\ren &=& -\frac{\Gamma_{6;2;\rho\rho'} (0,\dots,0)}{Z^3} \crcr
&=&
\lambda_{6;2;\rho\rho'} - \lambda_{6;2;\rho\rho'} \Bigg[
2[ \lambda_{6;1;\rho} +  \lambda_{6;1;\rho'} ]S^1
+
\Big[ \sum_{\bar\rho \in \{1,2,3,4\} \setminus \{\rho\} }\lambda_{6;2;\rho \bar\rho}
 +  \sum_{\bar\rho \in \{1,2,3,4\} \setminus \{\rho'\} }\lambda_{6;2;\rho' \bar\rho}\Big]   [S^1 +  S^{12}] \Bigg] \crcr
&+&  3\lambda_{6;2;\rho\rho'} \Bigg[2\lambda_{6;1;1}S^1 + 
\Big[\sum_{\bar\rho=2,3,4}\lambda_{6;2;1\bar\rho}\Big]   [S^1 +  S^{12}] \Bigg]  +O(\lambda^3)
\cr\cr
&=&
\lambda_{6;2;\rho\rho'}  
- 2\lambda_{6;2;\rho\rho'} [ \lambda_{6;1;\rho} +  \lambda_{6;1;\rho'} - 3 \lambda_{6;1;1}]S^1 \crcr
&-& \lambda_{6;2;\rho\rho'}
\Bigg\{ \sum_{\bar\rho \in \{1,2,3,4\} \setminus \{\rho\} }\lambda_{6;2;\rho \bar\rho}
 +  \sum_{\bar\rho \in \{1,2,3,4\} \setminus \{\rho'\} }\lambda_{6;2;\rho' \bar\rho}  
-3 \sum_{\tilde\rho=2,3,4}\lambda_{6;2;1\tilde\rho}  \Bigg\}  [S^1 +  S^{12}] +O(\lambda^3)
\crcr
&&
\eea
from which \eqref{lamren62} becomes immediate. 
\qed

\noindent{\bf Discussion.} We can discuss now the UV behaviour
of the model by restricting the space of parameters. 
If the coupling constants are such that
\bea
\forall \rho,\rho'  \qquad 
\lambda_{6;1;\rho} = \lambda_{6;1} \qquad 
\lambda_{6;2;\rho\rho'} = \lambda_{6;2}
\eea
we are led to our initial model \eqref{action}, 
and then, from Theorem \ref{theobeta6}, the renormalized coupling constants  satisfy
\bea
&&
\lambda_{6;1}^\ren = \lambda_{6;1} + O(\lambda^3) \crcr
&& 
\lambda^\ren_{6;2} = \lambda_{6;2} 
+ 2\lambda_{6;2} \lambda_{6;1}S^1 
 +3 \lambda_{6;2}^2   [S^1 +  S^{12}] +O(\lambda^3)
\label{lam12}
\eea
Assuming positive coupling constants $\lambda_{6;1} >0$ and
$\lambda_{6;2}>0$, the second equation tells us that the 
$\phi^6_{(2)}$ model is asymptotically free (charge screening phenomenon). The UV free theory in the present situation is a theory
of non interacting spheres in 4D. 
Meanwhile, a cancellation occurs in the $\phi^6_{(1)}$ sector at two loops. Thus the model $\phi^6_{(1)}$ is safe at two loops
and we have 
\beq
\beta_{6;1} = 0
\eeq
However, one needs to go beyond the first order corrections to 
understand how actually behaves this sector. This study will 
be addressed in a forthcoming section.

Let us emphasize that it is not possible to perform a 
full identification of the coupling constants, i.e., 
that the above RG equations hold for different 
quantities $\lambda^i_{6;1;\rho}$ and $\lambda^i_{6;2;\rho\rho'}$,
at the scale $i$. In other words, the RG equations 
cannot be merged into a single one by assuming, 
for instance, that $\lambda_{6;1} = \alpha \lambda_{6;2}$
in \eqref{lam12}. Hence, this $\phi^6$ tensor model 
is the first of a new kind in the sense
that its $\beta$-functions cannot be discussed in 
a single coupling formulation\footnote{Such RG equations mixing several coupling constants occur in condensed matter for instance
in the theory of d-wave superconductivity \cite{Ftrub}.}. 
Note that this was not the case for other
nonlocal models, like the Grosse-Wulkenhaar matrix model and the rank 3 $\phi^4$ tensor model  treated in \cite{BenGeloun:2012pu} for which the RG equations can be reduced to a unique one. In the present
situation, a  peculiarity allows us to write two $\beta$-functions for the
same coupling constant in the $\phi^6_{(2)}$ sector  
\bea
\beta_{6;2;\; (2)} = 3 \quad \qquad 
\beta_{6;2;\; (12)} = 2
\eea

Another significant feature has to be discussed as well. 
Up to this order of perturbation, 
the RG equations for $\lambda_{6;1;\rho}$ 
involve $\lambda_{6;2;\rho\rho'}$ only through contributions
which have mixed vertices yielding always a product of
couplings as $\lambda_{6;1;\rho}\lambda_{6;2;\rho\rho'}$ 
and vice-versa. Hence, at this order of perturbation, we did not
find any 1PI graphs built uniquely in one sector 
(for instance $\phi^6_{(1)}$) which could generate a relevant
contribution in the other sector (say $\phi^6_{(2)}$). 
This can be accidental or really a hint of something
worthy to be analyzed in greater details.

\subsection{$\beta_{6;1;\rho}$-functions at four loops}
\label{sect:beta6phi}

Since the $\beta_{6;1}$-function is vanishing by summing 
two-loop diagrams and merging all the coupling constants $\lambda_{6;1;\rho}=\lambda_{6;1}$, we need to go at third order of perturbation theory in order to determine the UV behaviour of the $\phi^6_{(1)}$ sector. This order of perturbation generates four-loop
diagrams.
  Once again, the calculation requires the determination of the 
four-loop contributions to the self-energy and, from 
this, the wave function renormalization. 
We also need to compute the $\Gamma_{6;1;\rho}(0,\dots,0)$ function.
The following fact will be used in order to simply achieve 
the calculation of the $\beta$-functions: since the $\phi^6_{(2)}$
sector is asymptotically free at large scale, this means that $\lambda^i_{6;2} \simeq 0$ for $i>>1$, we will directly 
use a vanishing expression for all $\lambda_{6;2;\rho\rho'}$
in the next calculations.

The following statement holds
\begin{lemma}
\label{lem:wphigam61}
At four loops, the wave function renormalization of the $\phi^6_{(1)}$ model 
is given by 
\bea
Z = 1 - 2 \lambda_{6;1;1} S^1 +  2 \lambda_{6;1;1}^2  \Big[2 \cS^{1}_{(1)} +3\cS^{1}_{(2)} \Big]
 + 2 \lambda_{6;1;1}\Big(\sum_{\rho \in \{2,3,4\}}
\lambda_{6;1;\rho}\Big)
\Big[2\cS^{12}_{(1)}+\cS^{12}_{(2)}\Big]  +O(\lambda^3)
\label{wphi61}
\eea
and the truncated amputated six-point functions at four loops 
satisfy, for any $\rho=1,2,3,4$,
\bea
\Gamma_{6;1;\rho}(0,\dots,0) &=& - \lambda_{6;1;\rho} + 2\cdot 3\, \lambda_{6;1;\rho}^2 S^1
- 2\cdot 3\cdot 5\,\lambda_{6;1;\rho}^3 \,\cS^{1}_{(2)} 
 - 2\cdot 3\,\lambda_{6;1;\rho}^2
\Big[\sum_{\rho' \in \{1,2,3,4\}\setminus \{\rho\}}\lambda_{6;1;\rho'}\Big]  \cS^{12}_{(2)} \crcr
&& - 2^2\cdot 5\,\lambda_{6;1;\rho}^3 \,\cS^{1}_{(1)}
 - 2^2\cdot 3\,\lambda_{6;1;\rho}^2
\Big[\sum_{\rho' \in \{1,2,3,4\}\setminus \{\rho\}}\lambda_{6;1;\rho'}\Big]  \cS^{12}_{(1)} 
+O(\lambda^4)
\label{gam61pur}
\eea
\end{lemma}
\noindent{\bf Proof.} See Appendix \ref{app:fourloop}.

\noindent{\bf Proof of Theorem \ref{theo:fourloop}.} 
Using Lemma \ref{lem:wphigam61}, 
the $\beta_{6;1;\rho}$-functions are provided by the ratios
\bea
&&
-\frac{\Gamma_{6;1;\rho}(0,\dots,0)}{Z^3} 
= \crcr
&&- \Bigg\{ - \lambda_{6;1;\rho} + 6 \lambda_{6;1;\rho}^2 S^1
- 30\,\lambda_{6;1;\rho}^3 \,\cS^{1}_{(2)} 
 - 6\,\lambda_{6;1;\rho}^2
\Big[\sum_{\rho'\neq \rho}\lambda_{6;1;\rho'}\Big]  \cS^{12}_{(2)} - 20\,\lambda_{6;1;\rho}^3 \,\cS^{1}_{(1)} 
 - 12\,\lambda_{6;1;\rho}^2
\Big[\sum_{\rho'\neq \rho}\lambda_{6;1;\rho'}\Big]  \cS^{12}_{(1)}\Bigg\}
\crcr
&&
\Bigg\{ 
1 + 6\lambda_{6;1;1} S^1 -3 \Bigg[ 2 \lambda_{6;1;1}^2  [2 \cS^{1}_{(1)} +3\cS^{1}_{(2)} ]
 + 2 \lambda_{6;1;1}\Big[\sum_{\rho'\in \{2,3,4\}}\lambda_{6;1;\rho'}\Big]
\Big[2\cS^{12}_{(1)}+\cS^{12}_{(2)}\Big] \Bigg]
 +6(-2\lambda_{6;1;1})^2 (S^1)^2  
\Bigg\}  \crcr
&&
= \lambda_{6;1;\rho} + 6\lambda_{6;1;\rho}[\lambda_{6;1;1}-\lambda_{6;1;\rho}]S^1
+  \lambda_{6;1;\rho} 
\Bigg\{ 
2\Big[10\lambda_{6;1;\rho}^2 -6 \lambda_{6;1;1}^2\Big]    \cS^1_{(1)}  
+
\Big[30\lambda_{6;1;\rho}^2 + 6\lambda_{6;1;1}^2
-36\lambda_{6;1;1}\lambda_{6;1;\rho} \Big]    \cS^1_{(2)}  
\Bigg\}   \crcr
&&
+ 6  \lambda_{6;1;\rho}\Bigg[
\lambda_{6;1;\rho}\sum_{\rho'\neq \rho}\lambda_{6;1;\rho'}
 -  \lambda_{6;1;1}\sum_{\rho'=2,3,4}\lambda_{6;1;\rho'}
\Bigg] \Big[2\cS^{12}_{(1)} + \cS^{12}_{(2)}\Big]
\eea
where we use the fact that $(S^1)^2 = \cS^1_{(2)}$. 
 Theorem \ref{theo:fourloop} is then immediate. \qed

\noindent{\bf Discussion.}
Let us discuss the case of a unique coupling constant
such that $\lambda_{6;1;\rho}= \lambda_{6;1}$, the above 
equation yields
\beq
 \lambda^\ren_{6;1} = \lambda_{6;1} 
+ 
8 \lambda_{6;1}^3\cS^1_{(1)}    
\eeq 
Hence the $\beta$-function, at four loops, for the reduced single coupling model is given by 
\beq
\beta_{6;1} = 8
\eeq
showing that the model is asymptotically free in this sector
also. 

Note that an important cancellation occurs in the calculation
after identifying $\lambda_{6;1;\rho}=\lambda_{6;1}$. 
Many contributions match perfectly in the wave function
renormalization and the six-point functions. 
It could be interesting to look at these contributions
more closely because they might generate an asymptotically safe model
with a bounded RG flow relevant for a constructive program \cite{Rivasseau:1991ub}.  

Both this study and the former prove that 
the overall model described by \eqref{action} is asymptotically free in the UV. We mention that the above result is derived 
using connected 1PI graphs made only with $\phi^6_{(1)}$ vertices. 
The combinatorial study shows that the third order of
perturbation the $\phi^6_{(1)}$ does not generate any 1PI
graph with boundary of the form of $\phi^6_{(2)}$
and this even before having put $\lambda_{6;2;\rho\rho'}=0$
(see Appendix \ref{app:fourloop}). This strengthens a previous remark. 
The fact that we can set the bare value $\lambda_{6;2;\rho\rho'}=0$
(as if we were in the UV for this sector) is without consequence
on  the UV behavior of the second interaction
with coupling $\lambda_{6;1;\rho}$ and vice versa.
Indeed, for instance in \eqref{lam12}, putting $\lambda_{6;1}=0$
leads to the same UV behaviour of the model $\phi^6_{(2)}$
since a unique $\beta_{6;2;\;(2)}=2$ still remains and determines 
the asymptotic freedom in this sector.

\subsection{$\beta_{4;\xi;\rho}$-functions of the model}
\label{sect:4pt}

\begin{figure}[ht]
\centering
 \begin{minipage}{.8\textwidth}
 \centering
 \includegraphics[width=7cm, height=3cm]{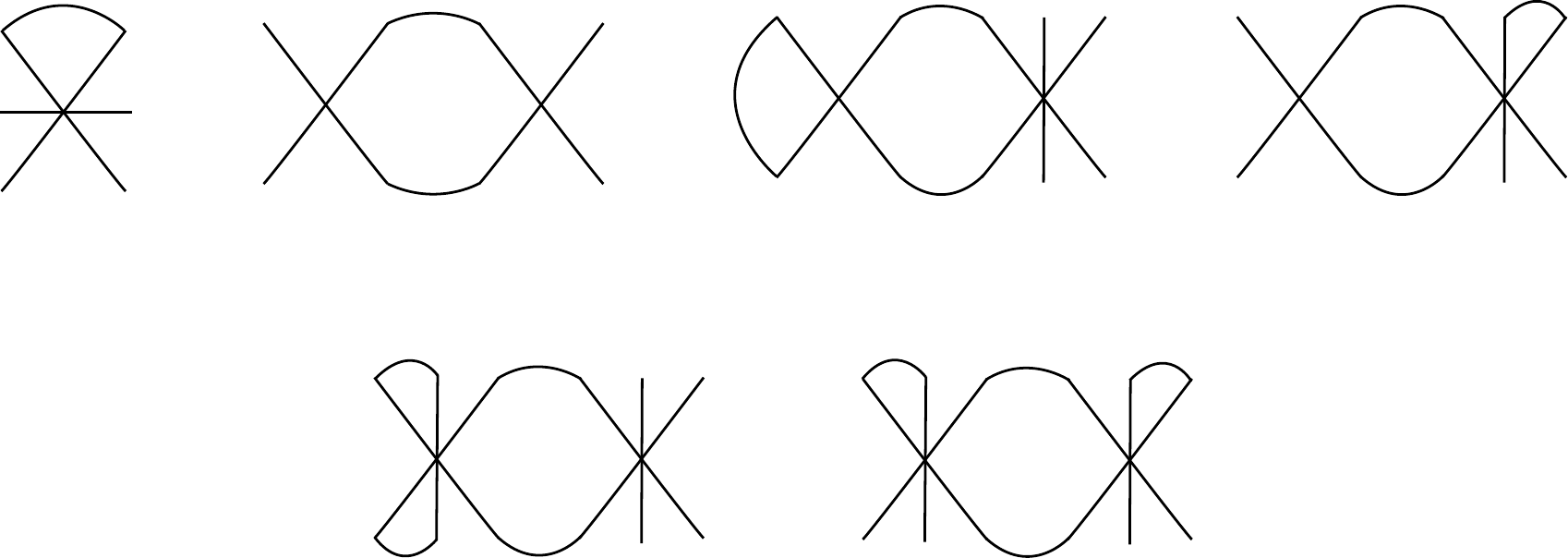}  
\vspace{0.3cm}
\caption{{\small Main four-point melonic graphs in simplified forms. }}
\label{fig:simpliphi4}
\end{minipage}
\put(-298,15){B}
\put(-252,15){B'}
\put(-188,15){D}
\put(-128,15){E}
\put(-240,-40){W}
\put(-178,-40){Y}
\end{figure}
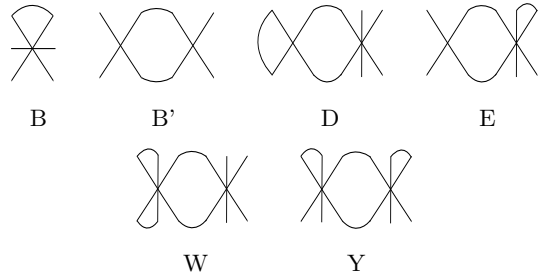

To start with, we will focus only on divergent contributions defined by
melonic graphs with melonic boundary  
having $V_4 = 0,1$ and four external legs with momenta of the form 
of  $\phi^4$.  Note that these are necessarily given
by one of the simplified diagrams as given in Figure \ref{fig:simpliphi4}.

Remark that graphs of the type B' are all convergent 
if they are built from $\phi^4_{(1)}$ vertices. Indeed, in this case,
such a graph will be convergent by the power counting.
Nevertheless, if both vertices are of the $\phi^4_{(2)}$ type, then
B' leads to a unique divergent contribution. 

\begin{lemma}
\label{lem:4ptfunc}
At two loops, the truncated amputated four-point functions 
at external momentum data set to zero are given by 
\bea
\Gamma_{4;1;\rho}(0,\dots,0) &=&
 - \lambda_{4;1;\rho}  + 2 \lambda_{4;2} 
\Big[\lambda_{6;1;\rho} + \sum_{\rho'\in \{1,2,3,4\}\setminus \{\rho\}} \lambda_{6;2;\rho\rho'}  \Big] S''^0 \crcr
&&
 + 
2 \Big[ 3\lambda_{6;1;\rho} \lambda_{4;1;\rho}+\sum_{\rho' \in \{1,2,3,4\} \setminus \{\rho\}} \lambda_{6;2;\rho\rho'}\lambda_{4;1;\rho'}
\Big] S^1  \crcr
&&
+2 \sum_{\rho'\in \{1,2,3,4\}\setminus \{\rho\}} \Big[ \lambda_{6;1;\rho}\lambda_{4;1;\rho' }  + \sum_{\rho''\in \{1,2,3,4\} \setminus \{\rho'\}} \lambda_{6;2;\rho\rho'}\lambda_{4;1;\rho''}\Big] 
 S^{12} \cr\cr
&&
+
2\lambda_{4;1;\rho}\Big[ \sum_{\rho'\in \{1,2,3,4\} \setminus \{\rho\}}  \lambda_{6;2;\rho\rho'} 
\Big] [S^1 +S^{12}]
 + \cF_{4;\rho}(\lambda_{6;1};\lambda_{6;2}) 
+ O(\lambda^3) \crcr
&&
S''^0 := \sum_{p_1,\dots,p_7}  
\frac{1}{(p_1^2+p^2_2+p^2_3+ m^2)^2} 
\frac{1}{(p_4^2+p^2_5+p^2_6+p^2_7 + m^2)^2}
\label{gam4rho}
\eea
where $\cF_{4;\rho}(\lambda_{6;1};\lambda_{6;2})$ is a function of the coupling constants  $\lambda_{6;1;\rho}$ and $\lambda_{6;2;\rho\rho'}$
and $O(\lambda^3)$ denotes a sum of O-functions of all possible
cubic monomials in the coupling constants. 
\end{lemma}
\noindent{\bf Proof.} See Appendix \ref{appsubsect:4pt}.

The proof of Corollary \ref{corobeta4} can be now worked out. 

\begin{figure}[ht]
\centering
 \begin{minipage}{.8\textwidth}
 \centering
 \includegraphics[width=3cm, height=1.5cm]{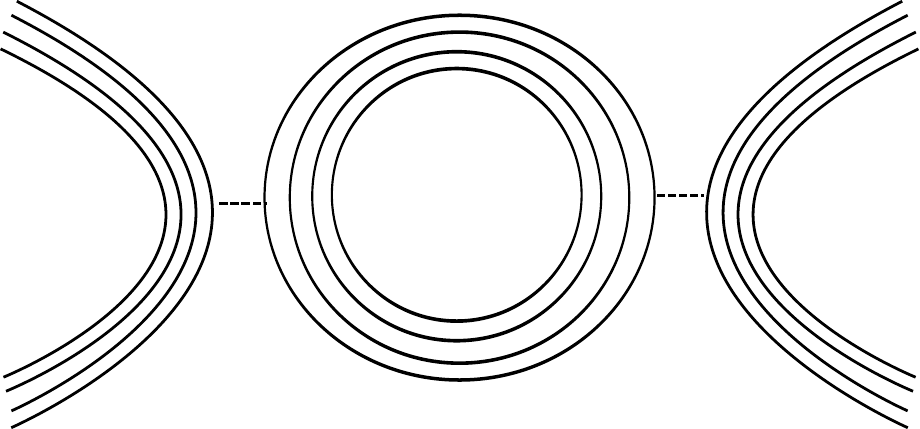}  
\caption{{\small Graph $F'$ of the type $B'$ is the unique contribution to $\Gamma_{4;2}$
for all $\lambda_{6;\xi}^i$ equal to zero for $i>>1$.}}
\label{fig:four42}
\end{minipage}
\end{figure}

\noindent{\bf Proof of Corollary \ref{corobeta4} and Discussion.}
Section \ref{sect:6pt} and \ref{sect:beta6phi} have shown 
that, at high scale, the bare values of $\lambda_{6;1;\rho}$ and of
$\lambda_{6;2;\rho\rho'}$ vanish. We simply modify 
Lemma \ref{lem:4ptfunc} and get the reduced $\Gamma_{4;\rho}$
as
\beq
\Gamma_{4;\rho}(0,\dots,0) =
 - \lambda_{4;\rho}   + O(\lambda^3)
\eeq
dividing by $Z=1$ leads to the expected result, namely
\beq
\lambda^\ren_{4;1;\rho} = \lambda_{4;1;\rho} + O(\lambda^3)
\eeq
In fact, the above equation holds at all orders 
$\lambda^\ren_{4;1;\rho} = \lambda_{4;1;\rho}$ 
for a sufficiently high scale enforcing $\lambda_{6;\xi;\rho/\rho\rho'}$ 
to be zero. 
Furthermore, the flow of $\lambda_{4;1;\rho}$ is not really driven 
by $\phi^4$ vertices  but  only by $\phi^6_{(1/2)}$ vertices.
Adding more than one $\phi^4_{(1)}$ leads to a convergence
 in any four-point functions due to the power counting. 
All these contributions indeed vanish in the UV.
Hence, the $\phi^4_{(1)}$ sector is safe at all loops and
we have
\beq
\beta_{4;1;\rho} =0 
\eeq
Let us discuss the anomalous term $\phi^4_{(2)}$.
In the same vein discussed above, we assume that
all $\lambda_{6;\xi;\rho/\rho\rho'}=0$ yielding $Z=1$. 
Nevertheless, in contrast with $\phi^4_{(1)}$, 
$\phi^4_{(2)}$ contribute to its own flow.
At one loop, the unique  contribution entering
in the 1PI four-point function with external data 
governed by the $\phi^4_{(2)}$ interaction is given by 
$F'$ (see Figure \ref{fig:four42}).

Note that the amplitude $A_{F'}$ is independent
of the external data after amputation. This is just a vacuum 
amplitude and we have
\bea
\Gamma_{4;2} = -\lambda_{4;2} + \lambda_{4;2}^2S'^0 +O(\lambda_{4;2}^3) 
\qquad \qquad 
S'^0 = \sum_{p_1,\dots,p_4} \frac{1}{(p_1^2+p^2_2+p^2_3+p^2_4+ m^2)^2}
\label{anom2}
\eea
Thus, \eqref{anom2} means that the anomalous 
term possesses a Landau ghost in the UV. One has
\beq
\beta_{4;2} = -1
\eeq 
This sector behaves like an ordinary $\phi^4$ model 
in $\mathbb{R}^4$.

The UV fixed manifold associated with all RG equations calculated
earlier is  $\lambda_{6;1;\rho}=0=\lambda_{6;2;\rho\rho'}$, $\lambda_{4;2}=0$ for any bare value for $\lambda_{4;1;\rho}$. 
The interacting theory is defined by a small perturbation around this UV fixed manifold by $\lambda_{6;\xi=1,2}=\epsilon$ and $\lambda_{4;2}=\delta$. The fact that the $\beta_{6;\xi}$-functions are positive and independent of 
any other coupling, immediately ensures that the perturbation yields $\lambda^{IR}_{6;\xi=1,2} > \epsilon \log \Lambda$
making both of these couplings growing in the IR. 
Let us focus now on $\lambda_{4;1;\rho}$ and the anomalous
coupling $\lambda_{4;2}$. The first order corrections
in $\epsilon$ are of the form  
\bea
\lambda^{IR}_{4;1;\rho} &=& \lambda^{UV}_{4;1;\rho} + 8\epsilon \Lambda 
\label{41ro} \\
\lambda^{IR}_{4;2} &=& \lambda^{UV}_{4;2;\rho} + 12\epsilon \log\Lambda 
- \delta \log\Lambda \label{42} 
\eea
where $8\epsilon$ comes from the contribution 
$2 \Big[ \lambda_{6;1;\rho} + \sum\lambda_{6;2;\rho\rho'}\Big]\, \Lambda $
(see the first order corrections in $\lambda_{6;\xi}$ in $\mathcal{F}_{4;\rho}$ \eqref{cf} in Appendix \ref{grap4w});
$12\epsilon$  are induced by the (6 possible $\times$ a factor of 2) tadpoles from $\phi^6_{(2)}$  
\begin{figure}[ht]
\centering
 \begin{minipage}{.8\textwidth}
 \centering
 \includegraphics[width=3.5cm, height=3cm]{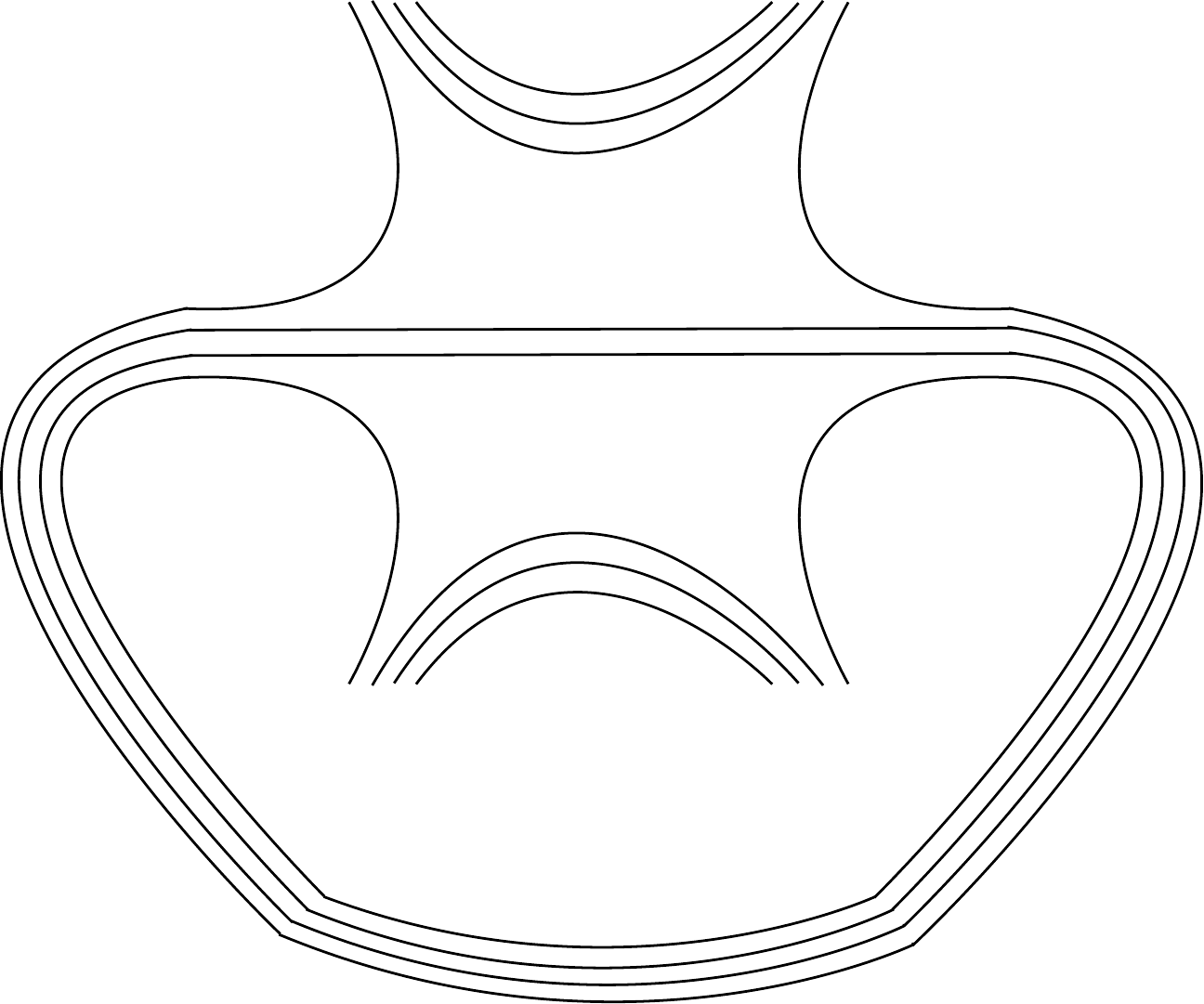}  
\caption{{\small Form of the first order correction to $\phi^4_{(2)}$ as a tadpole graph of $\phi^6_{(2)}$.}}
\label{fig:ano}
\end{minipage}
\end{figure}
(see Figure \ref{fig:ano});  such corrections are log-divergent;
finally, $\delta$ is the contribution of the anomalous vertex itself. 
Hence, from \eqref{41ro}, one notes that $\lambda^{IR}_{4;1;\rho}>\lambda^{UV}_{4;1;\rho} $ and so the coupling constants $\lambda_{4;1;\rho}$ increase in the IR whatever their initial value.
A look at the anomalous coupling equation \eqref{42} reveals that first order corrections between $\epsilon$ and $\delta$
can compete. Nevertheless, in the IR, given the negative sign of the $\beta_{4;2}$-function, the contribution 
in  $\lambda_{4;2}$ is in any way decreasing meanwhile
the contribution in $\lambda_{6;\xi}$ becomes larger. 
In conclusion,  $\lambda_{4;2}^{IR}>\lambda^{UV}_{4;2}$ and
the anomalous coupling is also increasing in the IR.

\section{Conclusion}

The $\beta$-functions of the $\phi^6$ tensor
model as introduced in \cite{arXiv:1111.4997} have been 
worked out. We find that the two main interactions of the 
$\phi^6$ form vanish in the UV and hence prove that 
the model is asymptotically free in the UV. 
The model incorporates also two  $\phi^4$ interactions.
One of these is safe at all loops and the other one yields a diverging
bare coupling. The fact that one coupling diverges in 
the UV is not of a particular significance for the model. 
Indeed, the said coupling is not associated with one of the main 
$\phi^6$ interactions which prove to drive the RG flow of all
remaining couplings. 
The calculations have been performed
at two loops in some cases, whereas an intriguing cancellation
in the $\phi^6_{(1)}$ sector has required to go beyond
 two-loop calculations. Third order corrections in the coupling constants 
up to four loops have to be determined in order to probe the UV behaviour in this sector.
 We have found that there exists a UV fixed manifold associated with the model determined for $[\lambda_{6;\xi=1,2}=0;  \lambda_{4;1}; \lambda_{4;2}=0]$ and that all coupling constants increase in the IR.
Interestingly, this result entails that it might exist a variety of models emerging from the present 4D model in the IR through a phase transition. 

This study validates the pertinence of the model \cite{arXiv:1111.4997}
for the point of view of renormalization and can be considered 
as a hint of a phase transition for some large renormalized coupling
constants towards new degrees of freedom. 
This is consistent with the geometrogenesis scenario advocated in 
\cite{oriti,Oriti:2006ar,Sindoni:2011ej}. 
Note that a phase transition has been discussed for the same type of model but  in the case of unbroken unitary invariant action without flow (without Laplacian in the kinetic term) in the work by Bonzom et al. \cite{Bonzom:2012hw}.

Another property which can be pointed out is that 
the sectors $\phi^6_{(1/2)}$ cannot be merged
into a single one. The underlying question is whether
or not this model can be restricted to a renormalizable model with 
unique coupling constant coming, for instance, from the Gurau colored model \cite{color} with one dynamical color. 
According to the above results, the answer is no. 
Even though one can combinatorially restore the colors
in the model, thereby making a combinatorial link between 
the coupling constants of this model and the colored one
(a little combinatorics shows that $\lambda_{6;1;\rho}$ can be viewed
as $3 \cdot 2^2 (\bar\lambda_{\text{color}}\lambda_{\text{color}})^3$, $\lambda_{\text{color}}$ and $\bar\lambda_{\text{color}}$ 
being the coupling constants of the bipartite colored model, meanwhile
$\lambda_{6;2;\rho}$ can be related to $3^2 \cdot 2^2  (\bar\lambda_{\text{color}}\lambda_{\text{color}})^3$), 
there is no clear way to reduce the RG
equations of all couplings into a single one by using 
just a coefficient between the two types of coupling 
constants (as the above could lead to $3\lambda_{6;1;\rho} =\lambda_{6;2;\rho}$). In summary, the four RG equations associated with 
the couplings $\lambda_{6;1;\rho}$ can be merged 
into one equation and the six RG equations associated with 
the couplings $\lambda_{6;2;\rho'\rho}$ can be merged 
into a unique and independent equation. 
 
It could be also valuable to scrutinize better the 
cancellation occurring in the $\phi^6_{(1)}$ sector which could 
 lead to asymptotic safety for this sector or, 
at least, for a particular subsector (some specific category of graphs)
in this sector.
 A first matrix model which has proved to be asymptotically safe is the Grosse-Wulkenhaar model \cite{Grosse:2004yu} 
(some recent developments on its solution can be found in \cite{Grosse:2012uv}). 
Remark that the meaning of UV and IR in that latter model
is drastically different as the ordinary one. Nevertheless,
this lead us to the natural question: Is there a 
tensor model generalizing faithfully this safeness feature?
The above mentioned  cancellation might be a hint towards
an answer to this question. 
Another straightforward attempt would be to define a model
like the one presented here by just replacing the group $U(1)$ by $\mathbb{R}^4$ and to use the Mehler kernel as propagator
in order to avoid the issue of UV/IR mixing. 
This study fully deserves to be performed.

\section*{Acknowledgements}
Discussions with  R. Gurau and V. Rivasseau are gratefully acknowledged. 
Research at Perimeter Institute is supported by the Government of Canada through Industry
Canada and by the Province of Ontario through the Ministry of Research and Innovation.

\section*{ Appendix}
\label{app}

\appendix

\renewcommand{\theequation}{\Alph{section}.\arabic{equation}}
\setcounter{equation}{0}

\section{Proof of Lemma \ref{lemgam6}}
\label{appsect:6pt}

We prove  Lemma \ref{lemgam6} by computing all 
1PI amputated six-point functions at two loops in this section. But, first,
let us discuss some general features and notations 
valid in all cases.

Consider a graph and the different contractions contributing
to a given $\Gamma_{6;\xi;\rho/\rho\rho'}$ or
$\Gamma_{4;\xi;\rho}$.   Note that these graphs
can be parametrized by a collection of permutation indices, $\rho$
or $\rho\rho'$, of their vertices. Nevertheless, this index notation is often not enough to 
capture the  features of graphs one is dealing with.
In this particular situation,  extra symbols $(\pm)$ are used. Any graph is always
considered as the same under permutation of its indices, namely, 
$\cG_{\rho\rho'}= \cG_{\rho\rho'}$. In case of multiple index notation,
this also holds but only in each sector, 
i.e. $\cG_{\rho\rho';\rho''\rho'''} = \cG_{\rho'\rho;\rho''\rho'''} = 
\cG_{\rho\rho';\rho'''\rho''}= \cG_{\rho'\rho;\rho'''\rho''}$.
Moreover, in the following, an amplitude of a graph 
$\cG$ will be written formally $A_{\cG}(b_\rho,b_\rho')$
or $A_{\cG}(b_\rho,b_\rho',b_\rho'')$ where 
the arguments $(b_\rho,b_\rho')$ or $(b_\rho,b_\rho',b_\rho'')$ mean
 all external (not summed) momenta involved in the graph.

We introduce the formal sums
\bea
S^{3}(b,b') &:=&
 \sum_{p_1,\dots, p_6}  \Big[
\frac{1}{(b^2 + p_1^2+ p_2^2+ p_3^2+ m^2)}
\frac{1}{(b^2 +  p_4^2 + p_5^2+ p_6^2 + m^2)}
\frac{1}{(b'^2+ p_1^2+ p_2^2+ p_3^2 +m^2)} \Big] 
\crcr
S^4(b,b',b'')&:=&
 \sum_{p_1,\dots, p_6} \Big[ 
\frac{1}{(b^2 + p_1^2+ p_2^2+ p_3^2+ m^2)}
\frac{1}{(b'^2+ p_1^2+ p_2^2+ p_3^2 +m^2)} 
\frac{1}{(b''^2 +  p_4^2 + p_5^2+ p_6^2 + m^2)}\Big] \crcr
S^{14}(b,b')&:=&
 \sum_{p_1,\dots, p_6} \Big[
\frac{1}{(b^2 + p_1^2+ p_2^2+ p_3^2+ m^2)}
\frac{1}{(b'^2+ p_1^2+ p_2^2+ p_3^2 +m^2)} 
\frac{1}{(p^2_{1} +  p_4^2 + p_5^2+ p_6^2 + m^2)}\Big] 
\eea
Note that $S^3(0,0) =S^4(0,0,0) =S^1$ and $S^{14}(0,0)=S^{12}$.

\subsection{Graph F}
 \label{gra6F}
 
\begin{figure}[ht]
\centering
 \begin{minipage}{.8\textwidth}
 \centering
 \includegraphics[width=7cm, height=2cm]{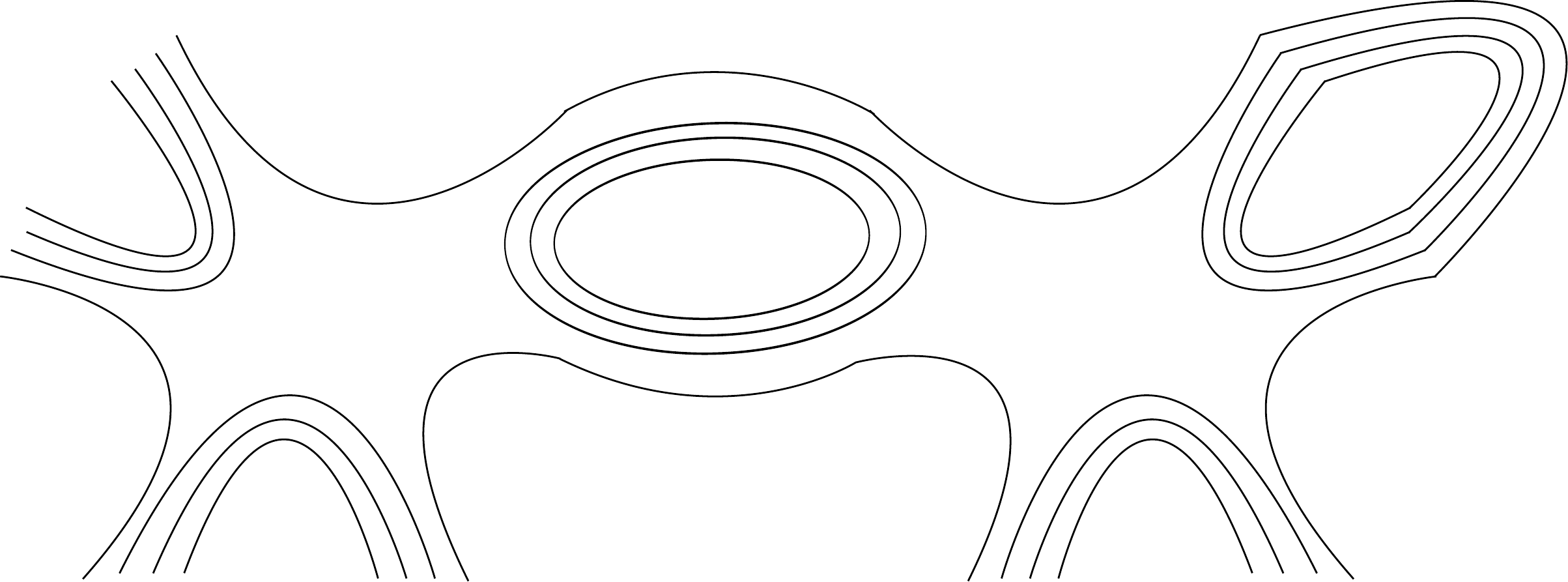}  
\vspace{0.5cm}
\caption{{\small Graph of type $F$: $F_{\rho}$ with coinciding permutation
index $\rho$ for $\phi^6_{(1)}$ vertices; here $\rho=\rho'=1$.}}
\label{fig:F1}
\end{minipage}
\put(-350,25){$\phi^6_{(1);\rho=1}$}
\put(-302,0){$b_{1}$}
\put(-305,35){$b_{4}$}
\put(-273,45){$b_{1}'$}
\put(-275,-5){$b_{4}''$}
\put(-240,-5){$b_{1}''$}
\put(-160,-5){$b_{4}'$}
\put(-130,-5){$b_{1}'$}
\put(-90,25){$\phi^6_{(1);\rho'=1}$}
\end{figure}
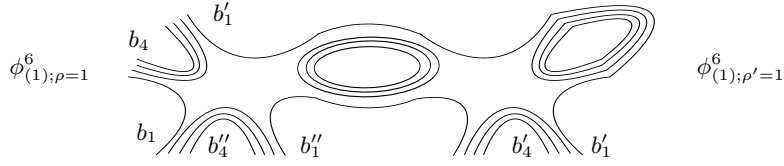

Graphs of type F are six-point function  configurations 
 described by the gluing of two vertices
of the type $\phi^6_{(1)}$. Call these graphs $F_{\rho}$ 
because they are parametrized by a unique permutation
index (for instance $F_1$ is depicted in Figure \ref{fig:F1}).

Given $\rho$, each graph $F_{\rho}$ contributes to 
the corresponding $\Gamma_{6;1;\rho}$ as
\bea
A_{F_{\rho}}(b_\rho,b_\rho',b_\rho'') &=& \frac1{2!}\Big[-\frac{\lambda_{6;1;\rho}}{3}\Big]^2
\Big[K_{F;\rho}\Big]   \Big[ S^3(b_{\rho},b_{\rho}')
\quad + \quad \text{circular permutations} \quad [b_{1}\to b_{1}' \to b_{1}'' ] \Big]   
\eea
where any combinatorial factor is given by 
$K_{F;\rho} = 3^2 \cdot 2^2$.
Setting external momenta to zero, for each $\rho$, the contribution 
becomes
\beq
A_{F_{\rho}}(0,\dots,0) = 2 \cdot 3\, \lambda_{6;1;\rho}^2\, S^1
\label{frho}
 \eeq

\subsection{Graphs H, G and I}
 \label{gra6B}

We now discuss another configuration defined by $H^\pm_{\rho\rho'}$, $G^\pm_{\rho\rho'}$ and $I^\pm_{\rho\rho'}$.  These graphs appear as the contraction of one vertex of the type $\phi^6_{(1)}$ and one of type $\phi^6_{(2)}$. They are parametrized by the index of the second type of vertex $\phi^6_{(2)}$. 

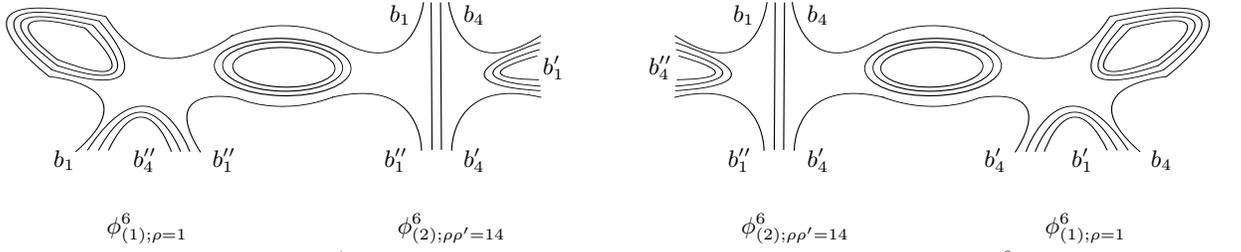
\begin{figure}[ht]
\centering
 \begin{minipage}{.8\textwidth}
 \centering
 \includegraphics[width=16cm, height=2cm]{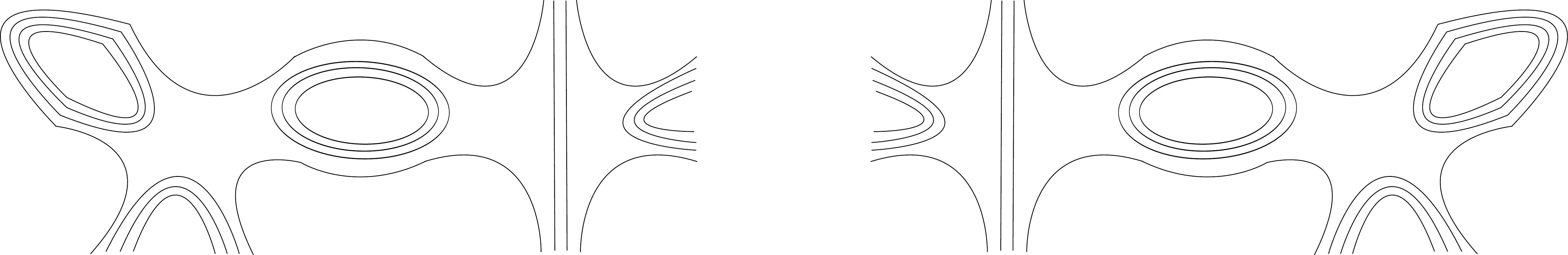}  
\vspace{0.5cm}
\caption{{\small Graphs of type $H$: $H^+_{\rho\rho'}$ (left) and $H^-_{\rho\rho'}$
(right) 
 are parameterized by $\rho\rho'$ indices of $\phi^6_{(2)}$ 
(and $\rho$ of $\phi^6_{(1)}$ becomes redundant); here $\rho\rho'=14$.}}
\label{fig:H}
\end{minipage}
\put(-370,-25){$\phi^6_{(1);\rho=1}$}
\put(-390,0){$b_{1}$}
\put(-360,0){$b_{4}''$}
\put(-263,55){$b_{1}$}
\put(-330,0){$b_{1}''$}
\put(-265,0){$b_{1}''$}
\put(-205,35){$b_{1}'$}
\put(-235,0){$b_{4}'$}
\put(-235,55){$b_{4}$}
\put(-260,-25){$\phi^6_{(2);\rho\rho'=14}$}
\put(-15,-25){$\phi^6_{(1);\rho=1}$}
\put(25,0){$b_{4}$}
\put(-133,55){$b_{1}$}
\put(-105,55){$b_{4}$}
\put(-5,0){$b_{1}'$}
\put(-135,0){$b_{1}''$}
\put(-165,35){$b_{4}''$}
\put(-38,0){$b_{4}'$}
\put(-105,0){$b_{4}'$}
\put(-130,-25){$\phi^6_{(2);\rho\rho'=14}$}

\end{figure}

Graphs of type $H$ for which the tadpole is on the vertex $\phi^6_{(1)}$ ($H^+_{14}$ and $H^-_{14}$ are drawn in Figure \ref{fig:H}) are now discussed and we separate them in different sector.

Given $\rho\rho'$, to $\Gamma_{6;2;\rho\rho'}$ contribute $H^+_{\rho\rho'}$ and
$H^-_{\rho\rho'}$.
Thus,  $\Gamma_{6;2;\rho\rho'}$ includes the amplitudes such that 
\bea
A_{H;6;2;\rho\rho'}(b_\rho,b'_\rho,b''_\rho) &=&  A_{H^+_{\rho\rho'}} + A_{H^-_{\rho\rho'}}  \\
&= &
\Big[-\lambda_{6;2;\rho\rho'}\Big]\Big[-\frac{\lambda_{6;1;\rho}}{3}\Big]
\Big[K_{H;\rho\rho'}^+\Big] S^3(b_\rho,b_\rho'')
+\Big[-\lambda_{6;2;\rho\rho'}\Big] \Big[-\frac{\lambda_{6;1;\rho'}}{3}\Big]
\Big[K_{H;1}^-\Big] S^3(b_{\rho'},b_{\rho'}')
\nonumber 
\eea
where the combinatorial factors are given by 
$K_{H;\rho\rho'}^\pm = 3 \cdot 2$. At low external momenta, the above
formula finds the form
\beq
A_{H;6;2;\rho\rho'}(0,\dots,0) = 2 \lambda_{6;2;\rho\rho'}\Big[
\lambda_{6;1;\rho} +  \lambda_{6;1;\rho'} \Big] S^1 
\label{hrhorho}
\eeq

\begin{figure}[ht]
\centering
 \begin{minipage}{.8\textwidth}
 \centering
 \includegraphics[width=16cm, height=2cm]{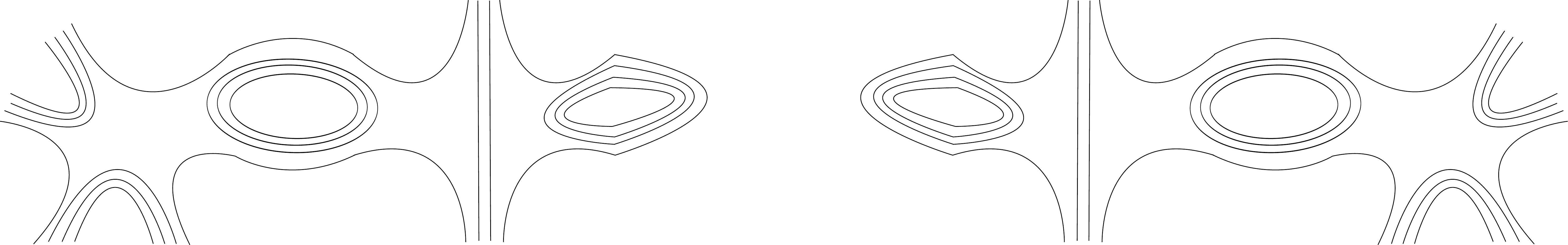}  
\vspace{0.5cm}
\caption{{\small  Graphs of type $G$: $G^+_{\rho\rho'}$ (left) and $G^-_{\rho\rho'}$
(right) 
 are parameterized by $\rho\rho'$ indices of $\phi^6_{(2)}$ 
(and $\rho$ of $\phi^6_{(1)}$ becomes redundant);  here $\rho\rho'=14$.}}
\label{fig:G}
\end{minipage}
\put(-387,-25){$\phi^6_{(1);\rho=1}$}
\put(-385,60){$b_{1}$}
\put(-283,60){$b_{1}$}
\put(-260,60){$b_{4}$}
\put(-283,0){$b_{1}'$}
\put(-258,0){$b_{4}$}
\put(-350,0){$b_{1}'$}
\put(-378,0){$b_{4}'$}
\put(-408,0){$b_{1}''$}
\put(-410,45){$b_{4}''$}
\put(-285,-25){$\phi^6_{(2);\rho\rho'=14}$}
\put(0,-25){$\phi^6_{(1);\rho=1}$}
\put(-109,60){$b_{1}$}
\put(-85,60){$b_{4}$}
\put(15,60){$b_{4}$}
\put(-109,0){$b_{1}$}
\put(-85,0){$b_{4}'$}
\put(-18,0){$b_{4}'$}
\put(10,0){$b_{1}'$}
\put(40,0){$b_{4}''$}
\put(38,45){$b_{1}''$}
\put(-115,-25){$\phi^6_{(2);\rho\rho'=14}$}
\end{figure}

\begin{figure}[ht]
\centering
 \begin{minipage}{.8\textwidth}
 \centering
 \includegraphics[width=16cm, height=3.5cm]{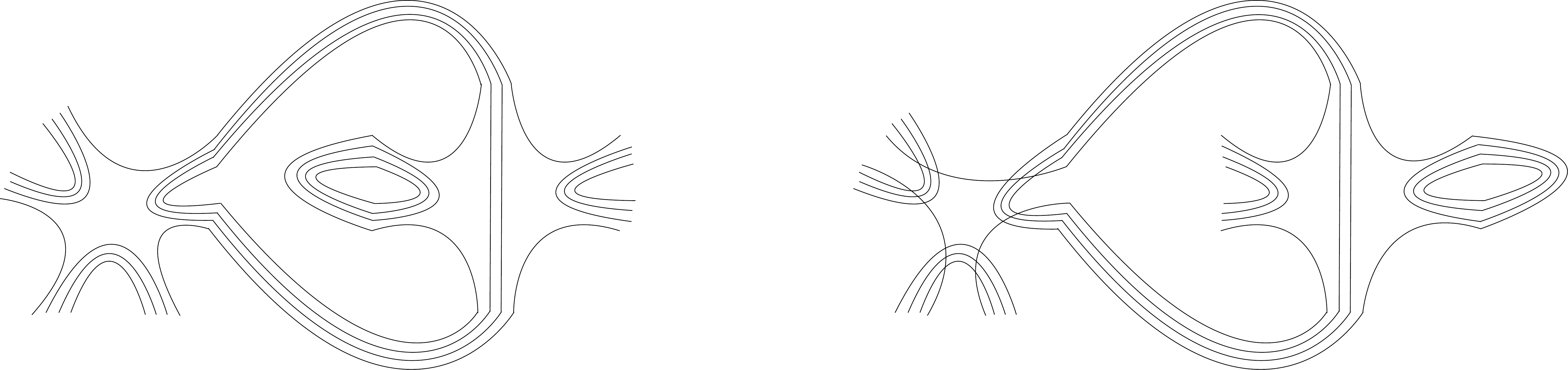}  
\vspace{0.5cm}
\caption{{\small  Graphs of type $I$: $I^+_{\rho\rho'}$  (left) and $I^-_{\rho\rho'}$ 
(right)  are parameterized by $\rho\rho'$ index of $\phi^6_{(2)}$
(and $\rho$ or $\rho'$ index of $\phi^6_{(1)}$ becomes redundant); here $\rho\rho'=14$.}}
\label{fig:I}
\end{minipage} 
\put(-387,-30){$\phi^6_{(1);\rho=1}$}
\put(-385,55){$b_{1}$}
\put(-235,50){$b_{1}$}
\put(-225,31){$b_{4}$}
\put(-235,12){$b_{1}'$}
\put(-355,0){$b_{1}'$}
\put(-380,0){$b_{4}'$}
\put(-408,0){$b_{1}''$}
\put(-410,45){$b_{4}''$}
\put(-295,-30){$\phi^6_{(2);\rho\rho'=14}$}
\put(-140,-30){$\phi^6_{(1);\rho'=4}$}
\put(-50,50){$b_{4}$}
\put(-55,31){$b_{1}$}
\put(-50,12){$b_{4}'$}
\put(-140,50){$b_{1}''$}
\put(-160,50){$b_{4}$}
\put(-170,40){$b_{4}''$}
\put(-130,-10){$b_{4}'$}
\put(-110,-10){$b_{1}'$}
\put(-50,-30){$\phi^6_{(2);\rho\rho'=14}$}
\end{figure}
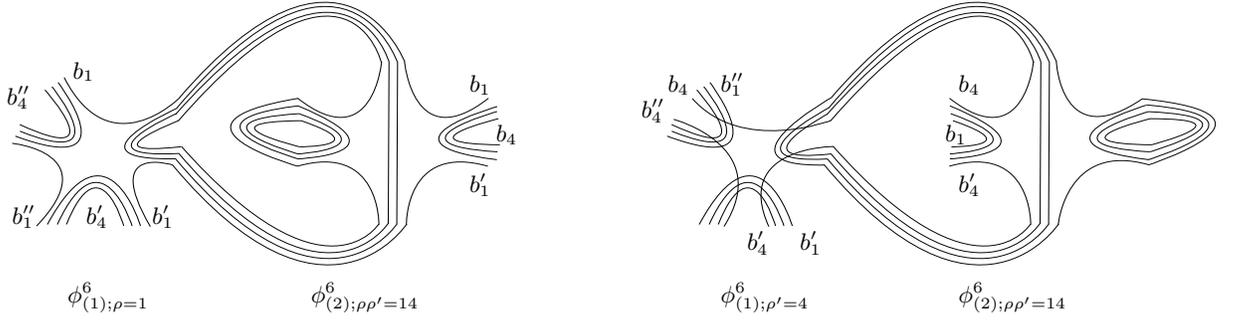

Consider now  graphs with the tadpole on the vertex $\phi^{6}_{(2)}$.
They appear in two forms,  $G$ and $I$, and possess indices of the vertex $\phi^{6}_{(2)}$ ($G^\pm_{14}$ and $I^\pm_{14}$ are given in Figure \ref{fig:G} and Figure \ref{fig:I}, respectively)

 Separating  the contributions 
in terms of the different six-point functions, one obtains:

\vspace{0.3cm}

$\bullet$ To $\Gamma_{6;1;1}$ contribute $G^+_{1\rho}$ and $I^+_{1\rho}$, for $\rho=2,3,4$;

$\bullet$ To $\Gamma_{6;1;2}$ contribute $G^+_{2\rho}$ and $I^+_{2\rho}$, for $\rho=3,4$, and
$G^-_{12}$ and $I^-_{12}$;

$\bullet$ To $\Gamma_{6;1;3}$ contribute
$G^-_{3\rho }$ and $I^-_{3\rho }$, for $\rho=1,2$,  
and $G^+_{34}$ and $I^+_{34}$;

$\bullet$ To  $\Gamma_{6;1;4}$ contribute $G^-_{4\rho}$ and
$I^-_{4\rho}$, for $\rho=1,2,3$.
\vspace{0.3cm}

Then, for instance, the following contribute to $\Gamma_{6;1;1}$: 
\bea
&&
A_{GI;6;1;1}(b_\rho,b'_\rho,b''_\rho) = \sum_{\rho=2,3,4} \Big[ A_{G^+_{1\rho}} + A_{I^+_{1\rho}}\Big]  \crcr
&&= 
\sum_{\rho=2,3,4}
\Big[-\lambda_{6;2;1\rho}\Big]\Big[-\frac{\lambda_{6;1;1}}{3}\Big]
\Big[K_{G;1\rho}^+\Big]  \Big[ S^4(b_1,b_1',b_\rho)
\quad + \quad \text{circular permutations} \quad [b_{1,\rho}\to b_{1,\rho}' \to b_{1,\rho}'' ]  \Big]\crcr
&&
+ \Big[-\frac{\lambda_{6;1;1}}{3}\Big] \Bigg\{
\sum_{\rho=2,3,4}
\Big[-\lambda_{6;2;1\rho}\Big] \Big[K_{I;1\rho}^+ \Big]
\Bigg\}  \Big[ S^{14}(b_1,b_1')
\quad + \quad \text{circular permutations} \quad [b_{1}\to b_{1}' \to b_{1}'' ]  \Big]
\eea
with combinatorial factors given by 
$K^+_{G;1\rho} =K^+_{I;1\rho} = 3$. The same yields at zero external data
\beq
A_{GI;6;1;1}(0,\dots,0) = 3 \lambda_{6;1;1}\Big[
\sum_{\rho=2,3,4} \lambda_{6;2;1\rho}\Big]  [S^1+S^{12}]
\label{gi1}
\eeq
In the same way, it can be shown that, for all $\rho\in \{1,2,3,4\}$,
to $\Gamma_{6;1;\rho}$ contribute
\beq
A_{GI;6;1;\rho}(0,\dots,0) = 3 \lambda_{6;1;\rho}\Big[
\sum_{\rho' \in \{1,2,3,4\}\setminus \{\rho\} } \lambda_{6;2;\rho\rho'}\Big]  [S^1+S^{12}]
\label{girho}
\eeq

\subsection{Graphs J, L, M and N}
 \label{gra6E}

Configurations defined by contractions of two 
vertices of the type $\phi^6_{(2)}$ have to be discussed finally. 
In this case, because of numerous relevant configurations,
we will use compact notations for vertices $\phi^6_{(2)}$ in the following
form 
\begin{figure}[ht]
\centering
 \begin{minipage}{.8\textwidth}
 \centering
 \includegraphics[width=3cm, height=2cm]{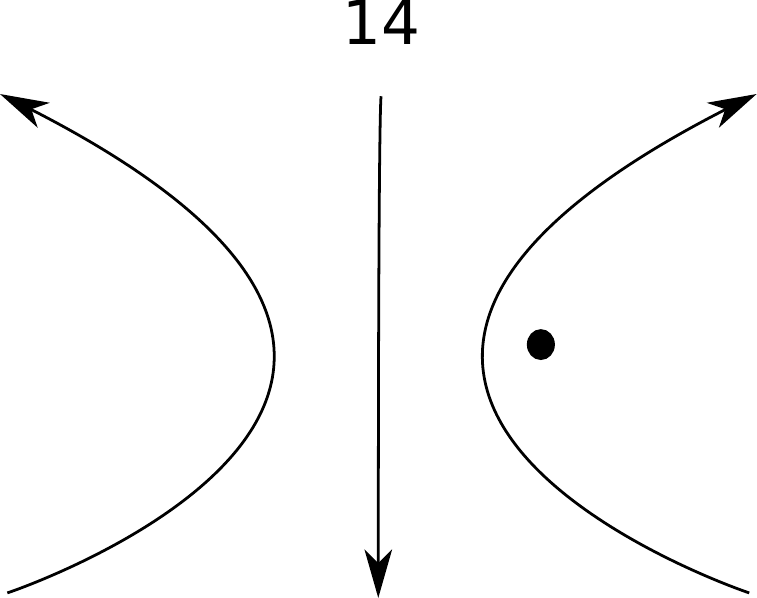}  
\caption{{\small Simplified notation of $\phi^6_{(2)}$ vertex for $\rho\rho'=14$. }}
\label{fig:simpliphi}
\end{minipage}
\end{figure}

Figure \ref{fig:simpliphi} displays all features of the vertex $\phi^6_{(2)}$:
arrows show how the vertex is oriented (positions of $\varphi$ and $\bar\varphi$),
the point underlines the fact that the left and right part of the vertex are not 
symmetric and, last, $\rho\rho'$. Omitting  the latter indices
means that the vertex $\phi^6_{(2)}$ is considered in general. 

Significant graphs can be described by six different configurations themselves
divided into two further cases as represented in Figure \ref{fig:phi6phi6}:

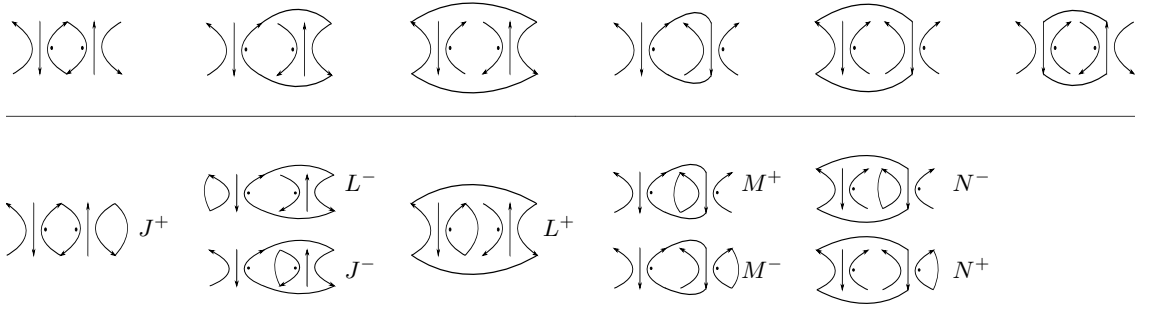
\begin{figure}[ht]
\centering
 \begin{minipage}{.8\textwidth}
 \centering
 \includegraphics[width=15cm, height=4cm]{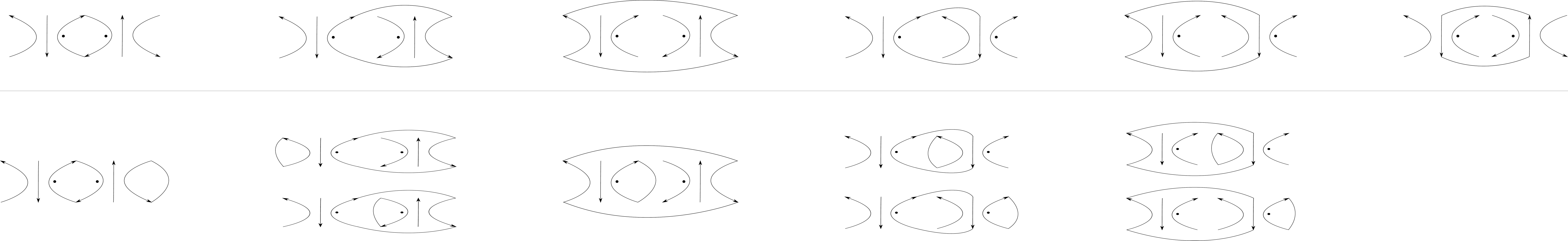}  
\caption{{\small Graphs $J^\pm, L^\pm,M^\pm$ and $N^\pm$. }}
\label{fig:phi6phi6}
\end{minipage}
\put(-358,-17){$J^+$}
\put(-280,0){$L^-$}
\put(-280,-33){$J^-$}
\put(-205,-17){$L^+$}
\put(-130,0){$M^+$}
\put(-130,-33){$M^-$}
\put(-50,0){$N^-$}
\put(-50,-33){$N^+$}
\end{figure} 

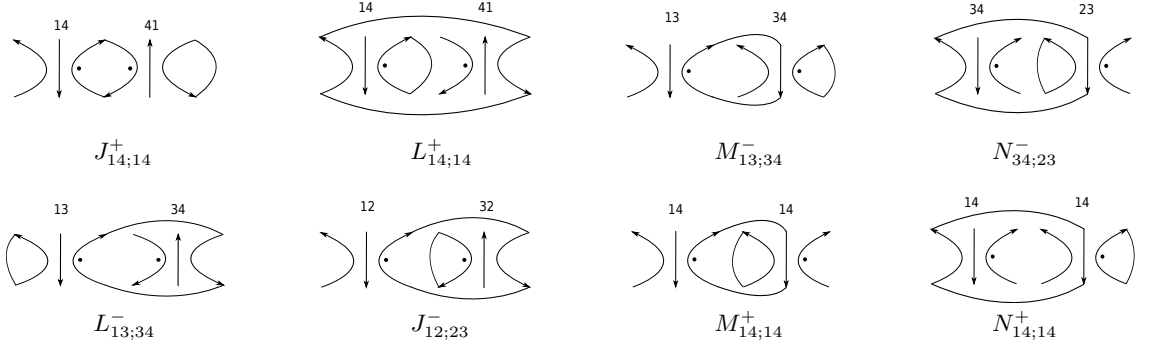
\begin{figure}[ht]
\centering
 \begin{minipage}{.8\textwidth}
 \centering
 \includegraphics[width=15cm, height=4cm]{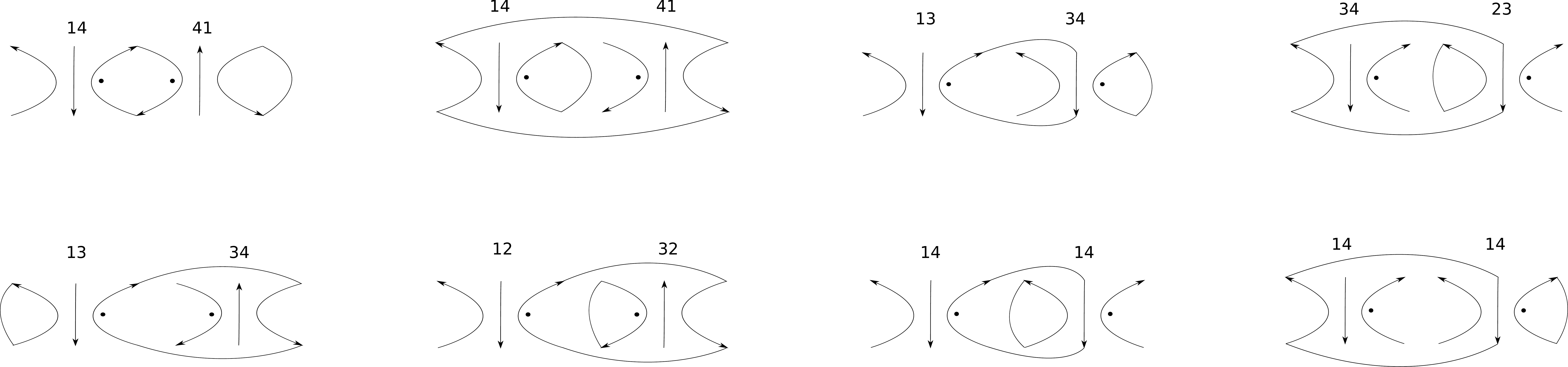}  
\vspace{0.5cm}
\caption{{\small Particular graphs $J^+_{14;14}$, $J^-_{12;23}$, 
$L^-_{13;34}$, $L^+_{14;14}$,  
$M^+_{14;14}$, $M^-_{13;34}$, $N^+_{14;14}$ and $N^-_{34;23}$. }}
\label{fig:JLMN}
\end{minipage}
\put(-375,25){$J^+_{14;14}$}
\put(-375,-40){$L^-_{13;34}$}
\put(-255,25){$L^+_{14;14}$}
\put(-255,-40){$J^-_{12;23}$}
\put(-140,25){$M^-_{13;34}$}
\put(-140,-40){$M^+_{14;14}$}
\put(-35,25){$N^-_{34;23}$}
\put(-35,-40){$N^+_{14;14}$}
\end{figure}

Note that, in the following, we have excluded many convergent situations
(for instance, all configurations coming from the sixth graph in Figure \ref{fig:phi6phi6} 
are all convergent) and have merged many combinatorially equivalent graphs (in Figure \ref{fig:phi6phi6}, $J^+$ and $J^-$ should have each 
a partner combinatorially equivalent to themselves). 
Graphs are now indexed by twice a pair $\rho\rho';\bar\rho\bar\rho'$, 
one pair for each vertex. 
Only graphs of the form $J^\pm_{\rho\rho';\bar\rho\bar\rho'}$, $L^\pm_{\rho\rho';\bar\rho\bar\rho'}$, 
$M^\pm_{\rho\rho';\bar\rho\bar\rho'}$ and $N^\pm_{\rho\rho';\bar\rho\bar\rho'}$ might
lead to divergence 
(graphs $J^+_{14;14}$, $J^-_{12;23}$, $L^-_{13;34}$, $L^+_{14;14}$,  
$M^+_{14;14}$, $M^-_{13;34}$, $N^+_{14;14}$ and $N^-_{34;23}$ are given in Figure \ref{fig:JLMN}). 

The following decomposition is valid:
 
\vspace{0.3cm}
$\bullet$ To $\Gamma_{6;2;14}$ contribute $J^+_{14; 4\rho }$ and
$M^+_{14; 4 \rho }$, for $\rho=1,2,3$,
 $L^+_{14;1 \rho}$ and $N^+_{14;1\rho}$, for $\rho=2,3,4$;

$\bullet$ To $\Gamma_{6;2;13}$ contribute 
$J^+_{13; 3\rho}$ and $M^+_{13; 3\rho }$, for $\rho=1,2$, 
$J^-_{13;34}$ and $M^-_{13;34}$,   
$L^+_{13;1\rho}$ and $N^+_{13;1\rho}$, for $\rho=2,3,4$;

$\bullet$ To $\Gamma_{6;2;12}$ contribute
$J^-_{12;2\rho}$ and $M^-_{12;2\rho}$, for $\rho=3,4$,  
$J^+_{12;12}$ and $M^+_{12;12}$,   
$L^+_{12;1\rho}$ and $N^+_{12;1\rho}$, for $\rho=2,3,4$;

$\bullet$ To  $\Gamma_{6;2;23}$ contribute 
$J^+_{23;3\rho}$ and $M^+_{23;3\rho}$, for $\rho=1,2$, 
$J^-_{23;34}$ and $M^-_{23;34}$,   
$L^-_{12;23}$ and $N^-_{23;12}$,
$L^+_{23;2\rho}$ and $N^+_{23;2\rho}$, for $\rho=3,4$;

$\bullet$ To  $\Gamma_{6;2;24}$ contribute 
$J^+_{24;4\rho}$ and $M^+_{24;4\rho}$, for $\rho=1,2,3$,
$L^-_{12;24}$ and $N^-_{24;12}$,
$L^+_{24;2\rho}$ and $N^+_{24;2\rho}$, for $\rho=3,4$;

$\bullet$ To  $\Gamma_{6;2;34}$ contribute 
$J^+_{34;4\rho}$ and $M^+_{34;4\rho}$, for $\rho=1,2,3$,
$L^+_{34;34}$ and $N^+_{34;34}$, 
$L^-_{3\rho;34}$ and $N^-_{34;3\rho}$, for $\rho=1,2$;

\vspace{0.3cm}

Explicitly, we count the following contributions for $\Gamma_{6;2;14}$:
\bea
&&
A_{JLMN;6;2;14}(b_\rho,b_\rho',b_\rho'')  = 
 \sum_{\rho=1,2,3} [A_{J^+_{14;4\rho}} +A_{M^+_{14;4\rho}} ]
+ \sum_{\rho=2,3,4} [A_{L^+_{14;1\rho}} +A_{N^+_{14;1\rho}}  ]\crcr
&&
=
\Big[-\lambda_{6;2;14}\Big] \Bigg\{ 
 \frac{1}{2!}\Big[-\lambda_{6;2;14}\Big] \Big[K_{J;14;14}^+ \Big]
S^4(b_4,b_4',b_1')
+ \sum_{\rho=2,3}
\Big[-\lambda_{6;2;\rho 4}\Big] \Big[K_{J;14;4\rho }^+ \Big]
S^4(b_4,b_4',b_\rho')
  \crcr
&&+ \frac{1}{2!} \Big[-\lambda_{6;2;14}\Big] \Big[K^+_{L;14;14} \Big]
S^4(b_1,b_1'',b_4'')
+ \sum_{\rho=2,3}
\Big[-\lambda_{6;2;1\rho}\Big] \Big[K^+_{L;14;1\rho} \Big]
S^4(b_1,b_1'',b_\rho'')
  \crcr
 &&
+ \Bigg[ 
\frac{1}{2!}\Big[-\lambda_{6;2;14}\Big] \Big[K_{M;14;14}^+ \Big]
+ \sum_{\rho=2,3} \Big[-\lambda_{6;2;\rho 4}\Big] \Big[K_{M;1;4\rho}^+ \Big]
\Bigg] S^{14}(b_4,b_4')
\crcr
&&+\Bigg[ 
\frac{1}{2!}\Big[-\lambda_{6;2;14}\Big] \Big[K_{N;14;14}^+ \Big]
+ \sum_{\rho=2,3}  \Big[-\lambda_{6;2;1\rho}\Big] \Big[K_{N;14; 1\rho}^+ \Big]
\Bigg]S^{14}(b_1,b_1'') 
\Bigg\}
\eea
with  combinatorial factors 
$K_{\bullet;14;14}^+ = 2$ and, otherwise,
 $K_{\bullet;14;\rho\rho'}^+=1$ for $\rho, \rho' \neq 1,4$. 
At low external momenta, we find
\beq
A_{JLMN;6;2;14}(0,\dots,0) = \lambda_{6;2;14}
\Big[ \sum_{\rho \in\{2,3,4\}} \lambda_{6;2;1\rho} 
+ \sum_{\rho \in\{1,2,3\}} \lambda_{6;2;\rho 4} \Big] [S^{1} + S^{12}] 
\label{jm1}
\eeq
 By a similar calculation, the following holds, for all $\rho,\rho'$,
\beq
A_{JLMN;6;2;\rho\rho'}(0,\dots,0) = \lambda_{6;2;\rho\rho'}\Big[
 \sum_{\bar\rho \in\{1,2,3,4\} \setminus \{\rho\}} \lambda_{6;2;\rho\bar\rho} 
+ \sum_{\bar\rho \in\{1,2,3,4\}  \setminus \{\rho'\}} \lambda_{6;2;\rho' \bar\rho} \Big] [S^{1} +S^{12}] 
\label{jmrho}
\eeq

All six-point functions at low external
data $\Gamma_{6;\xi;\rho/\rho\rho'}(0,\dots,0)$ are now summed. 
By adding \eqref{frho} and \eqref{girho}, we have
\beq
\Gamma_{6;1;\rho} (0,\dots,0)=
- \lambda_{6;1;\rho} +  \lambda_{6;1;\rho}
\Bigg[ 6\,\lambda_{6;1;\rho}S^1 \,  + 3\Big[\sum_{\rho'\in \{1,2,3,4\}
\setminus \rho}\lambda_{6;2;1\rho'} \Big]  [S^1 +  S^{12}] \Bigg]
\label{gam61rho}
\eeq
Moreover, adding \eqref{hrhorho} and \eqref{jmrho} yields 
\beq
\Gamma_{6;2;\rho\rho'}(0,\dots,0) =
-\lambda_{6;2;\rho\rho'} +  \lambda_{6;2;\rho\rho'} \Bigg[
2[ \lambda_{6;1;\rho} +  \lambda_{6;1;\rho'} ]S^1
+
\Big[ \sum_{\bar\rho \in \{1,2,3,4\} \setminus \{\rho\} }\lambda_{6;2;\rho \bar\rho}
 +  \sum_{\bar\rho \in \{1,2,3,4\} \setminus \{\rho'\} }\lambda_{6;2;\rho' \bar\rho}\Big]   [S^1 +  S^{12}] \Bigg] 
\eeq
which achieves the proof of Lemma \ref{lemgam6}.

\section{Proof of Lemma \ref{lem:wphigam61}}
\label{app:fourloop}

We start the computation of the wave function renormalization
and the truncated amputated 1PI six-point functions
at four loops and at second and third order of perturbation theory,
respectively.  In this section, we set $\lambda_{6;2;\rho\rho'}=0$
as explained in Section \ref{sect:beta6phi}  and 
 focus on the contributions for $\Sigma$ 
and $\Gamma_{6;1;\rho}$ made only with $\phi^{6}_{(1)}$
vertices. 

We introduce the formal sums
\bea
\cS^1(b)&:=&
 \sum_{p_1,\dots, p_{12}} \Big[ 
\frac{1}{(b^2 + p_1^2+ p_2^2+ p_3^2+m^2)^2}
\frac{1}{(b^2 + p_4^2+ p_5^2+ p_6^2+m^2)}
\frac{1}{(b^2 + p_7^2+ p_8^2+ p_9^2+m^2)} \times \crcr
&& \qquad \quad 
\frac{1}{(b^2 + p_{10}^2+ p_{11}^2+ p_{12}^2+m^2)} \Big]  
\crcr
\cS^{12}(b)&:=&
 \sum_{p_1,\dots, p_{12}} \Big[ 
\frac{1}{(b^2 + p_1^2+ p_2^2+ p_3^2+m^2)^2}
\frac{1}{(p_1^2 + p_4^2+ p_5^2+ p_6^2+m^2)}
\frac{1}{(p_1^2 + p_7^2+ p_8^2+ p_9^2+m^2)} \times \crcr
&& \qquad \quad 
\frac{1}{(b^2 + p_{10}^2+ p_{11}^2+ p_{12}^2+m^2)} \Big] \crcr
\cS^2(b,b')&:=&
 \sum_{p_1,\dots, p_{12}} \Big[ 
\frac{1}{(b^2 + p_1^2+ p_2^2+ p_3^2+m^2)^2}
\frac{1}{(b^2 + p_4^2+ p_5^2+ p_6^2+m^2)}
\frac{1}{(b^2 + p_7^2+ p_8^2+ p_9^2+m^2)} \times \crcr
&& \qquad \quad 
\frac{1}{(b^2 + p_{10}^2+ p_{11}^2+ p_{12}^2+m^2)}
 \frac{1}{(b'^2 + p_{10}^2+ p_{11}^2+ p_{12}^2+m^2)} \Big]  
\crcr
\cS^{21}(b,b')&=&
 \sum_{p_1,\dots, p_{12}} \Big[ 
\frac{1}{(b^2 + p_1^2+ p_2^2+ p_3^2+m^2)^2}
\frac{1}{(p_1^2 + p_4^2+ p_5^2+ p_6^2+m^2)}
\frac{1}{(p_1^2 + p_7^2+ p_8^2+ p_9^2+m^2)} \times \crcr
&& \qquad \quad 
\frac{1}{(b^2 + p_{10}^2+ p_{11}^2+ p_{12}^2+m^2)}
\frac{1}{(b'^2 + p_{10}^2+ p_{11}^2+ p_{12}^2+m^2)} \Big]    
\crcr
\cS^{3}(b,b',b'') &:=& 
\sum_{p_1,\dots, p_{12}} \Big[ 
\frac{1}{(b^2 + p_1^2+ p_2^2+ p_3^2+m^2)}
\frac{1}{(b^2 + p_4^2+ p_5^2+ p_6^2+m^2)}
\frac{1}{(b''^2 + p_1^2+ p_2^2+ p_3^2+m^2)} \times \crcr
&& \qquad \quad 
\frac{1}{(b''^2 + p_{7}^2+ p_{8}^2+ p_{9}^2+m^2)}
\frac{1}{(b'^2 + p_{7}^2+ p_{8}^2+ p_{9}^2+m^2)}
 \frac{1}{(b'^2 + p_{10}^2+ p_{11}^2+ p_{12}^2+m^2)} \Big]  
\crcr
\cS^4(b,b')&:=&
 \sum_{p_1,\dots, p_{12}} \Big[ 
\frac{1}{(b^2 + p_1^2+ p_2^2+ p_3^2+m^2)^2}
\frac{1}{(b^2 + p_4^2+ p_5^2+ p_6^2+m^2)}
\frac{1}{(b^2 + p_7^2+ p_8^2+ p_9^2+m^2)} \times \crcr
&& \qquad \quad 
\frac{1}{(b^2 + p_{10}^2+ p_{11}^2+ p_{12}^2+m^2)}
 \frac{1}{(b'^2 + p_{1}^2+ p_{2}^2+ p_{3}^2+m^2)} \Big]  
\crcr
\cS^{41}(b,b')&:=&
 \sum_{p_1,\dots, p_{12}} \Big[ 
\frac{1}{(b^2 + p_1^2+ p_2^2+ p_3^2+m^2)^2}
\frac{1}{(b^2 + p_4^2+ p_5^2+ p_6^2+m^2)}
\frac{1}{(p_1^2 + p_7^2+ p_8^2+ p_9^2+m^2)} \times \crcr
&& \qquad \quad 
\frac{1}{(p_1^2 + p_{10}^2+ p_{11}^2+ p_{12}^2+m^2)}
\frac{1}{(b'^2 + p_{1}^2+ p_{2}^2+ p_{3}^2+m^2)} \Big]   
\crcr
\cS'^4(b,b')&:=&
 \sum_{p_1,\dots, p_{12}} \Big[ 
\frac{1}{(b^2 + p_1^2+ p_2^2+ p_3^2+m^2)}
\frac{1}{(b^2 + p_4^2+ p_5^2+ p_6^2+m^2)}
\frac{1}{(b'^2 + p_7^2+ p_8^2+ p_9^2+m^2)} \times \crcr
&& \qquad \quad 
\frac{1}{(b'^2 + p_{10}^2+ p_{11}^2+ p_{12}^2+m^2)}
 \frac{1}{(b'^2 + p_{1}^2+ p_{2}^2+ p_{3}^2+m^2)^2} \Big]  \crcr
\cS'^{41}(b,b')&:=&
 \sum_{p_1,\dots, p_{12}} \Big[ 
\frac{1}{(b^2 + p_1^2+ p_2^2+ p_3^2+m^2)}
\frac{1}{(b^2 + p_4^2+ p_5^2+ p_6^2+m^2)}
\frac{1}{(p_1^2 + p_7^2+ p_8^2+ p_9^2+m^2)} \times \crcr
&& \qquad \quad 
\frac{1}{(p_1^2 + p_{10}^2+ p_{11}^2+ p_{12}^2+m^2)}
\frac{1}{(b'^2 + p_{1}^2+ p_{2}^2+ p_{3}^2+m^2)^2} \Big]   
\crcr
\cS^{5}(b,b',b'') &:=& 
\sum_{p_1,\dots, p_{12}} \Big[ 
\frac{1}{(b^2 + p_1^2+ p_2^2+ p_3^2+m^2)}
\frac{1}{(b^2 + p_4^2+ p_5^2+ p_6^2+m^2)}
\frac{1}{(b''^2 + p_1^2+ p_2^2+ p_3^2+m^2)} \times \crcr
&& \qquad \quad 
\frac{1}{(b''^2 + p_{7}^2+ p_{8}^2+ p_{9}^2+m^2)}
\frac{1}{(b'^2 + p_{1}^2+ p_{2}^2+ p_{3}^2+m^2)}
 \frac{1}{(b'^2 + p_{10}^2+ p_{11}^2+ p_{12}^2+m^2)} \Big] 
 \crcr
\cS^{1}_{(1)}  &:=&  \sum_{p_1,\dots, p_{12}} \Big[ 
\frac{1}{(p_1^2+ p_2^2+ p_3^2+m^2)^3}
\frac{1}{(p_4^2+ p_5^2+ p_6^2+m^2)}
\frac{1}{(p_7^2+ p_8^2+ p_9^2+m^2)} 
\frac{1}{(p_{10}^2+ p_{11}^2+ p_{12}^2+m^2)} \Big]  
\crcr
\cS^{1}_{(2)}  &:=& \sum_{p_1,\dots, p_{12}} \Big[ 
\frac{1}{(p_1^2+ p_2^2+ p_3^2+m^2)^2}
\frac{1}{(p_4^2+ p_5^2+ p_6^2+m^2)^2}
\frac{1}{(p_7^2+ p_8^2+ p_9^2+m^2)} 
\frac{1}{(p_{10}^2+ p_{11}^2+ p_{12}^2+m^2)} \Big]  \crcr
\cS^{12}_{(1)}  &:=&  \sum_{p_1,\dots, p_{12}} \Big[ 
\frac{1}{(p_1^2+ p_2^2+ p_3^2+m^2)^3}
\frac{1}{(p_1^2+ p_4^2+ p_5^2+ p_6^2+m^2)}
\frac{1}{(p_1^2+ p_7^2+ p_8^2+ p_9^2+m^2)}\times  \crcr
&& 
\frac{1}{(p_{10}^2+ p_{11}^2+ p_{12}^2+m^2)} \Big]  \crcr
\cS^{12}_{(2)}  &:=&  \sum_{p_1,\dots, p_{12}} \Big[ 
\frac{1}{(p_1^2+ p_2^2+ p_3^2+m^2)^2}
\frac{1}{(p_1^2+ p_4^2+ p_5^2+ p_6^2+m^2)}
\frac{1}{(p_1^2+ p_7^2+ p_8^2+ p_9^2+m^2)} \times\crcr
&&
\frac{1}{(p_{10}^2+ p_{11}^2+ p_{12}^2+m^2)^2} \Big] 
\eea
Note that $\cS^2(0,0)=\cS^3(0,0,0)=\cS^1_{(2)}$, 
$\cS^{21}(0,0)=\cS^{12}_{(2)}$, 
 $\cS^4(0,0)=\cS'^4(0,0)=\cS^5(0,0,0)=\cS^1_{(1)} $ and $\cS^{41}(0,0)=\cS'^{41}(0,0)=\cS^{12}_{(1)}$.

\begin{figure}[ht]
\centering
 \begin{minipage}{.8\textwidth}
 \centering
 \includegraphics[width=2.5cm, height=0.8cm]{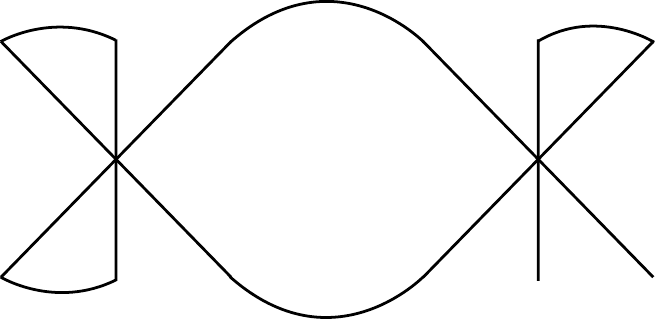}  
\caption{{\small General form of the second correction to $\Sigma$. }}
\label{fig:tad4th}
\end{minipage}
\end{figure}

  The new contributions at four loops which supplements $\Sigma(b_\rho,b_\rho')$ are given by graphs $T_{\rho;\rho'}$ of the form given in Figure \ref{fig:tad4th}
where $\rho$ and $\rho'$ are the two indices of the vertices. 

The sum of amplitudes of $T_{\rho;\rho'}$ contributing to $\Sigma$ 
at four loops is such that
\bea
A_{T}(b_1,b_2,b_3,b_4) &=& 
\sum_{\rho=1,\dots,4} \Big[ A_{T_{\rho;\rho}}(b_\rho)
+ \sum_{\rho'\neq \rho} A_{T_{\rho;\rho'}}(b_\rho) \Big] \crcr
&=&
\sum_{\rho=1,\dots,4}
\Bigg\{ \frac{1}{2!}\Big[-\frac{\lambda_{6;\rho}}{3}\Big]^2 
K_{T;\rho;\rho} \,\cS^1(b_\rho) + 
\Big(\sum_{\rho'\neq \rho} \Big[-\frac{\lambda_{6;\rho}}{3}\Big]\Big[-\frac{\lambda_{6;\rho'}}{3}\Big] 
K_{T;\rho;\rho'} \Big)\cS^{12}(b_\rho) \Bigg\}
\label{4lot}
\eea
with $K_{T;\rho;\rho} =3^2 \cdot 2^2$ and $K_{T;\rho;\rho'}=3^2 \cdot 2$, where $\rho'\neq \rho$. Differentiating \eqref{4lot} with respect 
to $b_1^2$ yields: 
\bea
&&
- \partial_{b_1^2} \; A_{T}\Big|_{b_\rho=0} =
 - \Bigg[ 2 \lambda_{6;1}^2  [-2 \cS^{1}_{(1)} -3\cS^{1}_{(2)} ]
 + 2 \lambda_{6;1}(\lambda_{6;2}+\lambda_{6;3}+\lambda_{6;4})
\Big[-2\cS^{12}_{(1)}-\cS^{12}_{(2)}\Big] \Bigg]
\eea
Using previous results on two-loop calculations  from Lemma \ref{lem:self}, the  expression $Z$ \eqref{wphi61} is recovered. 

Next, we evaluate the additional contributions at four loops  
to $\Gamma_{6;\rho}$. Those contributions are of the 
general form given by the following diagrams 

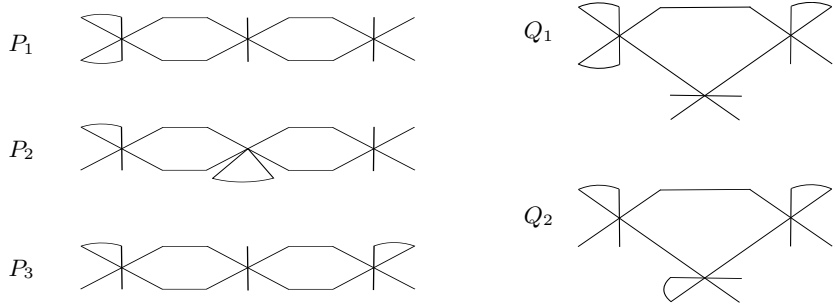
\begin{figure}[ht]
\centering
 \begin{minipage}{.8\textwidth}
 \centering
 \includegraphics[width=10cm, height=4cm]{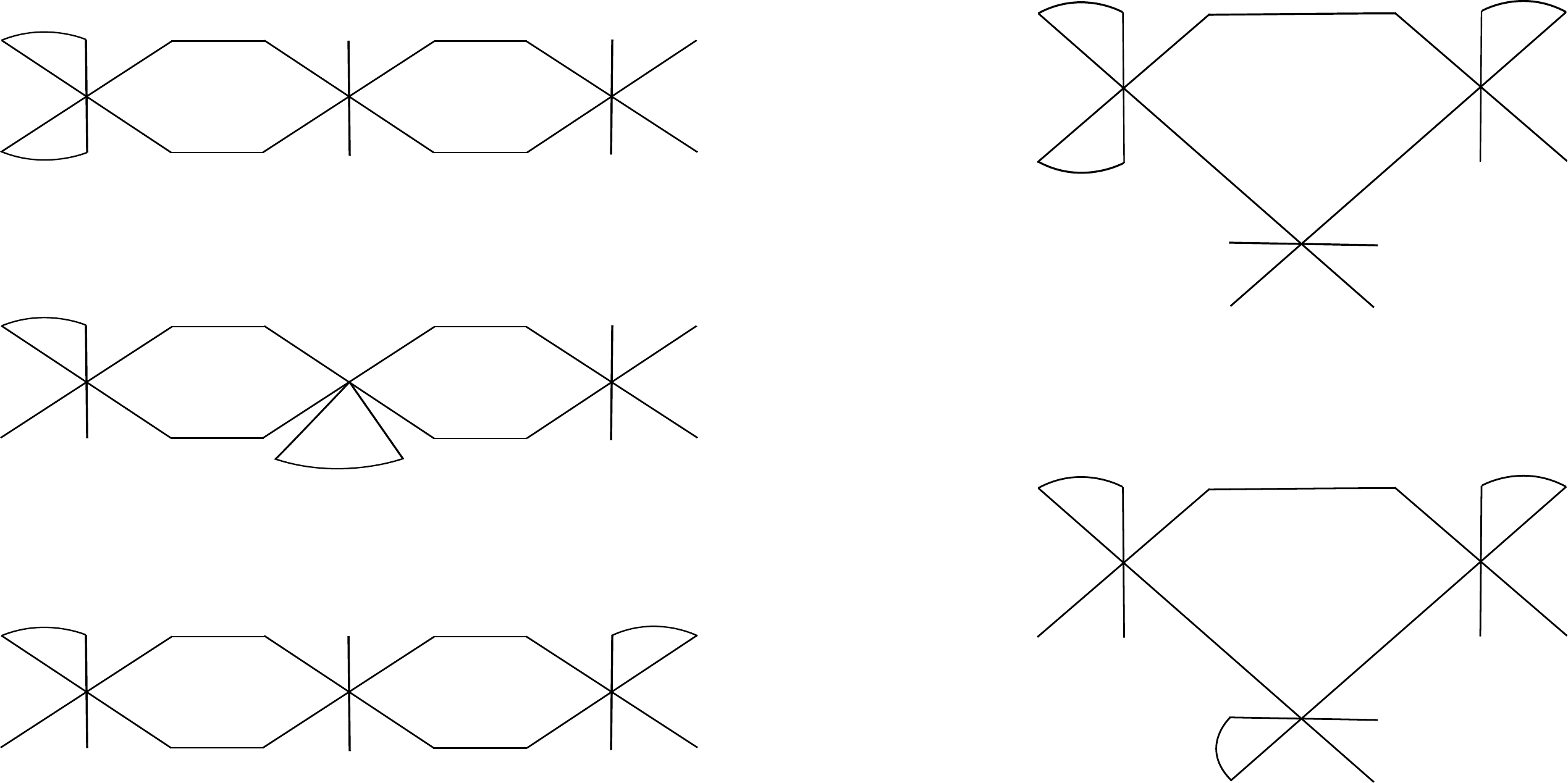}  
\vspace{0.5cm}
\caption{{\small Third order corrections to $\Gamma_{6;1}$. }}
\label{fig:PQ}
\end{minipage}
\put(-375,60){$P_1$}
\put(-375,20){$P_2$}
\put(-375,-25){$P_3$}
\put(-180,65){$Q_1$}
\put(-180,-5){$Q_2$}
\end{figure} 

The important graphs can be written as
$P_{1;\rho;\rho'}$, $P_{2;\rho}$,
$P_{3;\rho}$, $Q^\pm_{1;\rho;\rho'}$ and $Q^\pm_{2;\rho}$ and are not 
characterized by the three indices of their internal vertices but, at most,  by two of them. 
For instance, $P_{1;\rho;\rho'}$
can be fully represented by two indices of the three, whereas, for 
$P_{2;\rho}$ and $P_{3;\rho}$, a single index will be sufficient
to capture the relevant contribution. This index should be the same for
all internal vertices.

\begin{figure}[ht]
\centering
 \begin{minipage}{.8\textwidth}
 \centering
 \includegraphics[width=10cm, height=4cm]{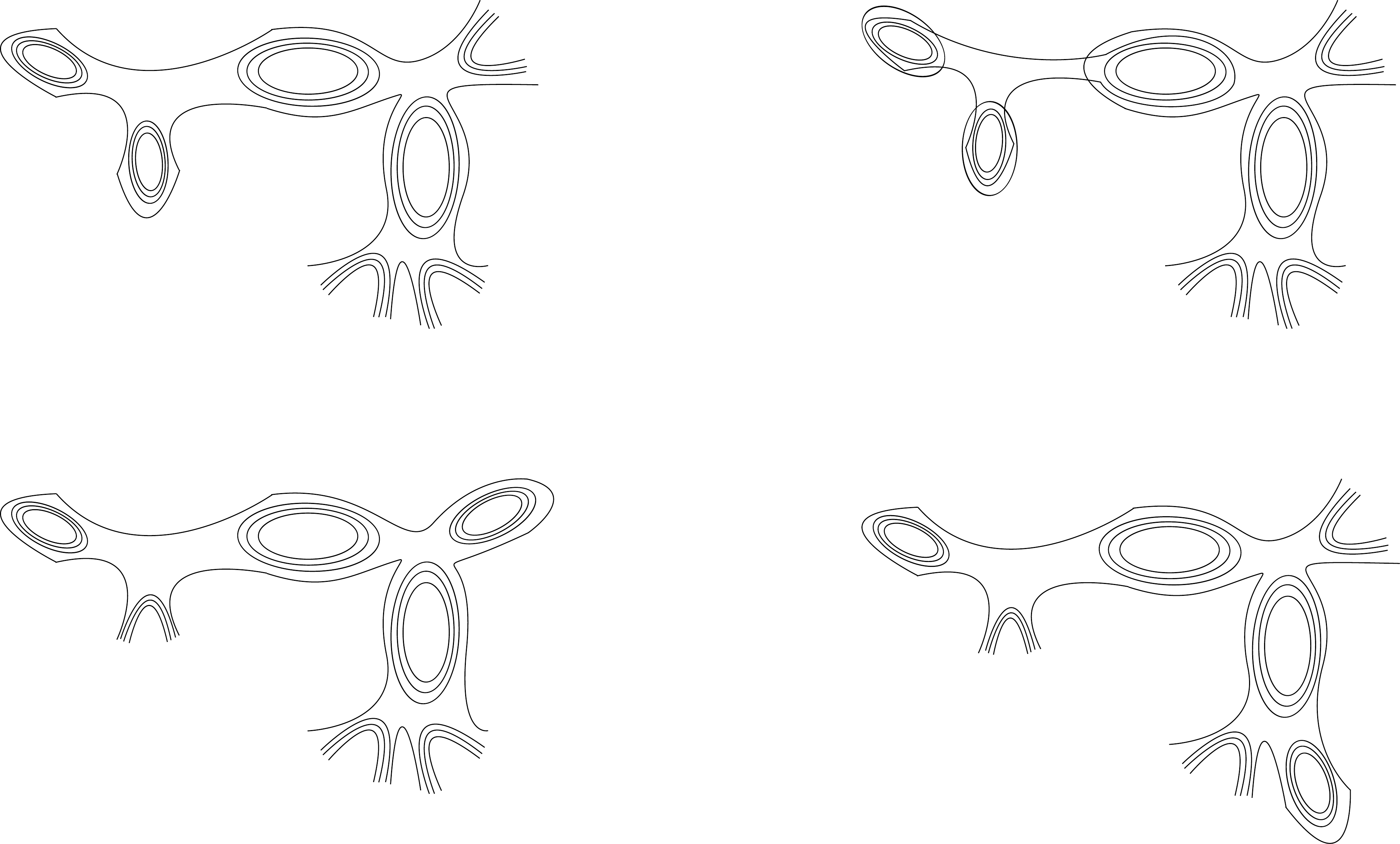}  
\vspace{0.5cm}
\caption{{\small Graphs of type $P$:  $P_{1;\rho;\rho'}$ 
graphs are parameterized by $\rho$ index of the bottom vertex and $\rho'$ index of the fully contracted vertex on the left
 (here $\rho=1$ and $\rho'=1,2$); $P_{2;\rho}$ and $P_{3;\rho}$ are parameterized by $\rho$
the unique and coinciding index of all vertices. }}
\label{fig:Prho}
\end{minipage}
\put(-375,55){$P_{1;\rho=1;\rho'=1}$}
\put(-260,90){$b_{1}$}
\put(-200,55){$P_{1;\rho=1;\rho'=2}$}
\put(-85,90){$b_{1}$}
\put(-375,-5){$P_{2:\rho=1}$}
\put(-245,-5){$b_{1}$}
\put(-200,-5){$P_{3:\rho=1}$}
\put(-85,25){$b_{1}$}
\end{figure}
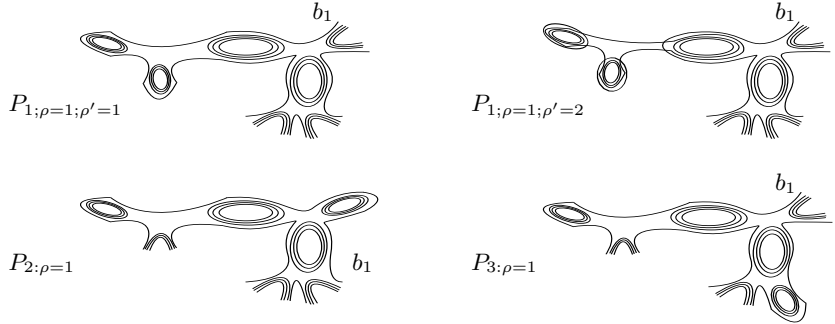

 Focusing on $P$ diagrams, 
to $\Gamma_{6;1}$ contribute $P_{1;1;\rho}$, $P_{2;1}$ and $P_{3;1}$
(the drawing of which is provided in Figure \ref{fig:Prho}) giving

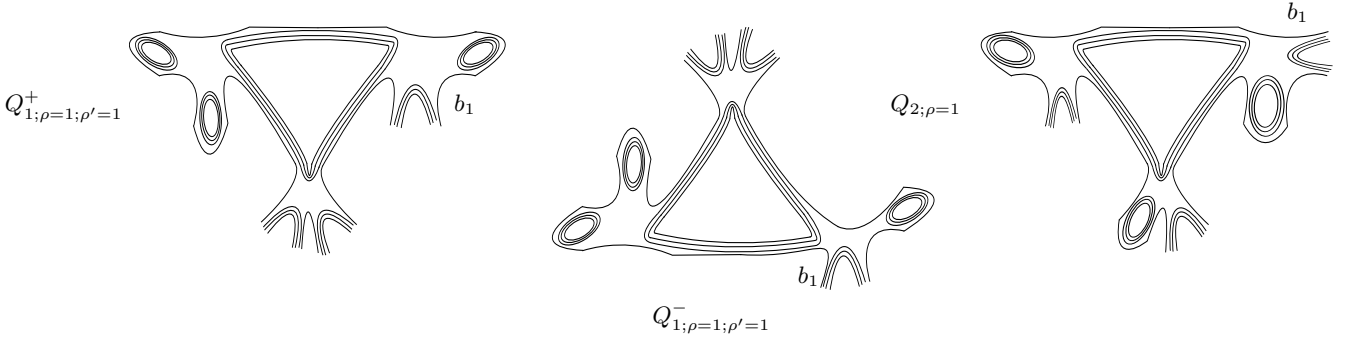
\begin{figure}[ht]
\centering
 \begin{minipage}{.8\textwidth}
 \centering
 \includegraphics[width=16cm, height=3.5cm]{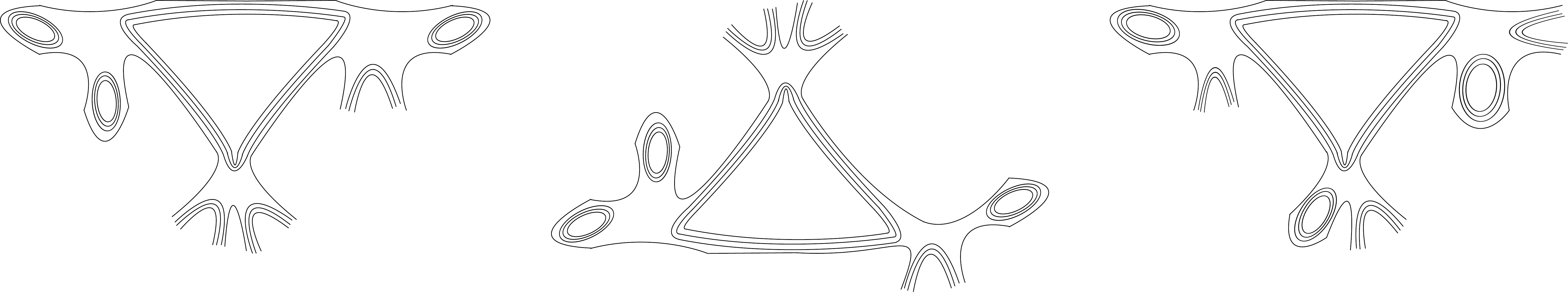}  
\vspace{0.5cm}
\caption{{\small Graphs of type Q: $Q^+_{1;\rho;\rho'}$, $Q^-_{1;\rho;\rho'}$
are parametrized by $\rho'$ index of the fully contracted vertex on the left and
$\rho$ unique and coinciding index of the two left vertices with external legs; 
$Q_{2;\rho}$ is parameterized by the unique index which should be common for all 
vertices. }}
\label{fig:Qrho}
\end{minipage}
\put(-455,55){$Q^+_{1;\rho=1;\rho'=1}$}
\put(-285,55){$b_{1}$}
\put(-210,-25){$Q^-_{1;\rho=1;\rho'=1}$}
\put(-155,-10){$b_{1}$}
\put(-120,55){$Q_{2;\rho=1}$}
\put(30,90){$b_{1}$}
\end{figure}

\bea
&&
A_{P;6;1}(b_\rho,b_\rho',b_\rho'') = \sum_{\rho=1,\dots,4} A_{P_{1;1;\rho}} + A_{P_{2;1}} + A_{P_{3;1}} \crcr
&&
= 
\frac{1}{3!}\Big[-\frac{\lambda_{6;1}}{3}\Big]^3 K_{P;1;1;1}
\Big[\cS^{2}(b_1,b_1')
+ [b_{1}\to b_{1}' \to b_{1}'' ]  \Big] \cr\cr
&& + 
\frac{1}{2!}\Big[-\frac{\lambda_{6;1}}{3}\Big]^2
\Big[\sum_{\rho=2,3,4}\Big[-\frac{\lambda_{6;\rho}}{3}\Big] K_{P;1;1;\rho}\Big] \Big[\cS^{21}(b_1,b_1') 
+    [b_{1}\to b_{1}' \to b_{1}'' ]  \Big]\crcr
&& 
+ \frac{1}{3!}\Big[-\frac{\lambda_{6;1}}{3}\Big]^3 K_{P;2;1}
\Big[\cS^{2}(b_1,b_1')
+ [b_{1}\to b_{1}' \to b_{1}'' ]  \Big] 
+ \frac{1}{3!}\Big[-\frac{\lambda_{6;1}}{3}\Big]^3 K_{P;3;1}
\Big[\cS^{3}(b_1,b_1',b_1'')
+ [b_{1}\to b_{1}' \to b_{1}'' ]  \Big] 
\eea
with
\beq
K_{P;1;1;1} = 3^4 \cdot 2^2 \qquad \quad 
K_{P;1;1;\rho\neq 1} = 3^3 \cdot 2^2\qquad \quad 
K_{P;2;1} = K_{P;3;1} =3^4 \cdot 2^3 
\eeq
Thus, we obtain at zero external momenta
\beq
A_{P;6;1}(0,\dots,0) 
= - 2\cdot 3\cdot 5\,\lambda_{6;1}^3 \,\cS^{1}_{(2)} 
 - 2\cdot 3\,\lambda_{6;1}^2
\Big[\sum_{\rho=2,3,4}\lambda_{6;\rho}\Big]  \cS^{12}_{(2)}
\eeq
Similarly, one infers, for $\rho=2,3,4$,
\beq
A_{P;6;\rho}(0,\dots,0) 
= - 2\cdot 3\cdot 5\,\lambda_{6;\rho}^3 \,\cS^{1}_{(2)} 
 - 2\cdot 3\,\lambda_{6;\rho}^2
\Big[\sum_{\rho'\neq \rho}\lambda_{6;\rho'}\Big]  \cS^{12}_{(2)}
\eeq
We concentrate now on contributions induced by $Q_{1}$
and $Q_{2}$. 
For $\Gamma_{6;1}$, $Q^\pm_{1;1;\rho}$ and $Q^\pm_{2;1}$ (a picture of these is given by Figure \ref{fig:Qrho}) contribute and the following sum is relevant

\bea
&&
A_{Q;6;1}(b_\rho,b_\rho',b_\rho'') =
 \sum_\pm \Big[ \sum_{\rho=1,\dots,4} A_{Q^\pm_{1;1;\rho}} + A_{Q^\pm_{2;1}} \Big] \crcr
&&
= 
\frac{1}{3!}\Big[-\frac{\lambda_{6;1}}{3}\Big]^3 \Big\{ 
K^+_{Q;1;1;1}
\Big[\cS^{4}(b_1,b_1')
+ [b_{1}\to b_{1}' \to b_{1}'' ]  \Big] +
K^-_{Q;1;1;1}
\Big[\cS'^{4}(b_1,b_1')
+ [b_{1}\to b_{1}' \to b_{1}'' ]  \Big]  \Big\} \cr\cr
&& + 
\frac{1}{2!}\Big[-\frac{\lambda_{6;1}}{3}\Big]^2\Bigg\{ 
\Big[\sum_{\rho=2,3,4}\Big[-\frac{\lambda_{6;\rho}}{3}\Big] K^+_{Q;1;1;\rho}\Big] \Big[\cS^{41}(b_1,b_1') 
+    [b_{1}\to b_{1}' \to b_{1}'' ]  \Big]\crcr
&&
+ \Big[\sum_{\rho=2,3,4}\Big[-\frac{\lambda_{6;\rho}}{3}\Big] K^-_{Q;1;1;\rho}\Big] \Big[\cS'^{41}(b_1,b_1') 
+    [b_{1}\to b_{1}' \to b_{1}'' ]  \Big] \Bigg\} \crcr 
&& 
+ \frac{1}{3!}\Big[-\frac{\lambda_{6;1}}{3}\Big]^3
\Big\{  K^+_{Q;2;1}
\Big[\cS^{5}(b_1,b_1',b''_1)
+ [b_{1}\to b_{1}' \to b_{1}'' ]  \Big]  +  K^-_{Q;2;1}
\Big[\cS^{5}(b_1,b_1',b''_1)
+ [b_{1}\to b_{1}' \to b_{1}'' ]  \Big] \Big\}
\eea
with
\beq
K^\pm_{Q;1;1;1} = 3^4 \cdot 2^2 \qquad \quad 
K^\pm_{Q;1;1;\rho\neq 1} = 3^3 \cdot 2^2 \qquad \quad 
K^\pm_{Q;2;1} =3^3 \cdot 2^3 
\eeq
At zero external momenta, we get
\beq
A_{Q;6;1}(0,\dots,0) 
= - 2^2\cdot 5\,\lambda_{6;1}^3 \,\cS^{1}_{(1)} 
 - 2^2\cdot 3\,\lambda_{6;1}^2
\Big[\sum_{\rho=2,3,4}\lambda_{6;\rho}\Big]  \cS^{12}_{(1)} 
\eeq
Similarly, for $\rho=2,3,4$, contributions to $\Gamma_{6;\rho}$
can be summed as
\beq
A_{Q;6;\rho}(0,\dots,0) 
= - 2^2\cdot 5\,\lambda_{6;\rho}^3 \,\cS^{1}_{(1)} 
 - 2^2\cdot 3\,\lambda_{6;\rho}^2
\Big[\sum_{\rho'\neq \rho}\lambda_{6;\rho'}\Big]  \cS^{12}_{(1)}
\eeq
Therefore, adding all contributions in each $\rho$-sector, 
we have, at four loops, $\Gamma_{4;1;\rho}(0,\dots,0)$
given by \eqref{gam61pur}.

\section{Proof of Lemma \ref{lem:4ptfunc}}
\label{appsubsect:4pt}

Lemma \ref{lem:4ptfunc} is proved in this appendix
by scrutinizing all contributions to $\Gamma_{4;1;\rho}$.
The following formal sums will be useful
\bea
S^0(b) &:=&  \sum_{p_1,p_2, p_3}
\frac{1}{(b^2 + p_1^2+ p_2^2+ p_3^2+m^2)}  \crcr
S^{2}(b,b') &:=& 
\sum_{p_1,\dots, p_6}
\frac{1}{(b^2 + p_1^2+ p_2^2+ p_3^2+m^2)^2}
\frac{1}{(b'^2 + p_4^2+ p_5^2+ p_6^2+m^2)} \crcr
S^{21}(b) &:=&
\sum_{p_1,\dots, p_6} 
\frac{1}{(b^2 + p_1^2+ p_2^2+ p_3^2+m^2)^2}
\frac{1}{(p_1^2 + p_4^2+ p_5^2+ p_6^2+m^2)} 
\label{s021} \crcr
S''^0 &:=&  \sum_{p_1,\dots,p_7}
\frac{1}{(p_1^2+ p_2^2+ p_3^2+m^2)^2} 
\frac{1}{(p_4^2+ p_5^2+ p_6^2 + p_7^2+m^2)}  \crcr
S^6 &:=&
 \sum_{p_1,\dots, p_9} 
\frac{1}{(p_1^2+ p_2^2+ p_3^2+ m^2)^2}
\frac{1}{(p_4^2 + p_5^2+ p_6^2 + m^2)} 
\frac{1}{(p_7^2+ p_8^2+ p_9^2 +m^2)} \crcr
S^{16}&:=&
 \sum_{p_1,\dots, p_9} 
\frac{1}{(p_1^2+ p_2^2+ p_3^2+ m^2)^2}
\frac{1}{(p_1^2 +  p_4^2 + p_5^2+ p_6^2 + m^2)} 
\frac{1}{(p_1^2+ p_7^2+ p_8^2+ p_9^2 +m^2)} \crcr
S^7&:=&
 \sum_{p_1,\dots, p_9} 
\frac{1}{(p_1^2+ p_2^2+ p_3^2+ m^2)^2}
\frac{1}{(p_4^2 + p_5^2+ p_6^2 + m^2)} 
\frac{1}{(p_4^2+ p_7^2+ p_8^2+ p_9^2 +m^2)} \crcr
S^{17}&:=&
 \sum_{p_1,\dots, p_9} 
\frac{1}{(p_1^2+ p_2^2+ p_3^2+ m^2)^2}
\frac{1}{(p_1^2 +  p_4^2 + p_5^2+ p_6^2 + m^2)} 
\frac{1}{(p_4^2+ p_7^2+ p_8^2+ p_9^2 +m^2)} \crcr
S^8&:=&
 \sum_{p_1,\dots, p_9} 
\frac{1}{(p_1^2+ p_2^2+ p_3^2+ m^2)^2}
\frac{1}{(p_4^2 + p_5^2+ p_6^2 + m^2)} 
\frac{1}{(p_1^2+ p_7^2+ p_8^2+ p_9^2 +m^2)} \crcr
S^{18}&:=&
 \sum_{p_1,\dots, p_9} 
\frac{1}{(p_1^2+ p_2^2+ p_3^2+ m^2)^2}
\frac{1}{(p_2^2 +  p_4^2 + p_5^2+ p_6^2 + m^2)} 
\frac{1}{(p_1^2+ p_7^2+ p_8^2+ p_9^2 +m^2)} 
\eea
We denote $S^0(0)=:S^0$ and note that 
$S^2(0,0)=S^1$ and $S^{21}(0)=S^{12}$.

\subsection{Graph B}  
\label{gra4B}

\begin{figure}[ht]
\centering
 \begin{minipage}{.8\textwidth}
 \centering
 \includegraphics[width=10cm, height=2.5cm]{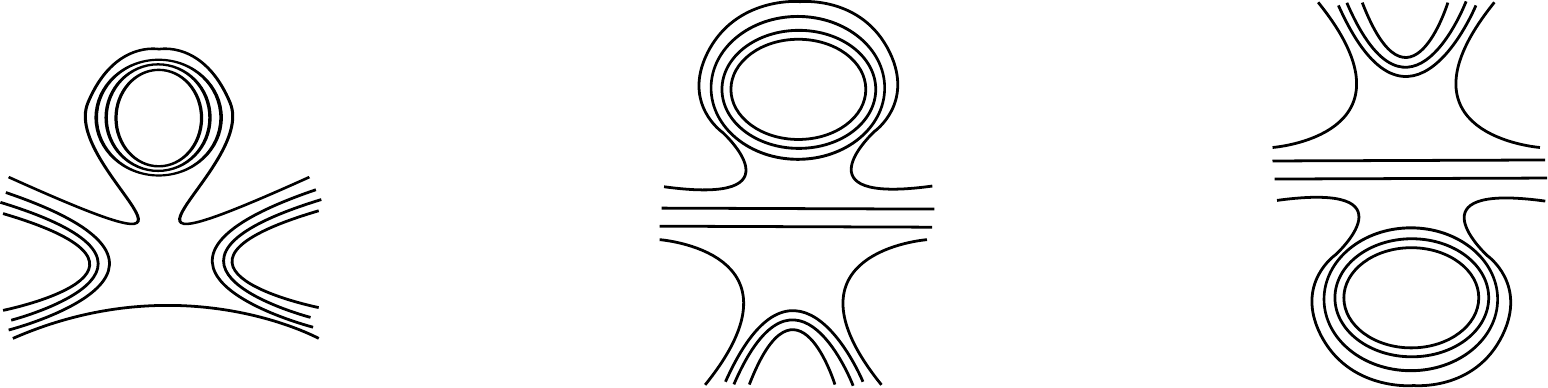}  
\vspace{0.5cm}
\caption{{\small Four-point graphs $B_{1;\rho=1}$, $B^\pm_{2;\rho\rho'=14}$. }}
\label{fig:B1B2}
\end{minipage}
\put(-330,-30){$B_{1;\rho=1}$}
\put(-350,30){$b_{1}$}
\put(-220,-30){$B^+_{2;\rho\rho'=14}$}
\put(-230,30){$b_{4}$}
\put(-230,2){$b_{1}$}
\put(-105,-30){$B^-_{2;\rho\rho'=14}$}
\put(-117,37){$b_{4}$}
\put(-117,13){$b_{1}$}
\end{figure}
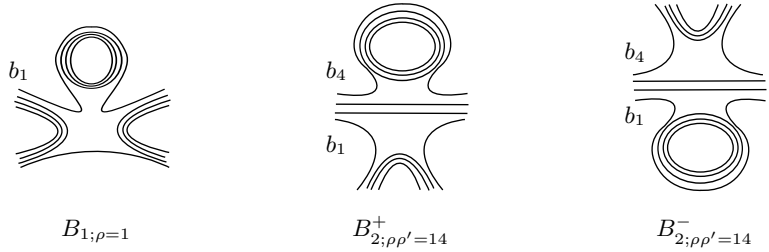

We start the analysis by tadpole graphs coined B. 
Graph $B_{1;\rho}$ is a  made with one $\phi^6_{(1)}$ vertex
meanwhile $B^{\pm}_{2;\rho\rho'}$ is made with one $\phi^6_{(2)}$ vertex
(see, in particular, $B_{1;1}$ and $B^{\pm}_{2;14}$ in Figure \ref{fig:B1B2}).

The calculation of
$\Gamma_{4;1;\rho}$ involves amplitudes of the graphs 
$B_{1;\rho}$ and of $B^{\pm}_{2;\rho\rho'}$.
Given $\rho=1,2,3,4$, to $\Gamma_{4;1;\rho}$ contribute the following amplitude
\beq
A_{B_{1;\rho}}(b_\rho,b_\rho')  
= \Big[-\frac{\lambda_{6;1;\rho}}{3} \Big]
\Big[K_{B;1;\rho}\Big]  \Big[S^0(b_\rho) \;+\;  (b_{\rho} \lra b_{\rho}')\Big]   
\label{411} 
\eeq
where all weight factors are fixed to $K_{B;1;\rho}=3$. At low external momenta, 
one infers
\beq
A_{B_{1;\rho}}(0,0)  = -2 \,\lambda_{6;1;\rho} \,S^0 \qquad
\label{b1rho}
\eeq
Meanwhile, the amplitudes corresponding to $B^\pm_{2;\rho\rho'}$ may contribute to different $\Gamma_{4;1;\rho}$. We have
\vspace{0.3cm}

$\bullet$ To $\Gamma_{4;1;1}$ contribute $B^+_{2;1\rho}$, for $\rho=2,3,4$;

$\bullet$ To $\Gamma_{4;1;2}$ contribute $B^-_{2;12}$ and, for $\rho=3,4$, $B^+_{2;2\rho}$;

$\bullet$ To $\Gamma_{4;1;3}$ contribute
$B^+_{2;34}$ and, for $\rho=1,2$,  $B^-_{2;3\rho}$;

$\bullet$ To  $\Gamma_{4;1;4}$ contribute $B^-_{2;4\rho}$, for $\rho=1,2,3$.
\vspace{0.3cm}
 
We sum these amplitudes such that to $\Gamma_{4;1;1}$ contribute: 
\beq
A_{B;4;1}(b_\rho,b_\rho') = \sum_{\rho=2,3,4} 
A_{B^+_{2;1\rho}}  
= \sum_{\rho=2,3,4}\Big[-\lambda_{6;2;1\rho}\Big]
\Big[K^+_{B;2;1\rho}\Big] 
\Big[ S^0(b_{\rho}) \;+\;  
(b_{\rho} \lra b_{\rho}')\Big]
\eeq
with  combinatorial weights fixed to $K^+_{B;2;1\rho}=1$. 
Putting zero to external momenta yields
\beq
A_{B;4;1}(0,\dots, 0)  = 
-  2[\lambda_{6;2;14}+\lambda_{6;2;13}+\lambda_{6;2;12}]\, S^0
\label{b41}
\eeq
Similarly, using the above graph repartition, one shows that 
to each $\Gamma_{4;1;\rho}$ contribute
\beq
A_{B;4;\rho}(0,\dots, 0)  =
-  2\Big[ \sum_{\rho'\in \{1,2,3,4\} \setminus\{\rho\}}\lambda_{6;2;\rho\rho'}\Big]\, S^0
\label{b4rho} 
\eeq

\subsection{Graph D}  
\label{gra4D} 

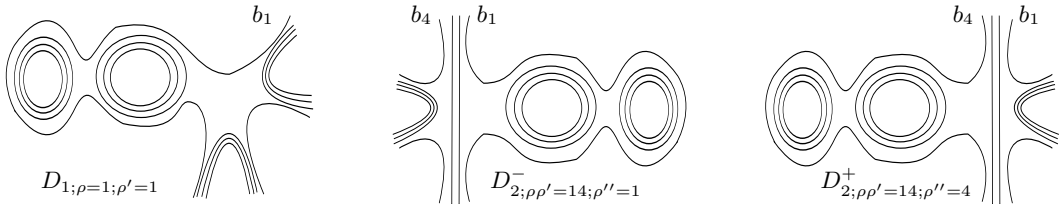
\begin{figure}[ht]
\centering
 \begin{minipage}{.8\textwidth}
 \centering
 \includegraphics[width=14cm, height=2.5cm]{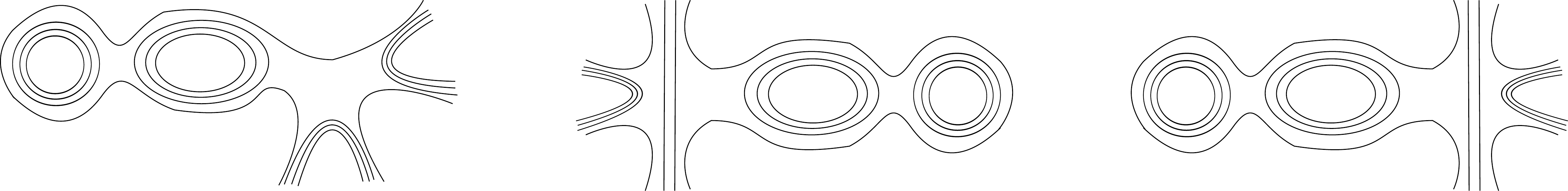}  
\caption{{\small Graphs of type D: $D_{1;\rho;\rho'}$ 
is parametrized by both indices, $\rho$ for the left $\phi^6_{(1)}$ and
$\rho'$ for the right $\phi^4_{(1)}$;
 $D^\pm_{2;\rho\rho';\rho''}$ is parametrized by $\rho\rho'$ 
indices of the $\phi^6_{(2)}$ vertex and $\rho''$ index of the $\phi^4_{(1)}$ vertex. }}
\label{fig:D}
\end{minipage}
\put(-390,-8){$D_{1;\rho=1;\rho'=1}$}
\put(-310,55){$b_{1}$}
\put(-220,-8){$D^-_{2;\rho\rho'=14;\rho''=1}$}
\put(-225,55){$b_{1}$}
\put(-250,55){$b_{4}$}
\put(-95,-8){$D^+_{2;\rho\rho'=14;\rho''=4}$}
\put(-45,55){$b_{4}$}
\put(-20,55){$b_{1}$}
\end{figure}

Type D graphs are formed with one $\phi^4$ vertex 
and one $\phi^6$ vertex. This type of graph can be expanded as 
$D_{1;\rho;\rho'}$ (a $\phi^6_{(1)}-\phi^4_{(1)}$ contraction) and  
$D^\pm_{2;\rho\rho';\rho''}$ (a $\phi^6_{(2)}-\phi^4_{(1)}$ contraction)
and are characterized by their vertex indices
(see for instance graphs
$D_{1;1;1}$ and $D^\pm_{2;14;1}$  given by Figure \ref{fig:D}).
Furthermore, by replacing $\phi^4_{(1)}$ by $\phi^4_{(2)}$ above, 
we get another category of graphs called 
$D'_{1;\rho}$ and  $D'^\pm_{2;\rho\rho'}$, respectively.
We emphasize the fact that these graphs contribute to $\Gamma_{4;1;\rho}$ even though they do not involve explicitly the $\phi^4_{(1)}$
interactions.

The contribution to $\Gamma_{4;1;\rho}$ coming from $D_{1;\rho;\rho'}$
writes
\bea
A_{D;4;\rho}(b_\rho,b_\rho') &=&  
\sum_{\rho'=1,\dots,4} A_{D_{1;\rho;\rho'}} 
= \Big[-\frac{\lambda_{6;1;\rho}}{3}\Big] 
\Bigg\{\Big[-\frac{\lambda_{4;1;\rho}}{2}\Big] 
\Big[K_{D;1;\rho;\rho}\Big] \Big[ S^2(b_\rho,b_\rho)
+ (b_\rho \lra b_\rho') \Big] \label{ah411}\\
&&
+ \sum_{\rho'\in \{1,2,3,4\}\setminus \{\rho\}} \Big[-\frac{\lambda_{4;1;\rho}}{2} \Big]
\Big[K_{D;1;\rho;\rho'}\Big] \Big[ S^{21}(b_\rho)
+ (b_1 \lra b_1') \Big]  
\Bigg\} 
\nonumber
\eea 
with all $K_{D;1;\rho;\rho'}=2\cdot 3$. 
Putting external momenta to zero, the above \eqref{ah411} can be recast as
\beq
A_{D;4;\rho}(0,\dots,0) = 
2\lambda_{6;1;\rho} \Big[\lambda_{4;1;\rho} \,S^1
+\Big[\sum_{\rho'\in \{1,2,3,4\}\setminus \{\rho\}}\lambda_{4;1;\rho' } \Big]S^{12}
\Big]
\label{d1rho}
\eeq
Next, the amplitudes of $D^\pm_{2;\rho\rho';\rho''}$
can be understood in terms of different contributions for $\Gamma_{4;1;\rho}$. Hence,
\vspace{0.3cm}
 
$\bullet$ To $\Gamma_{4;1;1}$ contribute $D^+_{2;1\rho;\rho'}$, for $\rho=2,3,4$ and $\rho'=1,2,3,4$. 

$\bullet$ To $\Gamma_{4;1;2}$ contribute $D^+_{2;2\rho;\rho'}$, for $\rho=3,4$ and $\rho'=1,2,3,4$, and
$D^-_{2;12;\rho'}$, for  $\rho'=1,2,3,4$.

$\bullet$ To $\Gamma_{4;1;3}$ contribute
$D^+_{2;34;\rho'}$ for  $\rho'=1,2,3,4$, 
and $D^-_{2;3\rho;\rho'}$, for $\rho=1,2$ and $\rho'=1,2,3,4$.

$\bullet$ To  $\Gamma_{4;1;4}$ contribute $D^-_{2;4\rho;\rho'}$, for $\rho=1,2,3$ and $\rho'=1,2,3,4$. 

\vspace{0.3cm}

For $\Gamma_{4;1}$, we can compute the contribution as
\bea
A^2_{D;4;1}(b_\rho,b_\rho') &=& 
\sum_{\rho=2,3,4}\sum_{\rho'=1,\dots,4} A_{D^+_{2;1\rho;\rho'}} 
= \sum_{\rho=2,3,4} [-\lambda_{6;2;1\rho}]
\Bigg\{-\frac{\lambda_{4;1;\rho}}{2} 
\Big[K^+_{D;2;1\rho;\rho}\Big] 
 \Big[S^2(b_\rho,b_\rho) + (b_\rho \lra b_\rho') \Big] \crcr
&&
-\sum_{\rho'\in \{1,2,3,4\} \setminus \{\rho\}}\Big[\frac{\lambda_{4;1;\rho'}}{2} 
K^+_{D;2;1\rho;\rho'}  \Big] \Big[ S^{21}(b_\rho) + (b_\rho \lra b_\rho') \Big] \Bigg\} 
\nonumber
\eea
with all weight factors being fixed to $K^+_{D;2;1\rho;\rho'}=2$. Setting all external $b$'s 
to zero,  one comes to
\beq
A^2_{D;4;1}(0,\dots,0) 
= 2 \Big[\sum_{\rho=2,3,4} \lambda_{6;2;1\rho}\lambda_{4;1;\rho}
\Big] S^1 
+2 \Big[\sum_{\rho=2,3,4}  \sum_{\rho'\in \{1,2,3,4\} \setminus \{\rho\}} \lambda_{6;2;1\rho}\lambda_{4;1;\rho'}\Big] 
 S^{12} 
\label{d41}
\eeq
In a similar way,  one identifies
the following contribution for  $\Gamma_{4;1;\rho}$: 
\beq
A^2_{D;4;\rho}(0,\dots,0) 
= 
2 \Big[\sum_{\rho' \in \{1,2,3,4\} \setminus \{\rho\}} \lambda_{6;2;\rho\rho'}\lambda_{4;1;\rho'}
\Big] S^1 
+2 \Big[\sum_{\rho' \in \{1,2,3,4\} \setminus \{\rho\}}  \sum_{\rho''\in \{1,2,3,4\} \setminus \{\rho'\}} \lambda_{6;2;\rho\rho'}\lambda_{4;1;\rho''}\Big] 
 S^{12}
\label{d4rho}
\eeq

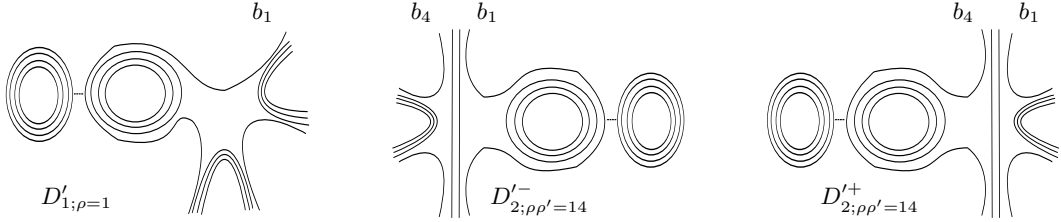
\begin{figure}[ht]
\centering
 \begin{minipage}{.8\textwidth}
 \centering
 \includegraphics[width=14cm, height=2.5cm]{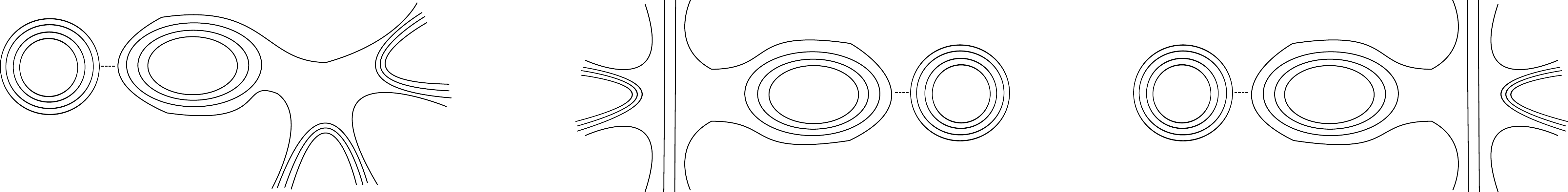}  
\caption{{\small Graphs of type D': $D'_{1;\rho}$ 
$D'^\pm_{2;\rho\rho'}$ are parametrized indices
of $\phi^6_{(1)}$ and $\phi^6_{(2)}$, respectively. }}
\label{fig:Dprim}
\end{minipage}
\put(-390,-15){$D'_{1;\rho=1}$}
\put(-310,55){$b_{1}$}
\put(-220,-15){$D'^-_{2;\rho\rho'=14}$}
\put(-225,55){$b_{1}$}
\put(-250,55){$b_{4}$}
\put(-95,-15){$D'^+_{2;\rho\rho'=14}$}
\put(-45,55){$b_{4}$}
\put(-20,55){$b_{1}$}
\end{figure}

Let us now focus on the second type of graphs $D'$. Examples
are given in Figure \ref{fig:Dprim}.

In terms of different contributions for $\Gamma_{4;1;\rho}$, one has
\vspace{0.3cm}
 
$\bullet$ To $\Gamma_{4;1;1}$ contribute $D'_{1;1}$ and $D'^+_{2;1\rho}$, for $\rho=2,3,4$. 

$\bullet$ To $\Gamma_{4;1;2}$ contribute $D'_{1;2}$ and 
$D^+_{2;2\rho}$, for $\rho=3,4$ and $D^-_{2;12}$.

$\bullet$ To $\Gamma_{4;1;3}$ contribute $D'_{1;3}$ and 
 $D^-_{2;3\rho}$, for $\rho=1,2$ and $D^+_{2;34}$.

$\bullet$ To  $\Gamma_{4;1;4}$ contribute $D'_{1;4}$ and 
$D^-_{2;4\rho}$, for $\rho=1,2,3$. 

\vspace{0.3cm}

A direct calculation as previously performed yields the contribution
to each $\Gamma_{4;1;\rho}$ as
\beq
A'_{D;4;\rho} = 2 \lambda_{4;2} 
\Big[\lambda_{6;1;\rho} + \sum_{\rho'\in \{1,2,3,4\}\setminus \{\rho\}} \lambda_{6;2;\rho\rho'}  \Big] S''^0
\label{dprim}
\eeq

\subsection{Graph E}
\label{grap4E}

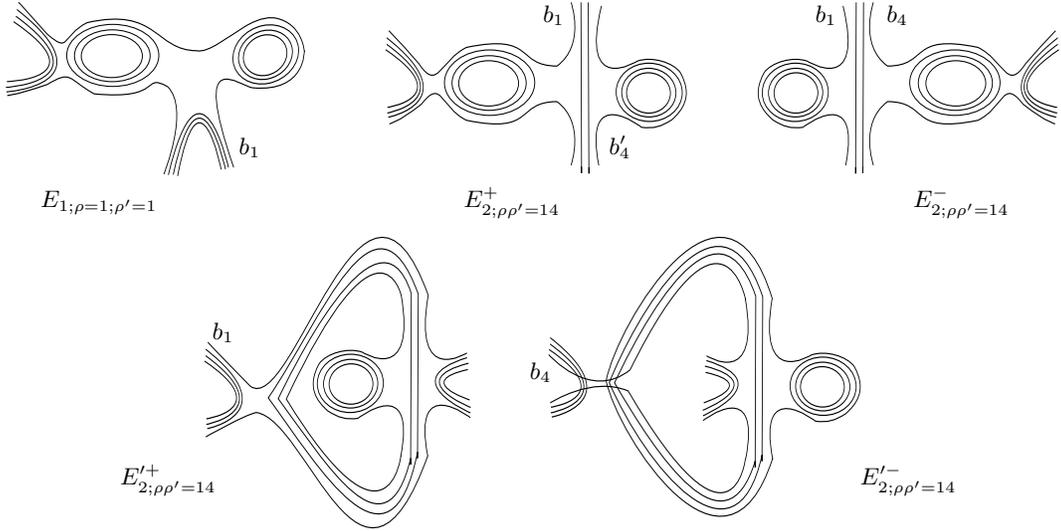
\begin{figure}[ht]
\centering
 \begin{minipage}{.8\textwidth}
 \centering
 \includegraphics[width=14cm, height=7cm]{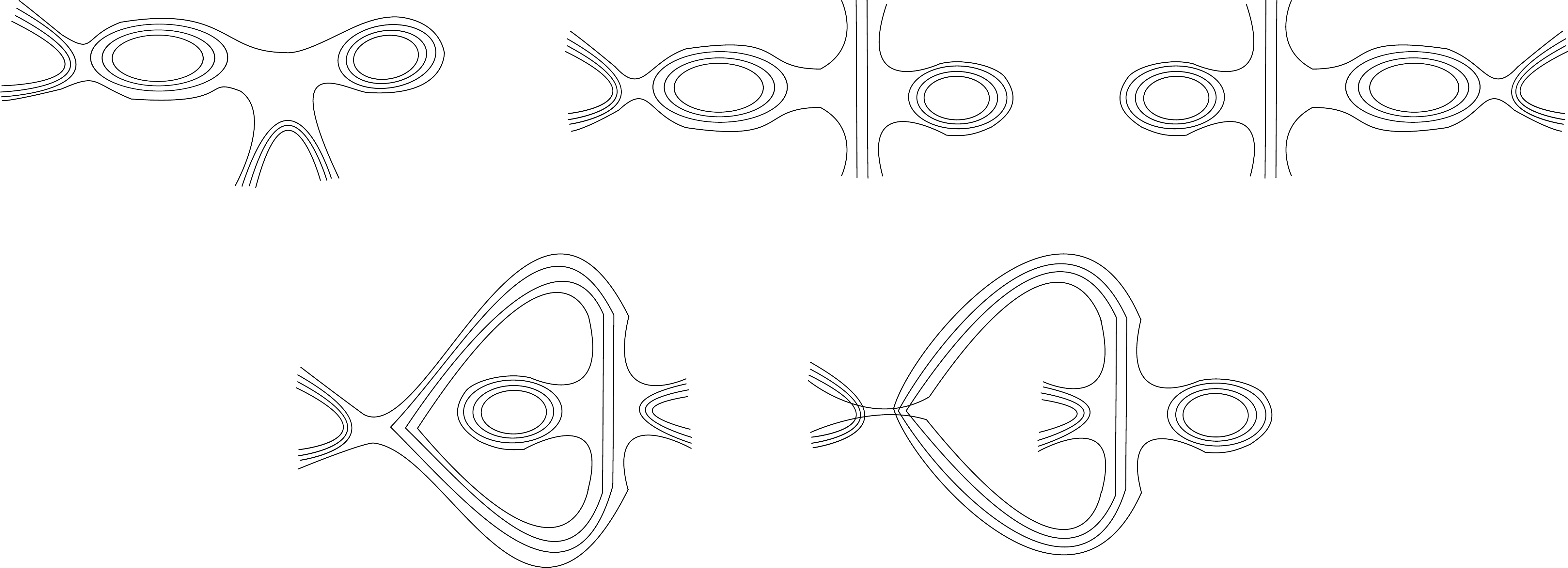}  
\caption{{\small Graphs of type E: $E_{1;\rho}$ 
is parametrized by $\rho$ index of both $\phi^6_{(1)}$ and $\phi^4_{(1)}$ vertices; 
 $E^\pm_{2;\rho\rho'}$ and 
$E'^\pm_{2;\rho\rho'}$ are parameterized by 
$\rho\rho'$ index of the $\phi^6_{(2)}$ vertex. Moreover,
for $E^\pm_{2;\rho\rho'}$ and $E'^\pm_{2;\rho\rho'}$, 
$\rho$ and $\rho'$ correspond to the index of the $\phi^4_{(1)}$ vertex, respectively. }}
\label{fig:E}
\end{minipage}
\put(-390,50){$E_{1;\rho=1;\rho'=1}$}
\put(-315,70){$b_{1}$}
\put(-230,50){$E^+_{2;\rho\rho'=14}$}
\put(-200,120){$b_{1}$}
\put(-175,70){$b_{4}'$}
\put(-60,50){$E^-_{2;\rho\rho'=14}$}
\put(-97,120){$b_{1}$}
\put(-70,120){$b_{4}$}
\put(-360,-55){$E'^+_{2;\rho\rho'=14}$}
\put(-325,0){$b_{1}$}
\put(-80,-55){$E'^-_{2;\rho\rho'=14}$}
\put(-205,-15){$b_{4}$}
\end{figure}

This is another configuration given by the contraction 
of one vertex $\phi^4$ and one vertex $\phi^6$.  
Graphs in this category are named $E_{1;\rho}$, 
$E^\pm_{2;\rho\rho'}$ and $E'^{\pm}_{2;\rho\rho'}$ (examples are given for $E_{1;1}$, 
$E^\pm_{2;14}$ and $E'^{\pm}_{2;14}$ in Figure \ref{fig:E}).

Given $\rho$, we start by the amplitude of $E_{1;\rho}$ as a
contribution to $\Gamma_{4;1;\rho}$: 
\beq
A_{E_{1;\rho}} (b_\rho,b_\rho') 
= \Big[ -\frac{\lambda_{6;1;\rho}}{3}\Big]
\Big[-\frac{\lambda_{4;1;\rho}}{2} \Big] 
\Big[K_{E;1;\rho}\Big] \Big[ S^3(b_\rho,b'_\rho) + (b_\rho \lra b_\rho') \Big]
\eeq
where $K_{E;1;\rho} = 2\cdot 3 \cdot 2$. Then, at zero
external data, the above amplitude takes the form
\beq
A_{E_{1;\rho}} (0,\dots,0) = 
2 \cdot 2 \; \lambda_{6;1;\rho}
\lambda_{4;1;\rho}\;  S^{1}
\label{e1}
\eeq

Next, we focus on configurations  $E^{\pm}_{2;\rho\rho';\rho''}$
and $E'_{2;\rho\rho';\rho''}$ that we divide in different sectors:
\vspace{0.3cm}

$\bullet$ To $\Gamma_{4;1;1}$ contribute $E^+_{2;1\rho}$ and
$E'^+_{2;\rho}$ for $\rho=2,3,4$;

$\bullet$ To $\Gamma_{4;1;2}$ contribute $E^+_{2;2\rho}$ and 
$E'^+_{2;2\rho}$, for $\rho=3,4$, in addition to
$E^-_{2;12}$ and $E'^-_{2;12}$;

$\bullet$ To $\Gamma_{4;1;3}$ contribute
 $E^-_{2;3\rho}$ and $E'^-_{2;3\rho}$, for $\rho=1,2$,
 in addition to  $E^+_{2;34}$ and $E'^+_{2;34}$;

$\bullet$ To  $\Gamma_{4;1;4}$ contribute $E^-_{2;4\rho}$ and
$E'^-_{2;4\rho}$, for $\rho=1,2,3$. 
\vspace{0.3cm}

Amplitudes included in $\Gamma_{4;1}$ can be summed as
\bea
A^2_{E;4;1} (b_\rho,b_\rho') &=& \sum_{\rho=2,3,4} 
\Big[A_{E^+_{2;1\rho}} + A_{E'^+_{2;1\rho}} \Big]
 = 
\sum_{\rho=2,3,4}  \Big[ -\lambda_{6;2;1\rho }\Big]
\Big[-\frac{\lambda_{4;1;1}}{2} \Big] 
\Big[K^+_{E;2;1\rho}\Big] \Big[ S^4(b_1,b'_1,b_\rho) + (b_\rho \lra b_\rho') \Big]  \cr\cr
&&
+ \Big[-\frac{\lambda_{4;1;1}}{2} \Big]  \Bigg\{ 
\sum_{\rho=2,3,4} 
\Big[ -\lambda_{6;2;1\rho}\Big]\Big[K'^+_{E;2;1\rho}\Big] 
 \Bigg\} 
 S^{14}(b_1,b_1') 
\eea
where the weights are such that $K^\pm_{E;2;1\rho'}= 2$ and $K'^\pm_{E;2;1\rho'}=2\cdot 2$. Setting external momenta to zero, 
it can be shown that
\beq
A^2_{E;4;1} (0,\dots,0) 
= 2\lambda_{4;1;1}\Big[ \sum_{\rho=2,3,4}  \lambda_{6;2;1\rho} 
\Big] [S^1 +S^{12}]
\label{e41}
\eeq
We can deduce the following contribution to any 
$\Gamma_{4;1;\rho}$ by an analogous technique
\bea
A^2_{E;4;\rho} (0,\dots,0) 
= 
2\lambda_{4;1;\rho}\Big[ \sum_{\rho'\in \{1,2,3,4\} \setminus \{\rho\}}  \lambda_{6;2;\rho\rho'} 
\Big] [S^1 +S^{12}]
\label{e4rho}
\eea

\subsection{ Graphs W and Y}
\label{grap4w} 

\begin{figure}[ht]
\centering
 \begin{minipage}{.8\textwidth}
 \centering
 \includegraphics[width=3cm, height=2cm]{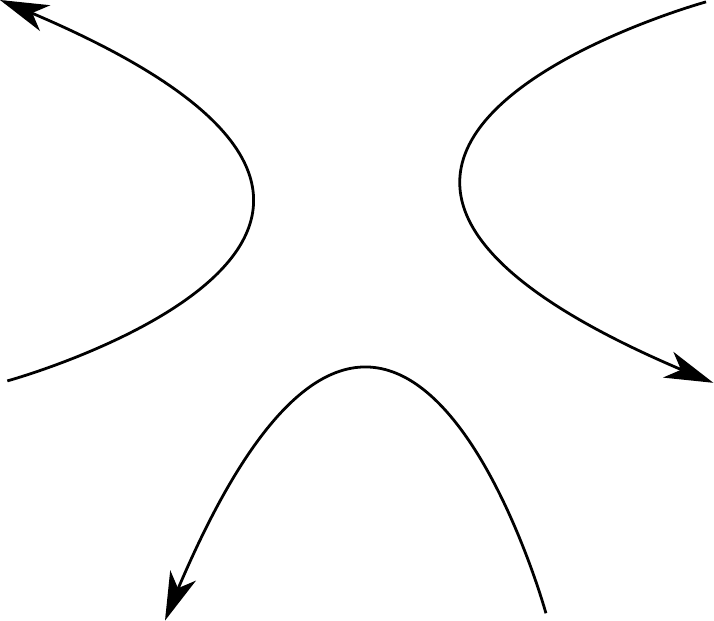}  
\caption{{\small Simplified notation of $\phi^6_{(1)}$ vertex for $\rho=1$. }}
\label{fig:simpliphi61}
\end{minipage}
\put(-260,10){1}
\end{figure}

 Y and W graphs are three loops diagrams of the rough form given
by Figure \ref{fig:simpliphi4}.  
Since these proliferate quickly, 
we cannot review them term by term
and only give some hints in order to achieve their
sum in different contribution to $\Gamma_{4;1;\rho}$.
Thus, in this section, in addition to the compact diagram
 for $\phi^6_{(2)}$ in Figure \ref{fig:simpliphi} introduced in Appendix \ref{gra6E}, 
we will use the  simplified picture for $\phi^6_{(1)}$ given in Figure \ref{fig:simpliphi61}.

The graphs of interest follow a decomposition according to the types of  vertices: $\phi^6_{(1)}-\phi^6_{(1)}$ ($W_A$ and $Y_A$), 
 $\phi^6_{(1)}-\phi^6_{(2)}$ ($W_B$ and $Y_B$)  and  $\phi^6_{(2)}-\phi^6_{(2)}$ ($W_C$ and $Y_C$).

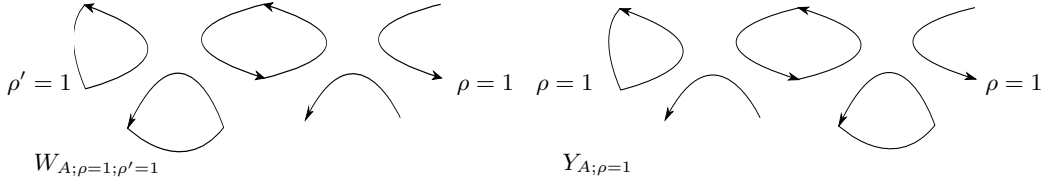
\begin{figure}[ht]
\centering
 \begin{minipage}{.8\textwidth}
 \centering
 \includegraphics[width=12cm, height=2cm]{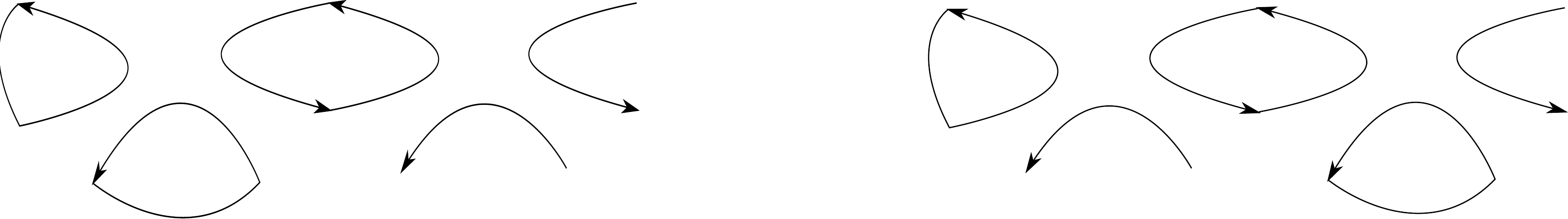}  
\caption{{\small Graphs of type $W_A$ and $Y_A$: 
$W_{A;\rho;\rho'}$ is parametrized by $\rho$ index
of the vertex with external legs and $\rho'$ index of the fully
contracted vertex;  $Y_{A;\rho}$ is parametrized by 
a unique index $\rho$ which should be coinciding for both 
vertices.}}
\label{fig:WAYA}
\end{minipage}
\put(-390,-10){$W_{A;\rho=1;\rho'=1}$}
\put(-400,20){$\rho'= 1$}
\put(-230,20){$\rho= 1$}
\put(-190,-10){$Y_{A;\rho=1}$}
\put(-30,20){$\rho= 1$}
\put(-200,20){$\rho= 1$}
\end{figure}

Let us focus on graph of the type $W_{A;\rho;\rho'}$ 
(a drawing of $W_{A;1;1}$ is given in Figure \ref{fig:WAYA})
contributing to $\Gamma_{4;1;\rho}$ by the amplitude (note that, in this paragraph, all amplitudes will be directly computed at zero external momenta)

\bea
A_{WA;4;\rho}(0,\dots,0) =
\frac{1}{2!}
\Big[ -\frac{\lambda_{6;1;\rho}}{3}\Big]^2
\Big[K_{WA;\rho;\rho}\Big] S^6
+ 
\Big[ \sum_{\rho'\neq \rho} \Big[-\frac{\lambda_{6;1;\rho}}{3}\Big]
\Big[-\frac{\lambda_{6;1;\rho'}}{2} \Big] \Big[K_{WA;\rho;\rho'}\Big] \Big] S^{16}
\eea
with $K_{WA;\rho;\rho}= 3^2 \cdot 2^2$ and $K_{WA;\rho;\rho'\neq \rho} = 3^2 \cdot 2$. 
Thus, the contribution to any $\Gamma_{4;1;\rho}$
is given by 
\beq
A_{WA;4;\rho}(0,\dots,0) =
2 \lambda_{6;1;\rho}^2 \,S^6
+2 \lambda_{6;1;\rho}\Big[ \sum_{\rho' \in \{ 1,2,3,4\} \setminus\{\rho\}} \lambda_{6;1;\rho'} \Big] S^{16}
\label{warho}
\eeq
Focusing now on $Y_{A;\rho}$ (see $Y_{A;1}$ in Figure \ref{fig:WAYA}),
we write each contribution to $\Gamma_{4;1;\rho}$ as
\beq
A_{YA;4;\rho}(0,\dots, 0)  =
\frac{1}{2!}
\Big[ -\frac{\lambda_{6;1;\rho}}{3}\Big]^2
\Big[K_{YA;\rho}\Big]  S^6 =
2^2  \lambda_{6;1;\rho}^2\; S^6 
\label{yarho}
\eeq
where $K_{YA;\rho}= 3^2 \cdot 2^3$. 

\begin{figure}[ht]
\centering
 \begin{minipage}{.8\textwidth}
 \centering
 \includegraphics[width=12cm, height=4cm]{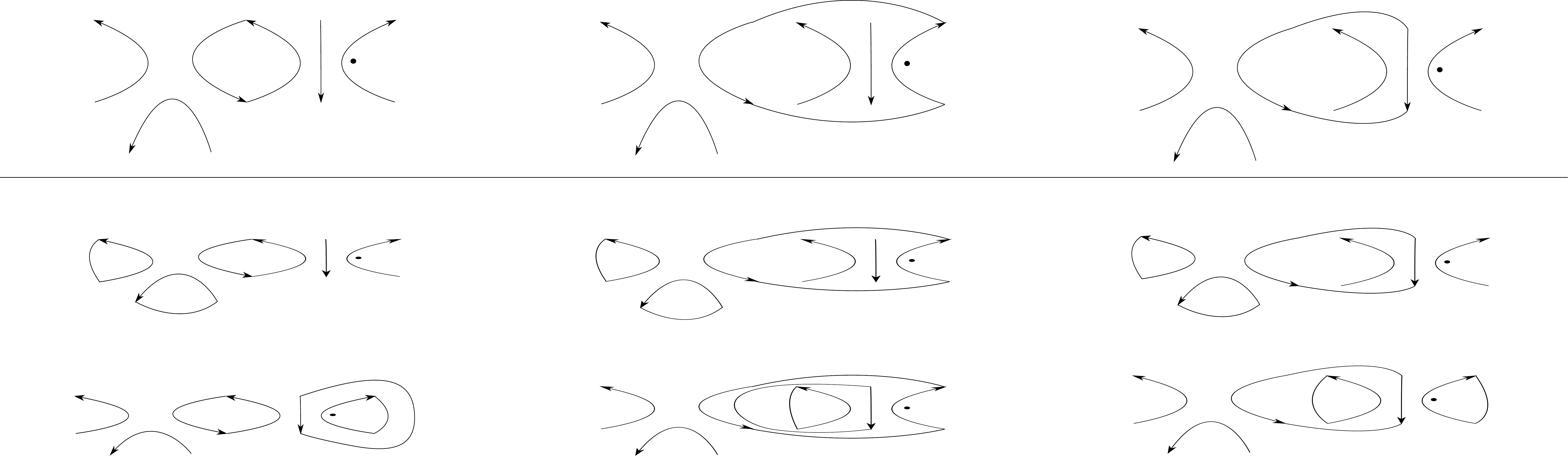}  
\caption{{\small Graphs of type $W_B$: 
$W^\pm_{Bi;\rho\rho';\rho''}$ are parametrized by $\rho\rho'$ index
of the vertex $\phi^6_{(2)}$ and $\rho''$ index of the $\phi^6_{(1)}$ vertex.}}
\label{fig:WB}
\end{minipage}
\put(-280,10){$W^+_{B1}$}
\put(-160,10){$W^-_{B1}$}
\put(-45,10){$W'_{B3}$}
\put(-280,-25){$W^+_{B2}$}
\put(-160,-25){$W^-_{B2}$}
\put(-45,-25){$W_{B3}$}
\end{figure}

\begin{figure}[ht]
\centering
 \begin{minipage}{.8\textwidth}
 \centering
 \includegraphics[width=12cm, height=4cm]{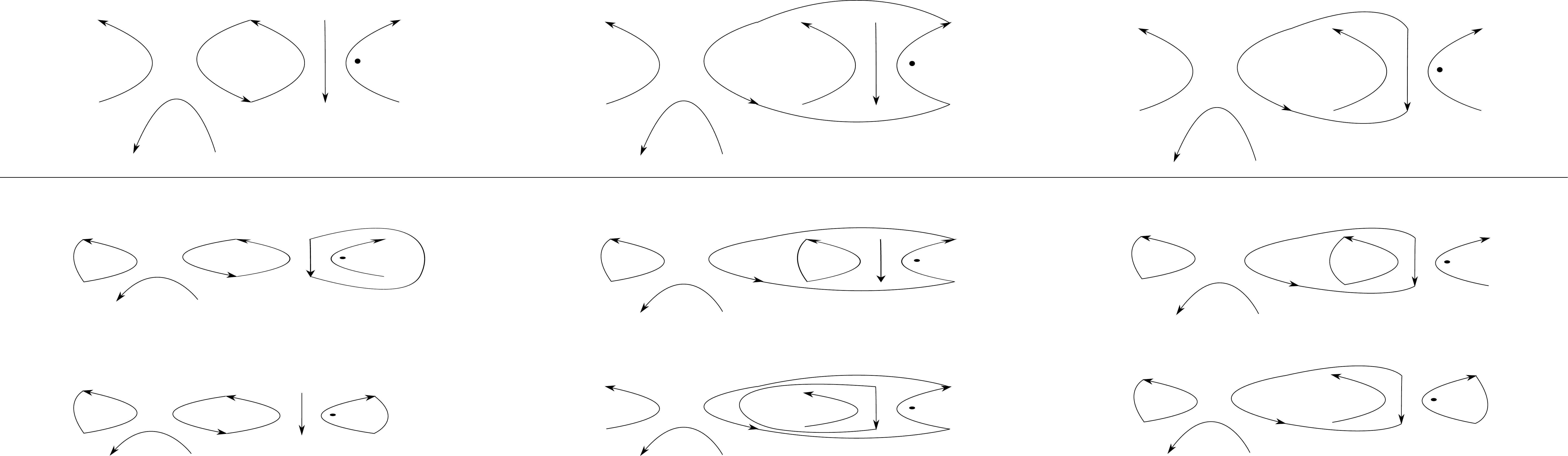}  
\caption{{\small Graphs of type $Y_B$: 
$Y^\pm_{Bi;\rho\rho';\rho''}$ are parametrized by $\rho\rho'$ index
of the vertex $\phi^6_{(2)}$ and $\rho''$ index of the $\phi^6_{(1)}$ vertex.}}
\label{fig:YB}
\end{minipage}
\put(-280,10){$Y'^+_{B1}$}
\put(-160,10){$Y^-_{B1}$}
\put(-45,10){$Y^-_{B2}$}
\put(-280,-25){$Y^+_{B1}$}
\put(-160,-25){$Y'^-_{B1}$}
\put(-45,-25){$Y^+_{B2}$}
\end{figure}

Next, let us evaluate $W_B$ and $Y_B$ graphs. 
They should be of the form given by the 
Figure \ref{fig:WB} and Figure \ref{fig:YB}, respectively. 
$W^{\bullet}_{Bi=1,2,3}$ and $Y^\bullet_{Bi=1,2,3}$ can be coined by their vertex indices:
the double $\rho\rho'$ comes from the $\phi^6_{(2)}$ vertex 
and the unique index $\rho''$ from $\phi^6_{(1)}$. 
Remark that classes of $W_{B3}'$ and $Y'^\pm_{B1}$ never contribute
to $\Gamma_{4;1;\rho}$.

The calculation of $\Gamma_{4;1;1}$ involves
\bea
&&
A_{WYB;4;1}(0,\dots,0) = 
\sum_{\rho=2,3,4}\sum_{\rho'=1,\dots,4}A_{W^+_{B1;1\rho,\rho'}}+ \crcr
&&
\sum_{\rho\rho'=12,\dots,34}\Big[ A_{W^+_{B2;\rho\rho';1}} +
 A_{W^-_{B2;\rho\rho';1}} +
A_{W_{B3;\rho\rho';1}}  \Big] 
+ 
\sum_{\rho=2,3,4}\Big[ A_{Y^+_{B1;1\rho}}+ A_{Y^+_{B2;1\rho}}\Big] 
\eea
yielding, after some algebra,
\bea
&&
A_{WYB;4;1}(0,\dots,0) = 
2 \Big[\sum_{\rho=2,3,4} \lambda_{6;2;1\rho} \lambda_{6;1;\rho} \Big] S^6  + 2 \Big[ \sum_{\rho=2,3,4} 
\lambda_{6;2;1\rho}  \sum_{\rho'\in \{1,2,3,4\} \setminus \{\rho\}}  \lambda_{6;1;\rho'}  \Big] S^{16} 
\crcr
&& 
+ 2\lambda_{6;1;1} \Bigg\{  
\Big[ \sum_{\rho\rho'=12,\dots,34} \lambda_{6;2;\rho\rho'} \Big] S^{17} +
\Big[\sum_{\rho=2,3,4}\lambda_{6;2;1\rho} \Big][ S^7 +S^8 ] + 
\Big[ \sum_{\rho\rho'=23,24,34}\lambda_{6;2;\rho\rho'}\Big][S^{17} + S^{18} ] \crcr
&&
+ 
2 \Big[  \sum_{\rho=2,3,4} \lambda_{6;2;1\rho} \Big][ S^6 +S^8 ] 
\Bigg\} 
\label{wyb1}
\eea
A similar calculation gives, in each sector, the contribution to $\Gamma_{4;1;\rho}$ as
\bea
&&
A_{WYB;4;\rho}(0,\dots,0) = 
2 \Big[\sum_{\rho'\in \{1,2,3,4\} \setminus \{\rho\}} \lambda_{6;2;\rho\rho'} \lambda_{6;1;\rho'} \Big] S^6  
+ 2 \Big[ \sum_{\rho'\in \{1,2,3,4\} \setminus \{\rho\}} \lambda_{6;2;\rho\rho'}  \sum_{\rho''\in \{1,2,3,4\} \setminus \{\rho'\}}  \lambda_{6;1;\rho''}  \Big] S^{16} 
\crcr
&& 
+ 2\lambda_{6;1;\rho} \Bigg\{  
\Big[ \sum_{\rho'\rho''=12,\dots,34} \lambda_{6;2;\rho'\rho''} \Big] S^{17} +
\Big[\sum_{\rho'\in \{1,2,3,4\} \setminus \{\rho\}}\lambda_{6;2;\rho\rho'} \Big][ S^7 +S^8 ] \crcr
&& + 
\Big[ \sum_{\rho'\rho''=\{12,\dots,34\} / \; \rho' \neq \rho\; \text{and} \;\rho'' \neq \rho }\lambda_{6;2;\rho'\rho''}\Big][S^{17} + S^{18} ]
+ 
2 \Big[  \sum_{\rho'\in \{1,2,3,4\} \setminus \{\rho\}} \lambda_{6;2;\rho\rho'} \Big][ S^6 +S^8 ] 
\Bigg\} 
\label{wybrho}
\eea

\begin{figure}[ht]
\centering
 \begin{minipage}{.8\textwidth}
 \centering
 \includegraphics[width=14cm, height=15cm]{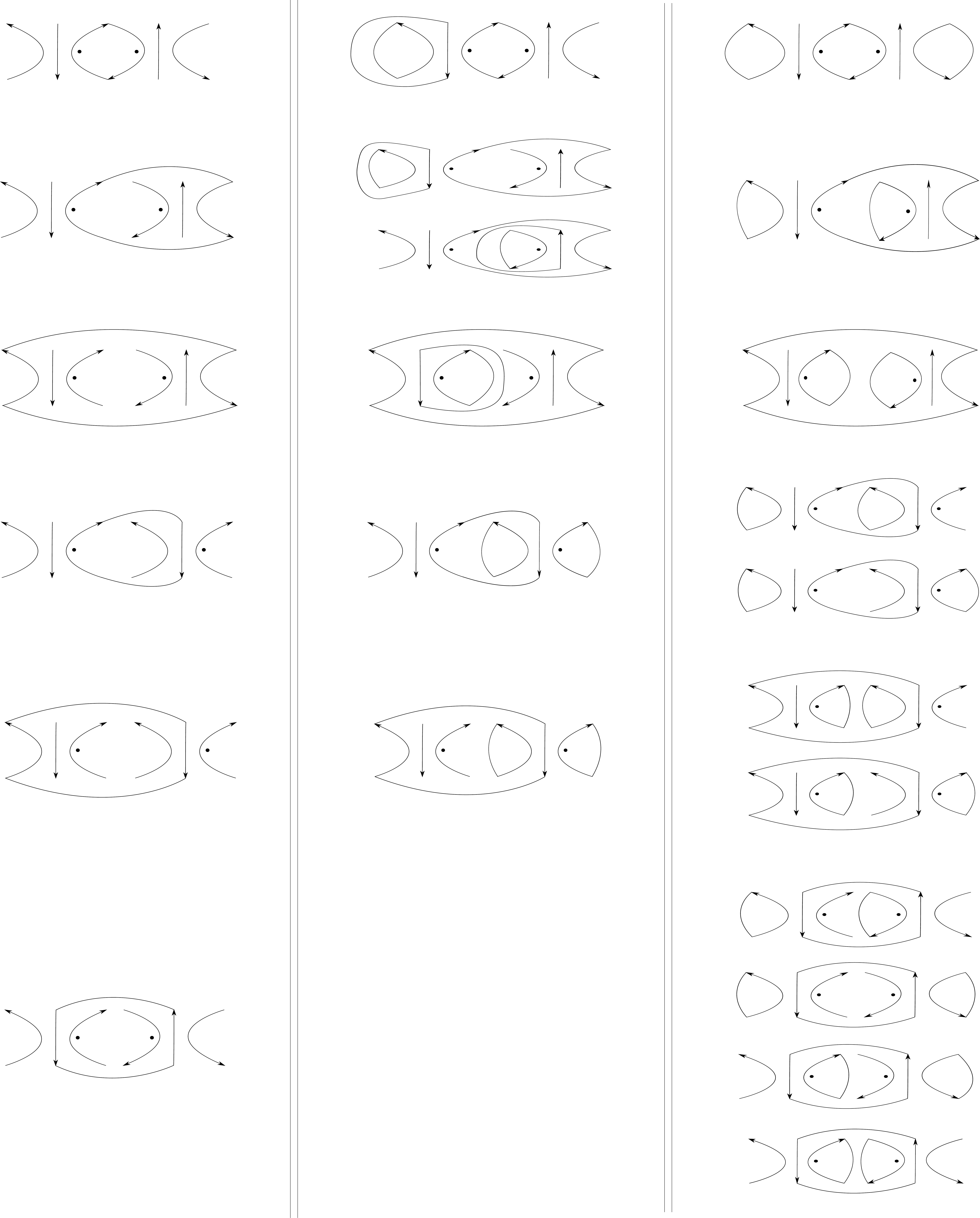}  
\caption{{\small List of all divergent graphs of type $W_C$ and $Y_C$.}}
\label{fig:WCYC}
\end{minipage}
\put(-220,235){$W_{C}$}
\put(-65,235){$Y_{C}$}
\end{figure}

Last,  $W_C$ and $Y_C$ graphs
of the form given by the Figure \ref{fig:WCYC} should contribute. 
A lengthy  but straightforward  calculation yields the contribution $A_{WYC;4;\rho}$ to $\Gamma_{4;1;\rho}$ as
\bea
&&
A_{WYC;4;\rho}(0,\dots,0) =
\Big[ \sum_{\rho' \in \{1,2,3,4\}\setminus \{\rho\}}\lambda_{6;2;\rho\rho'} \Big]^2  [S^6 + 2  S^8] 
 \crcr
&&
+  \Big[ \sum_{\rho' \in \{1,2,3,4\}\setminus \{\rho\}}\lambda_{6;2;\rho\rho'}^2\Big]
S^{16}
+2 \Big[ \sum_{\rho',\rho''\in \{1,2,3,4\}\setminus \{\rho\};\;\;  \rho'' < \rho'} 
\lambda_{6;2;\rho\rho'} \lambda_{6;2;\rho\rho''} \Big]  S^{18} 
\crcr
&+& \Bigg\{
\sum_{\rho' \in \{1,2,3,4\}\setminus \{\rho\}} \lambda_{6;2;\rho\rho'} 
\sum_{\rho''\{1,2,3,4\} \setminus \{\rho'\}}\lambda_{6;2;\rho'\rho''} 
\Bigg\}[ S^7 +S^8 ] \crcr
&+&  \Bigg\{
\sum_{\rho' \in \{1,2,3,4\}\setminus \{\rho\}}  \lambda_{6;2;\rho\rho'}
\sum_{\rho''\rho'''=\{12,\dots,34\} / \; \rho'' \neq \rho'\; \text{and} \;\rho'''\neq \rho'} \lambda_{6;2;\rho''\rho'''} 
\Bigg\} [ S^{17} +S^{18} ]   \crcr
&+&
\Big[\sum_{\rho' \in \{1,2,3,4\}\setminus \{\rho\}}\lambda_{6;2;\rho\rho'}   \Big] \Big[\sum_{\rho'\rho''\in \{12,\dots,34\}}  \lambda_{6;2;\rho'\rho''}\Big] S^{17} 
\label{wyc}
\eea

We are now in position to compute each $\Gamma_{4;1;\rho}$ by collecting all different contributions from graphs B \eqref{b1rho},\eqref{b4rho}, D \eqref{d1rho}, \eqref{d4rho}, \eqref{dprim} , E \eqref{e1},\eqref{e4rho} and graphs  W and Y \eqref{warho},\eqref{yarho}, \eqref{wybrho} and \eqref{wyc}. Hence,  it can be deduced  
\bea
&&
\Gamma_{4;\rho}(0,\dots,0) =  
 - \lambda_{4;\rho} 
+ 2 \lambda_{4;2} 
\Big[\lambda_{6;1;\rho} + \sum_{\rho'\in \{1,2,3,4\}\setminus \{\rho\}} \lambda_{6;2;\rho\rho'}  \Big] S''^0 \crcr
&&
 + 2\lambda_{6;1;\rho} \Big[3\lambda_{4;1;\rho} \,S^1
+\Big[\sum_{\rho'\in \{1,2,3,4\}\setminus \{\rho\}}\lambda_{4;1;\rho' } \Big]S^{12}
\Big]
+ 
2 \Big[\sum_{\rho' \in \{1,2,3,4\} \setminus \{\rho\}} \lambda_{6;2;\rho\rho'}\lambda_{4;1;\rho'}
\Big] S^1  \crcr
&&
+2 \Big[\sum_{\rho' \in \{1,2,3,4\} \setminus \{\rho\}}  \sum_{\rho''\in \{1,2,3,4\} \setminus \{\rho'\}} \lambda_{6;2;\rho\rho'}\lambda_{4;1;\rho''}\Big] 
 S^{12}
+
2\lambda_{4;1;\rho}\Big[ \sum_{\rho'\in \{1,2,3,4\} \setminus \{\rho\}}  \lambda_{6;2;\rho\rho'} 
\Big] [S^1 +S^{12}]
 + \cF_{4;\rho}(\lambda_{6;1};\lambda_{6;2}) \cr\cr
&&
=  - \lambda_{4;\rho}  
 + 
2 \Big[ 3\lambda_{6;1;\rho} \lambda_{4;1;\rho}+\sum_{\rho' \in \{1,2,3,4\} \setminus \{\rho\}} \lambda_{6;2;\rho\rho'}\lambda_{4;1;\rho'}
\Big] S^1  \cr\cr
&&
+2 \sum_{\rho'\in \{1,2,3,4\}\setminus \{\rho\}} \Big[ \lambda_{6;1;\rho}\lambda_{4;1;\rho' }  + \sum_{\rho''\in \{1,2,3,4\} \setminus \{\rho'\}} \lambda_{6;2;\rho\rho'}\lambda_{4;1;\rho''}\Big] 
 S^{12} \cr\cr\cr
&&
+
2\lambda_{4;1;\rho}\Big[ \sum_{\rho'\in \{1,2,3,4\} \setminus \{\rho\}}  \lambda_{6;2;\rho\rho'} 
\Big] [S^1 +S^{12}]
 + \cF_{4;\rho}(\lambda_{6;1};\lambda_{6;2})  \crcr
&&
\eea
where 
\bea
&&
\cF_{4;\rho}(\lambda_{6;1};\lambda_{6;2})  = 
  -2 \Big[ \lambda_{6;1;\rho} + \sum_{\rho'\in \{1,2,3,4\} \setminus\{\rho\}}\lambda_{6;2;\rho\rho'}\Big]\, S^0 
+ 2 \Big[3 \lambda_{6;1;\rho}^2 + \sum_{\rho'\in \{1,2,3,4\} \setminus \{\rho\}} \lambda_{6;2;\rho\rho'} \lambda_{6;1;\rho'} \Big] S^6  
\crcr
&+&  
\sum_{\rho'\in \{1,2,3,4\} \setminus \{\rho\}}
\Big[ 
\lambda_{6;2;\rho\rho'}^2
 +2\lambda_{6;1;\rho} \lambda_{6;1;\rho'} 
+ 2\lambda_{6;2;\rho\rho'}  \sum_{\rho''\in \{1,2,3,4\} \setminus \{\rho'\}}  \lambda_{6;1;\rho''}  \Big]S^{16} 
 \crcr
&+&
2 \Big[ \sum_{\rho',\rho''\in \{1,2,3,4\}\setminus \{\rho\};\;\;  \rho'' < \rho'} 
\lambda_{6;2;\rho\rho''} \lambda_{6;2;\rho'\rho''} \Big]  S^{18} 
+
\Big[
2\lambda_{6;1;\rho}   + \sum_{\rho' \in \{1,2,3,4\}\setminus \{\rho\}}\lambda_{6;2;\rho\rho'}    \Big]\Big[\sum_{\rho''\rho'''\in \{12,\dots,34\}}  \lambda_{6;2;\rho''\rho'''}\Big] S^{17} 
\crcr
&+&
\Big[ \sum_{\rho' \in \{1,2,3,4\}\setminus \{\rho\}}\lambda_{6;2;\rho\rho'} \Big]^2  [S^6 + 2  S^8] 
+ 
4\lambda_{6;1;\rho} \Big[  \sum_{\rho'\in \{1,2,3,4\} \setminus \{\rho\}} \lambda_{6;2;\rho\rho'} \Big][ S^6 +S^8 ] 
\crcr
&+&
\sum_{\rho'\in \{1,2,3,4\} \setminus \{\rho\}}
 \lambda_{6;2;\rho\rho'} \Big[ 2\lambda_{6;1;\rho} +
\sum_{\rho''\{1,2,3,4\} \setminus \{\rho'\}}\lambda_{6;2;\rho'\rho''} 
\Big] [ S^7 +S^8 ] \crcr
&+& \Bigg\{
\sum_{\rho' \in \{1,2,3,4\}\setminus \{\rho\}}  \lambda_{6;2;\rho\rho'}
\sum_{\rho''\rho'''=\{12,\dots,34\} / \; \rho'' \neq \rho'\; \text{and} \;\rho'''\neq \rho'} \lambda_{6;2;\rho''\rho'''} 
\crcr
&+& 
2\lambda_{6;1;\rho} \Big[ \sum_{\rho'\rho''=\{12,\dots,34\} / \; \rho' \neq \rho\; \text{and} \;\rho'' \neq \rho }\lambda_{6;2;\rho\rho'}\Big] 
\Bigg\} [ S^{17} +S^{18} ] 
\label{cf}
\eea

\end{document}